\documentclass[12pt,3p,sort&compress]{elsarticle}

\usepackage{graphicx}
\usepackage{bm} 
\usepackage{epsfig}
\usepackage[latin1]{inputenc}
\usepackage{float}
\usepackage{amsmath}
\usepackage{amssymb}
\usepackage{wrapfig}

\def\lsim{\mathrel{\rlap{\lower4pt\hbox{$\sim$}} \raise1pt\hbox{$<$}}}       
\def\gsim{\mathrel{\rlap{\lower4pt\hbox{$\sim$}}\raise1pt\hbox{$>$}}}       
\newcommand{\beq}{\begin{equation}}
\newcommand{\eeq}{\end{equation}}
\newcommand{\bqa}{\begin{eqnarray}}
\newcommand{\eqa}{\end{eqnarray}}
\newcommand{\phard}{p_{\rm hard}}
\newcommand{\nn}{\nonumber}

\journal{Nuclear Physics A}

\begin{document}


\begin{frontmatter}

\title{Thermal Bottomonium Suppression at RHIC and LHC}

\author[gettysburg,fias]{Michael Strickland}
\address[gettysburg]{
  Physics Department, Gettysburg College\\
  Gettysburg, PA 17325 United States
}
\address[fias]{
Frankfurt Institute for Advanced Studies\\
Ruth-Moufang-Strasse 1\\
D-60438, Frankfurt am Main, Germany
}
\author[gettysburg]{Dennis Bazow}

\begin{abstract}

In this paper we consider the suppression of bottomonium states in ultrarelativistic heavy ion collisions.
We compute the suppression as a function of centrality, rapidity, and transverse momentum for 
the states $\Upsilon(1s)$, $\Upsilon(2s)$, $\Upsilon(3s)$, $\chi_{b1}$, and $\chi_{b2}$.  Using this information,
we then compute the inclusive $\Upsilon(1s)$ suppression as a function of centrality, rapidity, and transverse 
momentum including feed down effects.  Calculations are performed for both RHIC $\sqrt{s_{NN}}=$ 200 GeV 
Au-Au collisions and LHC $\sqrt{s_{NN}}=$ 2.76 TeV Pb-Pb collisions. 
From the comparison of our theoretical results with data available from the STAR and CMS Collaborations we
are able to constrain the shear viscosity to entropy ratio to be in the range $0.08 < \eta/{\cal S} < 0.24$.  Our
results are consistent with the creation of a high temperature quark-gluon plasma at both RHIC and LHC 
collision energies.
\end{abstract}



\end{frontmatter}


\section{Introduction}
\label{sec:introduction}

The goal of ultrarelativistic heavy ion collision experiments at the Relativistic Heavy Ion Collider 
at Brookhaven National Laboratory (RHIC) and the Large Hadron Collider (LHC) at CERN is to create a 
tiny volume of matter ($\sim$ 1000 fm$^3$) which has been heated to a temperature exceeding that necessary
to deconfine quarks and gluons.  Lattice quantum chromodynamics (lattice QCD) measurements of 
the equation of state of strongly interacting matter \cite{Cheng:2007jq,Petreczky:2009at,Bazavov:2009zn,%
Borsanyi:2010cj,Bazavov:2010pg} show that there is crossover from hadronic matter to a quark-gluon 
plasma at temperatures on the order of 175 MeV which corresponds to approximately two trillion degrees
Kelvin.  For RHIC $\sqrt{s_{NN}} = 200$ GeV Au-Au collisions, initial maximum central temperatures of 
$T_0 \sim 450$ MeV were generated and for current LHC $\sqrt{s_{NN}} = 2.76$ TeV collisions one obtains 
$T_0 \sim 550$ MeV \cite{Schenke:2011tv}.  For the upcoming full energy LHC heavy ion runs with 
$\sqrt{s_{NN}} = 5.5$ TeV one expects $T_0 \sim $ 700 - 800 MeV.

At such extremely high temperatures strongly interacting matter makes a phase transition to a deconfined 
plasma of quarks and gluons and, as a result,  one expects the emergence of Debye screening of the 
interaction between quarks and gluons.  This leads to the dissolution of hadronic bound states 
\cite{Shuryak:1980tp}.  A particularly interesting subset of hadronic states consists of those which are 
comprised of heavy quarks because the spectrum of such states can be found using potential-based 
non-relativistic treatments.  Based on such potential models there were early predictions 
\cite{ Matsui:1986dk,Karsch:1987pv} that $J/\Psi$ production would be suppressed in heavy ion
collisions relative to the corresponding production in proton-proton collisions scaled by the number
of nucleons participating in the collision.

As mentioned above, heavy quarkonium has received the most theoretical attention, since heavy quark
states are dominated by short distance physics and can be treated using heavy quark effective theory.
Based on such effective theories of QCD, non-relativistic quarkonium states can be reliably
described. Their binding energies are much smaller than the quark mass $m_Q\gg\Lambda_{\rm
  QCD}$ ($Q=c,b$), and their sizes are much larger than $1/m_Q$. At zero
temperature, since the velocity of the quarks in the bound state is
small ($v\ll c$), quarkonium can be understood in terms of
non-relativistic potential models such as the Cornell potential which can
be derived directly from QCD using effective field theory 
\cite{Eichten:1979ms,Lucha:1991vn,Brambilla:2004jw}.
Using such non-relativistic potential models
studies of quarkonium spectral functions and meson current correlators
have been performed
\cite{Mocsy:2004bv,Wong:2004zr,Mocsy:2005qw,Cabrera:2006wh,%
Mocsy:2007jz,Alberico:2007rg,Mocsy:2007yj}. 
The results have been compared to first-principles lattice QCD calculations
\cite{Umeda:2002vr,Asakawa:2003re,%
Datta:2003ww,Aarts:2007pk,Hatsuda:2006zz,Jakovac:2006sf,%
Aarts:2010ek} which rely on the maximum entropy method 
\cite{Nakahara:1999vy,Asakawa:2002xj}.  Additionally, there have been
some lattice developments using non-relativistic lattice QCD \cite{Aarts:2011sm}.

Additionally, in recent years there has been an important theoretical development,
namely the first-principles calculation of the thermal widths of heavy quarkonium states
which emerge from imaginary-valued contributions to the
heavy quark potential.  The first calculation of the leading-order
perturbative imaginary part of the potential due to gluonic Landau damping was performed by Laine et 
al.~\cite{Laine:2006ns,Laine:2007gj}.
Since then an additional imaginary-valued contribution to the potential coming from singlet to
octet transitions has also been computed using the effective field theory approach~\cite{Brambilla:2008cx},
and lattice calculations have been performed in order to determine the imaginary part of the
heavy quark potential \cite{Rothkopf:2011db}.
These imaginary contributions to the potential are related to quarkonium decay processes in the
plasma.  The consequences of such imaginary parts on heavy quarkonium spectral functions 
\cite{Burnier:2007qm,Miao:2010tk}, perturbative thermal widths \cite{Laine:2006ns,Brambilla:2010vq},
quarkonium suppression in a T-matrix approach \cite{Grandchamp:2005yw,Rapp:2008tf,Riek:2010py},
and in stochastic real-time dynamics \cite{Akamatsu:2011se} have recently been studied; however, these 
studies were restricted to the case of an isotropicthermal plasma, which is only the case if one assumes 
ideal hydrodynamical evolution.

The calculation of the heavy quark potential has since been extended to the case of a plasma
with finite momentum-space anisotropy.   Both the real \cite{Dumitru:2007hy,Dumitru:2009ni}
and imaginary \cite{Burnier:2009yu,Dumitru:2009fy,Philipsen:2009wg} parts have been computed
in this case.  Additionally, the
impact of the imaginary part of the potential on the thermal widths of the states in both isotropic
and anisotropic
plasmas was recently studied~\cite{Margotta:2011ta}.  The consideration of momentum-space
anisotropic plasmas is necessary since, for any finite shear viscosity, the quark-gluon plasma possesses
local momentum-space anisotropies
\cite{Israel:1976tn,Israel:1979wp,Baym:1984np,Muronga:2003ta,%
Florkowski:2010cf,Ryblewski:2010bs,Ryblewski:2011aq,Martinez:2010sc,Martinez:2010sd}.
Depending on the magnitude of the shear viscosity, these momentum-space anisotropies can persist
for a long time and can be quite large, particularly at early times or near the edges of
the plasma.  This is true for both strong and weak coupling values of the shear viscosity and
the magnitude of the maximal momentum-space anisotropies increases with increasing
shear viscosity.  In fact, the magnitude of these momentum space anisotropies can become
so large that they call into doubt the reliability of the viscous hydrodynamical treatment which
implicitly assumes a nearly isotropic state.

This has motivated the development of a new dynamical formalism called ``anisotropic
hydrodynamics'' ({\sc aHydro}) which extends traditional viscous hydrodynamical treatments 
to cases in which the local momentum-space anisotropy of the plasma can be large
\cite{Florkowski:2010cf,Martinez:2010sc,Ryblewski:2010bs,Ryblewski:2011aq,Martinez:2010sd}.
The result is a dynamical framework that reduces to
2nd order viscous hydrodynamics for weakly anisotropic plasmas, but can
better describe highly anisotropic plasmas.
For one-dimensional dynamics which is homogeneous in the transverse directions, the
{\sc aHydro} approach provides the temporal and spatial rapidity evolution of the typical hard 
momentum of the plasma partons, $p_{\rm hard}$, and the plasma
anisotropy, $\xi$.  In a previous paper one of
us \cite{Strickland:2011mw} computed the thermal suppression of the $\Upsilon(1s)$ and
$\chi_{b1}$ states at LHC energies by folding together the
{\sc aHydro} temporal evolution of Ref.~\cite{Martinez:2010sd} with results obtained in 
Ref.~\cite{Margotta:2011ta} for the real and imaginary parts of the binding energy.
In this paper, we extend this study to compute the suppression of  $\Upsilon(1s)$, 
$\Upsilon(2s)$, $\Upsilon(3s)$, $\chi_{b1}$, and $\chi_{b2}$ states at both RHIC
and LHC energies.

The structure of the paper is as follows.  In Section~\ref{sec:potmodel} we introduce
the model potential we will use in order to compute the real and imaginary parts of the
binding energies of the states under consideration.  The potential used herein is an
improved version of the one used in Refs.~\cite{Strickland:2011mw} and 
\cite{Margotta:2011ta} and includes the effects of running coupling and an
improved parameterization of the numerical results for the short-range anisotropic 
potential.  In Section~\ref{sec:schrodinger}
we briefly review the numerical method used to solve the Schr\"odinger equation.
In Section~\ref{sec:dynmodel} we review the {\sc aHydro} dynamical model we use
and discuss the qualitative features we expect to emerge based on the resulting dynamical
evolution.  In Section~\ref{sec:initialconditions}
we present the initial conditions we will use which include Glauber (or participant) scaling and
a two-component model in which we use a linear combination of participant and binary collision
scaling.  In Section~\ref{sec:supfac} we describe how we compute the nuclear modification
factor $R_{AA}$ from the spatial and proper-time dependence of the real and imaginary parts
of the binding energy.  In Section~\ref{sec:feeddown} we detail how one can include the
effect of feed down from excited states to compute the inclusive or ``full'' nuclear modification
factor for the $\Upsilon(1s)$.  In Section~\ref{sec:results} we present our final results
as a function of centrality, rapidity, and transverse momentum.  Finally, in Section~\ref{sec:conclusions}
we present our conclusions and outlook for future work.

\section{Setup and Model Potential}
\label{sec:potmodel}

In this section we specify the two potential models we consider in this paper.
We consider the general case of a quark-gluon plasma which is anisotropic in 
momentum space.  In the limit that the plasma is isotropic, the real 
part of the potentials used here reduces to the potential originally introduced by Karsch,
Mehr, and Satz (KMS) \cite{Karsch:1987pv} with or without an additional entropy contribution \cite{Dumitru:2009ni}
and the imaginary part reduces to the result originally obtained by Laine et al \cite{Laine:2006ns}.  
To begin the discussion we first introduce
our ansatz for the one-particle distribution function subject to a momentum-space
anisotropy.

\subsection{Momentum-space anisotropic plasma}
\label{subsec:aniso}

The phase-space distribution of gluons in the local rest frame is assumed to be given by the
following ansatz~\cite{Romatschke:2003ms, Mrowczynski:2004kv,Romatschke:2004jh,Schenke:2006fz,%
Dumitru:2007hy}
\begin{equation}
f(t,{\bf x},{\bf p}) = f_{\rm iso}\left(\sqrt{{\bf p}^2+\xi({\bf p}\cdot{\bf
n})^2 }  / p_{\rm hard} \right) ,  \label{eq:f_aniso}
\end{equation}
where $f_{\rm iso}$ is an isotropic distribution which in thermal equilibrium
is given by a Bose-Einstein distribution, $\xi$ is the momentum-space 
anisotropy parameter, and $p_{\rm hard}$ is a momentum scale which specifies the typical momentum
of the particles in the plasma and can be identified with the temperature
in the limit of thermal isotropic ($\xi\!=\!0$) equilibrium.\footnote{The only place that we will assume 
thermal equilibrium herein is in the value of the isotropic Debye mass used in the heavy quark potential 
in Section~\ref{sssec:fullpot}.  In principle, one could use another isotropic distribution
function, in which case one would need to recompute the isotropic Debye mass.}  
The two parameters $p_{\rm hard}$ and $\xi$
can, in general, depend on 
proper time and position; however, we do not indicate this explicitly for
compactness of the notation.  The ansatz above is
the simplest ansatz which allows for the breaking of symmetry in the $p_T$-$p_L$
plane while maintaining local azimuthal symmetry in the transverse directions
in momentum space.  
Note that one can use the same distribution to describe the quarks in the system 
\cite{Romatschke:2003ms, Mrowczynski:2004kv,Romatschke:2004jh} and the
quark self-energy in this case has been computed explicitly \cite{Schenke:2006fz}.  
For our purpose, we are primarily interested in the gluon distribution since this will
enter into the determination of the heavy quark potential; however, in
the section on dynamics we implicitly assume the same distribution for
quark degrees of freedom.

Such a breaking of symmetry in the $p_T$-$p_L$ plane arises
naturally in a heavy-ion collision due to the rapid longitudinal expansion of the 
matter along the beamline direction and the parameter $\xi$ quantifies 
the degree of momentum-space anisotropy,
\beq
\xi = \frac{1}{2} \frac{\langle {\bf p}_\perp^2\rangle}
{\langle p_z^2\rangle} -1~,
\eeq
where $p_z\equiv \bf{p\cdot n}$ and ${\bf p}_\perp\equiv {\bf{p-n
(p\cdot n)}}$ denote the particle momenta along and perpendicular to
the direction ${\bf n}$ of anisotropy, respectively.  For heavy ion collisions
the anisotropy vector, ${\bf n}$, lies along the beamline direction which
we generally choose to lie along the $z$-axis.

The energy-momentum tensor $T^{\mu\nu}(t,{\bf x},{\bf p})= 
(2\pi)^{-3}\,\int d^3{\bf p}/p^0\, p^\mu p^\nu f(t,{\bf x},{\bf p})$ for the 
distribution function~(\ref{eq:f_aniso}) is diagonal in the comoving frame and its components 
are~\cite{Martinez:2010sc,Martinez:2009ry}
\begin{subequations}
\label{momentsanisotropic}
\begin{align}
\label{energyaniso}
{\cal E}(p_{\rm hard},\xi) &= T^{\tau\tau} \;= \frac{1}{2}\left(\frac{1}{1+\xi}
+\frac{\arctan\sqrt{\xi}}{\sqrt{\xi}} \right) {\cal E}_{\rm iso}(p_{\rm hard}) \; , \\ \nonumber
&\equiv{\cal R}(\xi)\,{\cal E}_{\rm iso}(p_{\rm hard})\, ,\\
\label{transpressaniso}
{\cal P}_T(p_{\rm hard},\xi) &= \frac{1}{2}\left( T^{xx} + T^{yy}\right) 
= \frac{3}{2 \xi} 
\left( \frac{1+(\xi^2-1){\cal R}(\xi)}{\xi + 1}\right)
 {\cal P}_{\rm iso}(p_{\rm hard}) \, , 
\\ \nonumber
&\equiv{\cal R}_{\rm T}(\xi){\cal P}_{\rm iso}(p_{\rm hard})\, , \\
\label{longpressaniso}
{\cal P}_L(p_{\rm hard},\xi) &= - T^{\varsigma}_\varsigma= \frac{3}{\xi} 
\left( \frac{(\xi+1){\cal R}(\xi)-1}{\xi+1}\right) {\cal P}_{\rm iso}(p_{\rm hard}) \; ,\\ \nonumber
&\equiv {\cal R}_{\rm L}(\xi){\cal P}_{\rm iso}(p_{\rm hard})\, ,
\end{align}
\end{subequations}
where ${\cal P}_{\rm iso}(p_{\rm hard})$ is the isotropic 
pressure and ${\cal E}_{\rm iso}(p_{\rm hard})$ is the isotropic energy density.
In everything that follows we will use a conformal equation of state for 
which ${\cal E}_{\rm iso} = 3{\cal P}_{\rm iso}$.

\subsection{Model potential}

In this subsection we first review the derivation of the short range screened heavy-quark potential in the
presence of finite momentum-space anisotropy.  The full complex potential for an isotropic plasma was first 
obtained in Refs.~\cite{Laine:2006ns,Laine:2007gj}.  The calculation of the real part of the potential at finite
anisotropy was first obtained in Ref.~\cite{Dumitru:2009ni} and was later extended to include the imaginary 
part in  Refs.~\cite{Burnier:2009yu,Dumitru:2009fy,Philipsen:2009wg}.  After this brief review we construct an 
analytic approximation to the real part of the heavy quark potential which allows us to compute the potential 
efficiently without having to resort to complicated two-dimensional numerical integration.  As we will
show, the resulting analytic approximation for the real part can be cast into the form of a Debye-screened
Coulomb potential with a Debye mass which depends on the relative angle of the line connecting the quark
and antiquark to the anisotropy direction.

\subsubsection{Integral expression for the real part of the short range potential}

One can determine the real part of the heavy-quark potential in the non-relativistic limit 
from the Fourier transform of the static gluon propagator.  In an anisotropic plasma with a distribution
function given by Eq.~(\ref{eq:f_aniso}) at leading order in the strong coupling constant
one finds \cite{Dumitru:2009ni}
\begin{eqnarray}
V({\bf{r}},\xi) &=& -g^2 C_F\int \frac{d^3{\bf{p}}}{(2\pi)^3} \,
e^{i{\bf{p \cdot r}}}\Delta^{00}(\omega=0, \bf{p},\xi) \, , \\
&=& -g^2 C_F\int \frac{d^3{\bf{p}}}{(2\pi)^3} \,
e^{i{\bf{p \cdot r}}} \frac{{\bf{p}}^2+m_\alpha^2+m_\gamma^2}
 {({\bf{p}}^2 + m_\alpha^2 +
     m_\gamma^2)({\bf{p}}^2+m_\beta^2)-m_\delta^4}~, \label{eq:FT_D00}
\end{eqnarray}
where $g$ is the strong coupling constant and $C_F = (N_c^2-1)/(2 N_c)$ is the quadratic Casimir
of the fundamental representation of $SU(N_c)$.  The masses in (\ref{eq:FT_D00}) are given by 
\cite{Dumitru:2009ni}
\begin{eqnarray}
m_\alpha^2&=&-\frac{m_D^2}{2 p_\perp^2 \sqrt{\xi}}%
\left(p_z^2 {\rm{arctan}}{\sqrt{\xi}}-\frac{p_{z} {\bf{p}}^2}{\sqrt{{\bf{p}}^2+\xi p_\perp^2}}%
{\rm{arctan}}\frac{\sqrt{\xi} p_{z}}{\sqrt{{\bf{p}}^2+\xi p_\perp^2}}\right) \; , \\
m_\beta^2&=&m_{D}^2
\frac{(\sqrt{\xi}+(1+\xi){\rm{arctan}}{\sqrt{\xi}})({\bf{p}}^2+\xi p_\perp^2)+\xi p_z\left(%
p_z \sqrt{\xi} + \frac{{\bf{p}}^2(1+\xi)}{\sqrt{{\bf{p}}^2+\xi p_\perp^2}} %
{\rm{arctan}}\frac{\sqrt{\xi} p_{z}}{\sqrt{{\bf{p}}^2+\xi p_\perp^2}}\right)}{%
2  \sqrt{\xi} (1+\xi) ({\bf{p}}^2+ \xi p_\perp^2)} \; , \nonumber \\
&& \\
m_\gamma^2&=&-\frac{m_D^2}{2}\left(\frac{{\bf{p}}^2}{\xi p_\perp ^2+{\bf{p}}^2}%
-\frac{1+\frac{2p_z^2}{p_\perp^2}}{\sqrt{\xi}}{\rm{arctan}}{\sqrt{\xi}}+\frac{
p_z{\bf{p}}^2(2{\bf{p}}^2+3\xi p_\perp^2)}{\sqrt{\xi}(\xi
p_\perp^2+{\bf{p}}^2)^{\frac{3}{2}}
p_\perp^2}{\rm{arctan}}\frac{\sqrt{\xi}
p_{z}}{\sqrt{{\bf{p}}^2+\xi p_\perp^2}}\right) \; , \nonumber \\
&& \\
m_\delta^2&=&-\frac{\pi m_D^2\xi p_z p_\perp |{\bf{p}}|}{4(\xi
p_\perp^2+{\bf{p}}^2)^{\frac{3}{2}}}\, ,
\end{eqnarray}
with $m_D$ being the isotropic Debye mass 
\beq \label{eq:Debye}
m_D^2 = -\frac{g^2}{2\pi^2} \int_0^\infty d p \,
  p^2 \, \frac{d f_{\rm iso}}{d p} ~,
\eeq
and $p^2 \equiv {\bf p}^2 = p_\perp^2 +p_z^2$.  The above 
expressions apply when ${\bf n}=(0,0,1)$ points along the $z$-axis; in the general case, $p_z$ and 
${\bf p}_\perp$ get replaced by $\bf{p\cdot n}$ and $\bf{p-n (p\cdot n)}$, respectively.  One can
factorize the denominator of (\ref{eq:FT_D00}) by introducing 
\beq
2 m_{\pm}^2 \equiv M^2 \pm \sqrt{M^4-4(m_\beta^2(m_\alpha^2+m_\gamma^2)-m_\delta^4)} \; ,
\label{mpm}
\eeq
with $ M^2 \equiv m_\alpha^2+m_\beta^2+m_\gamma^2$ \cite{Romatschke:2003ms}.  This
allows us to write
\beq
V({\bf{r}},\xi) = -g^2 C_F\int \frac{d^3{\bf{p}}}{(2\pi)^3} \,
e^{i{\bf{p \cdot r}}} \frac{{\bf{p}}^2+m_\alpha^2+m_\gamma^2}
 {({\bf{p}}^2 + m_+^2)({\bf{p}}^2 + m_-^2)}~. \label{eq:FT_D00_factorized}
\eeq
In general one must evaluate (\ref{eq:FT_D00_factorized}) numerically.  The integration can be
reduced to a two-dimensional integral over a polar angle, $\theta$, and the length of the 
three-momentum, $p$.  However, there can be poles in the integration domain due to the 
fact that $m_-^2$ can be negative for certain polar angles and momenta 
\cite{Romatschke:2003ms}.\footnote{This is related to the presence of unstable collective modes in 
momentum-space anisotropic plasmas.}
These poles are first order and can dealt with using a principle-part prescription, 
however, evaluating this integral with the necessary precision requires on the order of 0.5 to 1 
seconds per point.  This presents a fundamental problem if one intends to evaluate the potential
when solving the Schr\"odinger equation on large spatial lattices with on the order of $512^3$ points.  
We are, therefore, motivated to find an efficient parametrization of the resulting potential 
based on a finite set of numerical evaluations.  In order to do so, it is necessary to first consider
various asymptotic limits of the potential.

\subsubsection{Asymptotic limits of the real part of the short range potential}

When $\xi=0$ then $m_\beta=m_+=m_D$  and all other mass scales are zero. As a consequence, we
recover the isotropic Debye-screened Coulomb potential
\beq
\lim_{\xi\to0} V({\bf{r}},\xi) = V_{\rm iso}(r) = -g^2 C_F\int \frac{d^3{\bf{p}}}{(2 \pi)^3}
\frac{e^{i{\bf{p\cdot r}}}}{{\bf{p}}^2+m_D^2} =
- \frac{g^2 C_F}{4 \pi r} \, e^{-\hat{r}}~,  \label{eq:V_iso}
\eeq
where $\hat{r}\equiv rm_D$.

In the limit $r\to0$ for arbitrary $\xi$ one finds that the potential reduces to the vacuum Coulomb 
potential \cite{Dumitru:2009ni}
\beq
\lim_{{\bf{r}}\to0} V({\bf{r}},\xi) =
V_{\rm vac}(r) = -g^2 C_F\int \frac{d^3{\bf{p}}}{(2 \pi)^3}
\frac{e^{i{\bf{p\cdot r}}}}{{\bf{p}}^2}
= - \frac{g^2 C_F}{4 \pi r} ~.  \label{eq:V_vac}
\eeq
The same potential emerges for extreme anisotropy since all $m_i\to0$
as $\xi\to\infty$:
\begin{equation}
\lim_{\xi\to\infty} V({\bf{r}},\xi)= V_{\rm vac}(r) ~.   \label{eq:V_xi=inf}
\end{equation}
This is due to the fact that at $\xi=\infty$ the phase
space density $f(\bf{p})$ from Eq.~(\ref{eq:f_aniso}) has support only
in a two-dimensional plane orthogonal to the direction $\bf{n}$ of
anisotropy. As a consequence, the density of the medium vanishes in
this limit.

\subsubsection{Subleading terms in the small $\xi$ limit}

Having discussed the leading terms in the limits show above, we now discuss the subleading
terms in the small $\xi$ limit.  In the limit of small $\xi$ one finds that \cite{Romatschke:2003ms}
\bqa
\hat m_+^2 &=& 1 + \frac{\xi}{6}(3 \cos 2\theta-1) \; , \nonumber \\
\hat m_-^2 &=& \hat m_\alpha^2 + \hat m_\gamma^2 = -\frac{\xi}{3}\cos2\theta \; ,
\eqa
where $\hat m \equiv m/m_D$ and $\theta$ is the angle with respect to the anisotropy vector ${\bf n}$.
As a result, one finds that 
\beq
\lim_{\xi \to 0} V({\bf{r}},\xi) = -g^2 C_F\int \frac{d^3{\bf{p}}}{(2\pi)^3} \,
\frac{e^{i{\bf{p \cdot r}}} }
 {{\bf{p}}^2 + \nu^2}~, \label{eq:FT_D00_factorized_smallxi}
\eeq
where $\nu \equiv m_D [ 1 + \frac{\xi}{6}(3 \cos 2\theta-1) ]$.  Expanding the
integrand to leading order in $\xi$ and evaluating the resulting integrals one finds
\cite{Dumitru:2009ni}
\beq \label{eq:anisoPotlin_xi}
\lim_{\xi \to 0} V({\bf{r}},\xi) = {V}_{\rm iso}(r) \left[1-\xi {\cal F}(\hat{r},\theta) \right]~,
\eeq
where ${V}_{\rm iso}(r)$ is the Debye-screened Coulomb potential
in an isotropic medium (\ref{eq:V_iso}), and the function
${\cal F}(\hat{r},\theta) \equiv f_0(\hat{r})+f_1(\hat{r})\cos(2\theta)$ with
\begin{eqnarray}
f_0(\hat{r}) &=&
  \frac{6(1-e^{\hat{r}})+\hat{r}[6-\hat{r}(\hat{r}-3)]}{12\hat{r}^2}
= -\frac{\hat{r}}{6}-\frac{\hat{r}^2}{48}+\cdots~,\label{eq:f0}\\
f_1(\hat{r}) &=&
  \frac{6(1-e^{\hat{r}})+\hat{r}[6+\hat{r}(\hat{r}+3)]}{4\hat{r}^2}
= -\frac{\hat{r}^2}{16}+\cdots~.
\label{eq:f1}
\end{eqnarray}
We can now define a $\theta$-dependent screening mass in an
anisotropic medium as the inverse of the distance scale $r_{\rm
med}(\theta)$ over which $|rV(r)|$ drops by a factor of $e$:
\beq
\log \frac{V_{\rm vac}(r_{\rm med})}{V(r_{\rm med},\theta;\xi,T)}
= 1~. \label{eq:logV_rmed} 
\eeq
To leading order in $\xi$ this leads to $\hat{r}_{\rm med} = 
1-\xi {\cal F}(\hat{r}_{\rm med},\theta) \label{eq:rmed_xi}$.  
An approximate solution to this last equation
gives \cite{Dumitru:2009ni}
\beq 
\lim_{\xi \to 0}  \frac{\mu}{m_D} \simeq 1-\xi
\frac{3+\cos 2\theta}{16}~, \label{eq:mu_smallxi}
\eeq
where $\mu = r_{\rm med}^{-1}$.

With this in hand we have an analytic approximation
for the potential in the limit of small $\xi$, namely, that it is approximately
a Debye-screened Coulomb potential with a $\theta$-dependent screening
mass given by Eq.~(\ref{eq:mu_smallxi})  such that
\beq
\lim_{\xi \to 0} V({\bf{r}},\xi) \simeq V_{\rm iso}(r) = - \frac{g^2 C_F}{4 \pi r} \, e^{-\mu r}~,  \label{eq:V_iso_smallxi_model}
\eeq

\subsubsection{Subleading terms in the large $\xi$ limit}

We now turn our attention to the limit of large $\xi$.  For general $\xi$ one can show 
that in an anisotropic plasma with a distribution function given by Eq.~(\ref{eq:f_aniso}) the 
particle number density can be factorized using a simple change of variables
\beq
n(\phard,\xi) = \int \frac{d^3{\bf{p}}}{(2\pi)^3} f(t,{\bf x},{\bf p}) = 
\frac{n_{\rm iso}(\phard)}{\sqrt{1+\xi}} ~ ,
\eeq
where $n_{\rm iso}$ is the number density that would be obtained using the isotropic
distribution function used in Eq.~(\ref{eq:f_aniso}).  Since in an isotropic system one can estimate the 
screening mass via $m_D^2 \sim n/T$, we expect that in the large-$\xi$ limit one can will obtain 
$\mu^2 \propto n(\phard,\xi)/\phard$ for the anisotropic screening mass, which leads to 
$\mu \sim \xi^{-1/4}m_D$ in the large-$\xi$ limit.  To see how this emerges analytically we
return to the defining equation for the potential given in Eq.~(\ref{eq:FT_D00}).  In the limit of
large $\xi$ one finds\,\footnote{Note that the second integral below is infrared divergent and needs
to be regulated; however, since we will only compare the coefficients of such integrals, we do not
need to specify how it is regulated as long as we regulate it in the same manner in each case.}
\beq
\lim_{\xi \to \infty} V({\bf{r}},\xi) = V_{\rm vac}(r)  -  \frac{\pi}{4} \frac{g^2 C_F m_D^2}{\sqrt{\xi}}
\int \frac{d^3{\bf{p}}}{(2\pi)^3} \,\frac{e^{i{\bf{p \cdot r}}}}{{\bf p}^4} \, .
\eeq
We can compare this to the small screening-mass expansion of the isotropic potential Debye-screened 
Coulomb potential 
\beq
\lim_{\mu \to 0} V_{\rm iso}({\bf{r}})\!\mid_{m_D \to \mu} = V_{\rm vac}(r)  -  g^2 C_F \mu^2
\int \frac{d^3{\bf{p}}}{(2\pi)^3} \,\frac{e^{i{\bf{p \cdot r}}}}{{\bf p}^4} \, .
\eeq
From the comparison we see that the anisotropic case can be obtained if we identify
\beq
\lim_{\xi \to \infty} \frac{\mu}{m_D} \simeq \frac{\sqrt{\pi}}{2} \xi^{-1/4} \, .
\label{eq:largeximu}
\eeq

\subsubsection{Model for the real part of the short range potential}

\begin{figure}[t]
\begin{center}
\includegraphics[width=16.4cm]{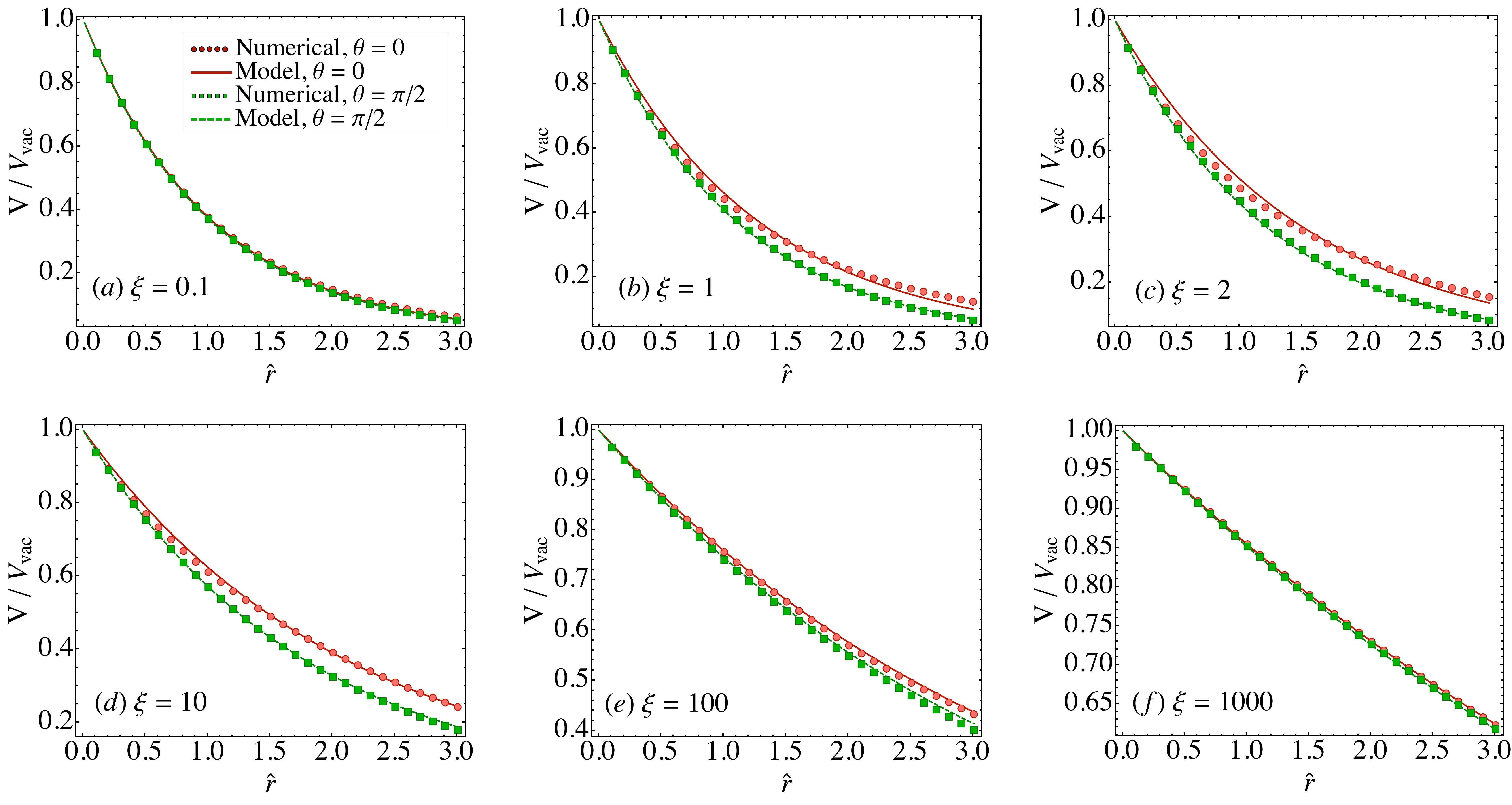}
\end{center}
\vspace{-6mm}
\caption{
Comparison of the real part of the short range potential obtained from the analytic model specified in 
Eq.~(\ref{eq:muparam}) and via direct numerical integration of Eq.~(\ref{eq:FT_D00_factorized}).  Panels
(a)-(f) show the potential for different values of the anisotropy parameter as indicated in the lower left
corner of each panel.  In each panel the potential has
been scaled by the vacuum Coulomb potential.  Note that the vertical scale changes between panels.
}
\label{fig:potcomp}
\end{figure}

\begin{figure}[t]
\begin{center}
\includegraphics[width=16.4cm]{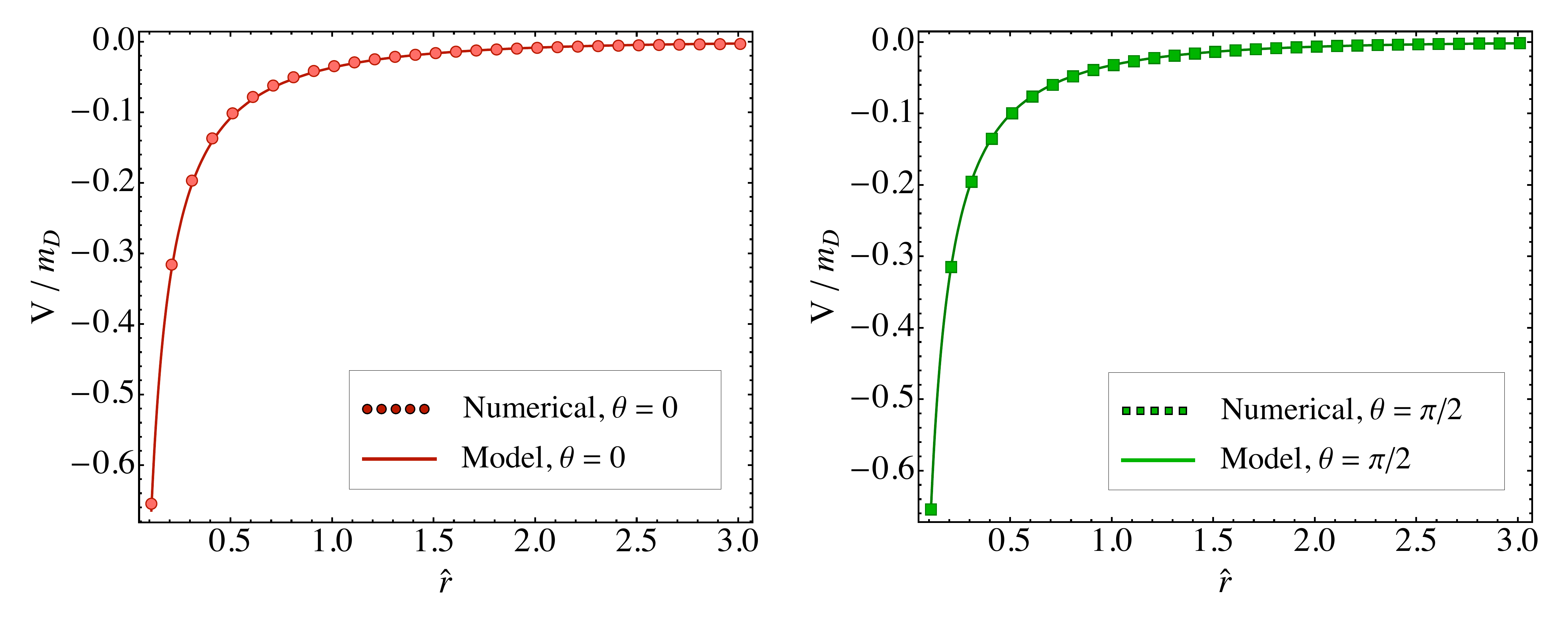}
\end{center}
\vspace{-6mm}
\caption{
Comparison of the real part of the short range potential obtained from the analytic model specified in 
Eq.~(\ref{eq:muparam}) and via direct numerical integration of Eq.~(\ref{eq:FT_D00_factorized}) for
$\xi=1$.
}
\label{fig:potcomp2}
\end{figure}

With both the small- and large-$\xi$ limits of the anisotropic screening mass in hand we can construct a model
for the real part and then compare to direct numerical evaluation of the potential via 
Eq.~(\ref{eq:FT_D00_factorized}).  We find that the following form works well to reproduce the $r$ 
dependence of the potential for all $\xi$.
\begin{equation}
\left(\frac{\mu}{m_D}\right)^{-4} =  
1 + \xi\left(a - \frac{2^b(a-1)+(1+\xi)^{1/8}}{(3+\xi)^b}\right) 
\left(1 + \frac{c(\theta) (1+\xi)^d}{(1+ e\xi^2)} \right) \, ,
\label{eq:muparam}
\end{equation}
with $a=16/\pi^2$, $b = 1/2$, $d=3/2$, $e=1/3$, and
\beq
c(\theta) = \frac{3 \pi ^2 \cos (2 \theta)+\left(9+4 \sqrt{3}-4 \sqrt{6}\right) \pi ^2+64 \left(\sqrt{6}-3\right)}{4
   \left(\sqrt{3} \left(\sqrt{2}-1\right) \pi ^2-16 \left(\sqrt{6}-3\right)\right)} \, .
\eeq
The value of the parameter $a$ in (\ref{eq:muparam}) guarantees that the large-$\xi$ form for the 
anisotropic screening mass (\ref{eq:largeximu}) is reproduced.  The expression for $c(\theta)$ is 
determined by requiring that the small-$\xi$ limit (\ref{eq:mu_smallxi}) is reproduced.  The 
coefficients $b$, $d$, and $e$ were fit by hand in order to optimally reproduce the anisotropic
short-range potential obtained by direct numerical integration.  In addition to reproducing these limits,
the form (\ref{eq:muparam})  also guarantees that $\mu/m_D \sim (1+\xi)^{-1/4}$ in the infinitely
prolate limit of $\xi \to -1$.  We emphasize that the form  (\ref{eq:muparam}) is only a parametrization
of the numerical results which is constructed in such a way as to guarantee the necessary asymptotic limits
and to efficiently reproduce the potential obtained via direct numerical evaluation in an efficient manner.  With
this parametrization of $\mu$ in hand we can construct a model for the real part of the 
short range potential for all $\xi$:
\beq
\Re[V(r)] = - \frac{g^2 C_F}{4 \pi r} \, e^{-\mu r}~,  \label{eq:V_iso_shortrange_model}
\eeq
with $\mu$ given by Eq.~(\ref{eq:muparam}).

In Fig.~\ref{fig:potcomp} we compare the model specified by Eq.~(\ref{eq:V_iso_shortrange_model}) 
with  results obtained by direct numerical integration for $\xi \in \{ 0.1,
1,2,10,100,1000\}$ by plotting the ratio of the potential over the vacuum Coulomb potential.  This
is a very sensitive test of whether or not the parametrization is a good one and as can been seen from
Fig.~\ref{fig:potcomp} works well over a very large range of possible plasma anisotropies.  
To see what the actual unscaled potential looks like in Fig.~\ref{fig:potcomp2} we show the potential
again; however, this time, we do not scale by the vacuum Coulomb potential.  As can be seen from this
figure, the model specified by  Eq.~(\ref{eq:V_iso_shortrange_model}) works extremely well allowing
us to express the short-range anisotropic quarkonium potential as a Debye-screened Coulomb 
potential with an anisotropic screening mass $\mu$.
In the following subsection we will discuss the fact that one needs to model the long-range potential and
construct a model for the potential at all scales.

\subsubsection{Model for the real part of the potential at all scales}
\label{sssec:fullpot}

In order to make a realistic phenomenological model for quarkonium states one must consistently describe 
both short and long distance scales.  Since heavy quark states are dominated by short distance physics at zero
temperature they can be treated using heavy quark effective theory; however, as the temperature increases
one expects the size of the states to increase causing the states to become sensitive to the long range part
of the potential.  At zero temperature, since the velocity of the quarks in 
the bound state is small, quarkonium can be understood in terms of non-relativistic potential 
models such as the Cornell potential which can be derived directly from QCD using effective field theory 
\cite{Eichten:1979ms,Lucha:1991vn,Brambilla:2004jw}.  A finite temperature extension of the Cornell 
potential might be provided by the KMS model~\cite{Karsch:1987pv} which describes the free energy of a static 
heavy quark-antiquark pair in an isotropic plasma via \cite{Dumitru:2009ni,Petreczky:2005bd}
\beq 
F(r,T) = -\frac{g^2 C_F }{4 \pi r} e^{-m_D r} + \frac{\sigma}{m_D}\left[1-e^{-m_D \, r  }\right]\, ,
\label{eq:KMS_free}
\eeq
where $g$ is the strong coupling constant, $\sigma$ is the string tension, and $m_D$ is the isotropic 
Debye screening mass.  Eq.~(\ref{eq:KMS_free}) is a model for the action of a Wilson loop of size $1/T$ and 
$r$ in the temporal and spatial directions, respectively (see~\cite{Petreczky:2005bd} and references 
therein).   In the 
interest of spanning the possibilities for the real  part of the potential we define
potential model A  by equating the real part of the potential with the free energy given in Eq.~(\ref{eq:KMS_free}).
However, in the general anisotropic case we must replace the isotropic screening mass by the anisotropic screening 
mass (\ref{eq:muparam}) to obtain
\beq
\Re[V_A]=  F = -\frac{a}{r} e^{-\mu r} + \frac{\sigma}{\mu}\left[1-e^{-\mu \, r  }\right]\, ,
\label{eq:real_pot_model_A}
\eeq
where we have replaced $g^2 C_F/4 \pi$ by a phenomenological parameter $a$ in the screened coulomb
contribution which will 
be adjusted to match lattice data. Here we take $a=0.385$ which is consistent with the short range part
of the heavy quark potential measured on the lattice \cite{Petreczky:2010yn}.  For the isotropic Debye mass, $m_D$, we use 
$m_D^2 = (1.4)^2 \cdot N_c (1+N_f/6)  \, 4  \pi \alpha_s  \, p_{\rm hard}^2/3$.  The isotropic leading-order Debye mass is adjusted
by a factor of $(1.4)^2$ in order to take into account higher-order corrections which have been measured in lattice simulations
\cite{Kaczmarek:2004gv}.  In
the isotropic Debye mass we use a three-loop running for $\alpha_s$ \cite{Nakamura:2010zzi} 
with $\Lambda_{\overline{MS}} =$ 344 MeV which gives $\alpha_s({\rm 5\;GeV}) = 0.2034$ in accordance 
with recent high precision lattice measurements of the running coupling constant~\cite{McNeile:2010ji}.
For the scale of the running coupling we use $2\pi T$ which is consistent with hard thermal loop calculations
of quark-gluon plasma thermodynamics \cite{Andersen:2010wu,Andersen:2011sf}.
Finally, for the string tension we use a value of $\sigma = 0.223\;{\rm GeV}^2$ which is again obtained from
fits to lattice data \cite{Petreczky:2010yn}.  In all cases we use $N_c=3$ since we are modeling QCD and 
take the number of contributing light quark flavors to be $N_f=2$, which is appropriate for the temperature range considered 
herein.\footnote{If one uses instead $N_f=3$ the isotropic Debye mass increases by $\sim$ 6\% which has only a small effect on the final results.}

As potential model B we will use the internal energy, $U$,  of the states which has an entropy contribution added to it.  
To achieve this we calculate the full entropy $S=- \partial F/\partial T$ using (\ref{eq:real_pot_model_A}) 
and add $T$ times this to the free energy (\ref{eq:real_pot_model_A}), which leads to the internal energy 
$U=F+TS$. This procedure gives model B for the real part of the heavy quark potential
\begin{eqnarray}
\Re[V_B] &=& U = F - T\frac{\partial F}{\partial T} \label{eq:KMSpotFE} \, ,\\
&=& -\frac{a}{r} \left(1+\mu \, r\right) e^{-\mu \, r }
+ \frac{2\sigma}{\mu}\left[1-e^{-\mu \, r }\right]
- \sigma \,r\, e^{-\mu \, r }\, , \label{eq:real_pot_model_B}
\end{eqnarray}
with $\mu$ given by Eq.~(\ref{eq:muparam}).
In potential model B, we use the same parameters and Debye mass 
prescription as used in potential model A.

\subsubsection{Model for the imaginary part of the potential}

The imaginary part of the potential $\Im[V]$ is obtained from a leading order perturbative calculation 
which was performed in the small anisotropy limit \cite{Dumitru:2009fy}.  The resulting imaginary part
of the potential is
\begin{equation} 
\Im[V] = - \alpha_s C_F T \biggl[ \phi(\hat{r}) - \xi \left(\psi_1(\hat{r},
\theta)+\psi_2(\hat{r}, \theta)\right)\biggr] ,
\label{impot}
\end{equation}
where $\hat r = m_D r$, $\alpha_s = g^2/(4 \pi)$, $C_F= (N_c^2-1)/(2N_c)$, and
\begin{eqnarray}
 \phi(\hat{r}) &=& 2\int_0^{\infty}dz \frac{z}{(z^2+1)^2} \left[1-\frac{\sin(z\, \hat{r})}{z\, \hat{r}}\right]~, \\
 \psi_1(\hat{r}, \theta) &=& \int_0^{\infty} dz
 \frac{z}{(z^2+1)^2}\left(1-\frac{3}{2}
 \left[\sin^2\theta\frac{\sin(z\, \hat{r})}{z\, \hat{r}}
 +(1-3\cos^2\theta)G(\hat{r}, z)\right]\right), \\
 \psi_2(\hat{r}, \theta) &=&- \int_0^{\infty} dz
\frac{\frac{4}{3}z}{(z^2+1)^3}\left(1-3 \left[
  \left(\frac{2}{3}-\cos^2\theta \right) \frac
 {\sin(z\, \hat{r})}{z\, \hat{r}}+(1-3\cos^2\theta)
 G(\hat{r},z)\right]\right),\nonumber\\
\label{funcs}
\end{eqnarray}
with $\theta$ being the angle from the beam direction and
\begin{equation}
 G(\hat{r}, z)= \frac{\hat{r} z\cos(\hat{r} z)- \sin(\hat{r} z)
 }{(\hat{r} z)^3}~.
\label{gdef}
\end{equation}
For numerical efficiency three separate analytic expressions for $\Im[V]$ which are 
valid in the small, medium, and large distance limits were determined and used in a piecewise fashion 
in their respective radii of convergence.

\subsubsection{Final Potential Models}

As mentioned above, here we consider two potential models, A and B, in which we identify the potential as
coming from the free energy or internal energy, respectively.  From both models discussed above
we will additionally subtract a temperature- and spin-independent finite quark mass correction taken from 
Ref.~\cite{Bali:1997am} which improves the description of charm quark states at low temperatures, but is
a small correction for bottom quarks.
The final result for potential model A is
\beq
V_A = \Re[V_A] + i \Im[V] -  \frac{0.8\,\sigma}{m_Q^2 r} \, , \quad \quad {\rm Model\;A}
\label{eq:potmodela}
\eeq
with $\Re[V_A]$ given by Eq.~(\ref{eq:real_pot_model_A}) and $\Im[V]$ given by Eq.~(\ref{impot}).
The final result for potential model B is
\beq
V_B = \Re[V_B] + i \Im[V] -  \frac{0.8\,\sigma}{m_Q^2 r} \, ,  \quad \quad {\rm Model\;B}
\label{eq:potmodelb}
\eeq
with $\Re[V_B]$ given by Eq.~(\ref{eq:real_pot_model_B}) and $\Im[V]$ given by Eq.~(\ref{impot}).  We
note that both  $\Re[V_A]$ and  $\Re[V_B]$ reduce to the Cornell potential at $T=0$ and the short range part
($r \ll 1/m_D$ and $r \ll 1/\sqrt{\sigma}$) of both reduces to the Coulomb potential, $V = -a/r$, at all 
temperatures, with $a$ constrained by lattice data \cite{Petreczky:2010yn}.

\section{Solving the 3d Schr\"odinger Equation}
\label{sec:schrodinger}

To solve the resulting Schr\"odinger equation we
use the finite difference time domain method~\cite{Sudiarta:2007,Strickland:2009ft} extended
to the case of a complex-valued potential~\cite{Margotta:2011ta}.  
Here we briefly review the technique.  
To determine the wave functions of bound quarkonium states, we must solve
the time-independent Schr\"odinger equation
\bqa
\hat{H} \phi_\upsilon({\bf x}) &=& E_\upsilon \, \phi_\upsilon({\bf
x})  ~, \nonumber \\
\hat{H} &=& -\frac{\nabla^2}{2 m_R} + V({\bf x}) + m_1 + m_2~,
\label{3dSchrodingerEQ}
\eqa
on a three-dimensional lattice in coordinate space with the potential
given by $V = \Re[V] + i \Im[V]$ where the real and imaginary
parts are specified in either Eqs.~(\ref{eq:potmodela}) and (\ref{eq:potmodelb}), respectively.  
Here, $m_1$ and $m_2$ are the masses
of the two heavy quarks and $m_R$ is the reduced mass, $m_R = m_1
m_2/(m_1+m_2)$. The index $\upsilon$ on the eigenfunctions,
$\phi_\upsilon$, and energies, $E_\upsilon$, represents a list of all
relevant quantum numbers, such as\ $n$, $l$, and $m$ for a radial Coulomb
potential. Due to the anisotropic screening mass, the wave functions
are no longer radially symmetric if $\xi \neq 0$. Nevertheless we still 
label the states as $1S$ (ground state) and $1P$ (first p-wave excited state), 
respectively.

To obtain the time-independent eigenfunctions 
we start with the time-dependent Schr\"odinger equation
\beq
i \frac{\partial}{\partial t} \psi({\bf x},t) = \hat H \psi({\bf x},t) \, ,
\label{3dSchrodingerEQminkowski}
\eeq
which can be solved by expanding in terms of the eigenfunctions,
$\phi_\upsilon$:
\beq \psi({\bf x},t) = \sum_\upsilon c_\upsilon \phi_\upsilon({\bf x})
e^{- i E_\upsilon t}~.
\label{eigenfunctionExpansionMinkowski}
\eeq
If one is only interested in the lowest energy states (ground state
and first few excited states) an efficient way to proceed is to
transform~(\ref{3dSchrodingerEQminkowski})
and~(\ref{eigenfunctionExpansionMinkowski}) to Euclidean time using a
Wick rotation, $\tau \equiv i t$:
\beq \frac{\partial}{\partial \tau} \psi({\bf x},\tau) = - \hat H
\psi({\bf x},\tau) \, ,
\label{3dSchrodingerEQeuclidean}
\eeq
and
\beq \psi({\bf x},\tau) = \sum_\upsilon c_\upsilon \phi_\upsilon({\bf
x}) e^{- E_\upsilon \tau} ~.
\label{eigenfunctionExpansionEuclidean}
\eeq
For details of the discretizations used etc. we refer the reader 
to Refs.~\cite{Sudiarta:2007,Strickland:2009ft}.

\subsection{Finding the ground state}

By definition, the ground state is the state with the lowest energy
eigenvalue, $E_0$. Therefore, at late imaginary time the sum
over eigenfunctions (\ref{eigenfunctionExpansionEuclidean}) is
dominated by the ground state eigenfunction
\beq \lim_{\tau \rightarrow \infty} \psi({\bf x},\tau) \rightarrow c_0
\phi_0({\bf x}) e^{- E_0 \tau}~.
\label{groundstateEuclideanLateTime}
\eeq
Due to this, one can obtain the ground state wavefunction,
$\phi_0$, and energy, $E_0$, by solving
Eq.~(\ref{3dSchrodingerEQeuclidean}) starting from a random
three-dimensional wavefunction, $\psi_{\text{initial}}({\bf x},0)$,
and evolving forward in imaginary time. The initial wavefunction
should have a nonzero overlap with all eigenfunctions of the
Hamiltonian; however, due to the damping of higher-energy
eigenfunctions at sufficiently late imaginary times we are left with
only the ground state, $\phi_0({\bf x})$. Once the ground state
wavefunction (or any other wavefunction) is found, we can
compute its energy eigenvalue via
\bqa
E_\upsilon(\tau\to\infty) = \frac{\langle \phi_\upsilon | \hat{H} |
\phi_\upsilon \rangle}{\langle \phi_\upsilon | \phi_\upsilon
\rangle} = \frac{\int d^3{\bf x} \, \phi_\upsilon^*
\, \hat{H} \, \phi_\upsilon }{\int d^3{\bf x} \, \phi_\upsilon^*
\phi_\upsilon} \; .
\label{bsenergy}
\eqa
%

\begin{figure}
\begin{center}
\includegraphics[width=8.1cm]{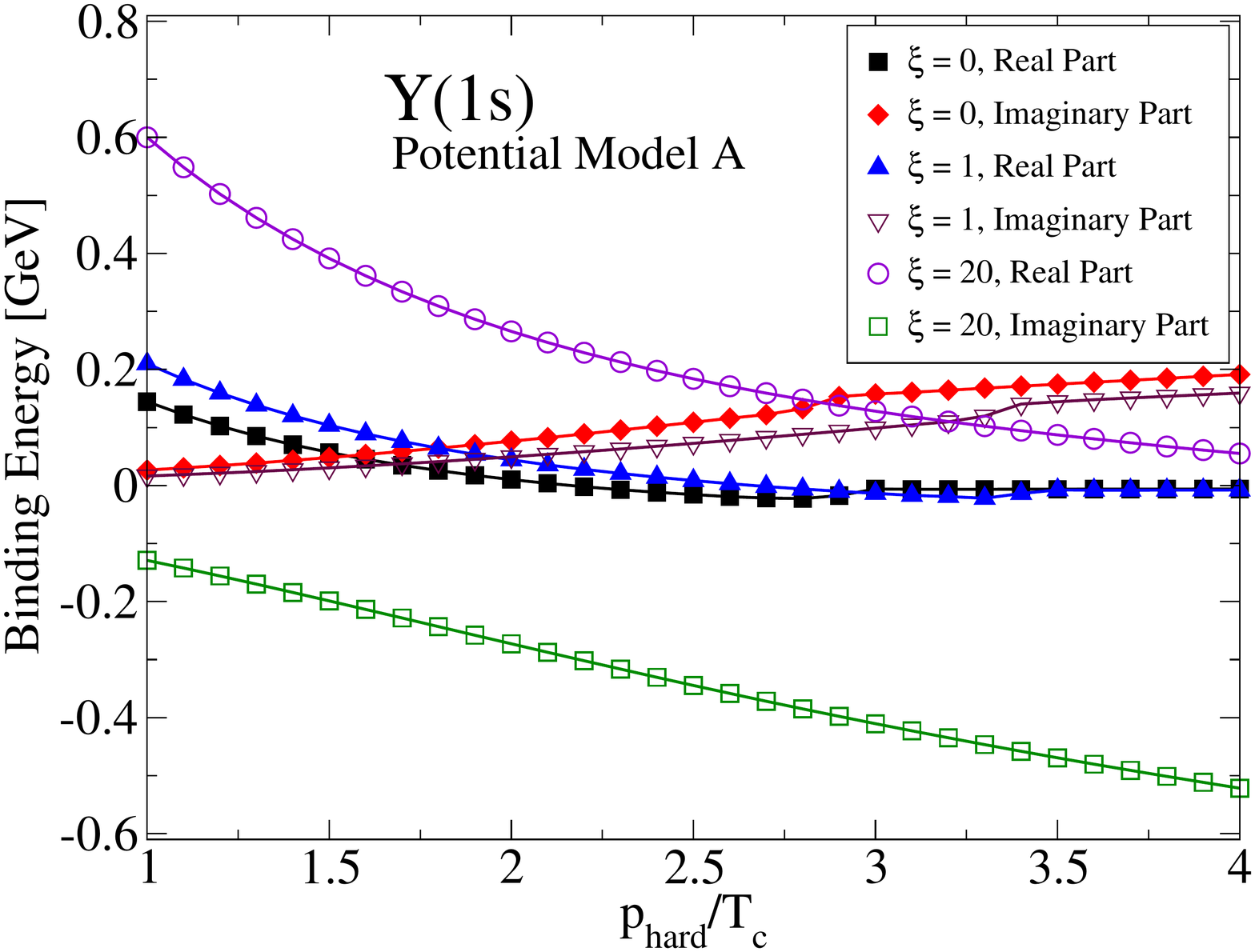}
\includegraphics[width=8.1cm]{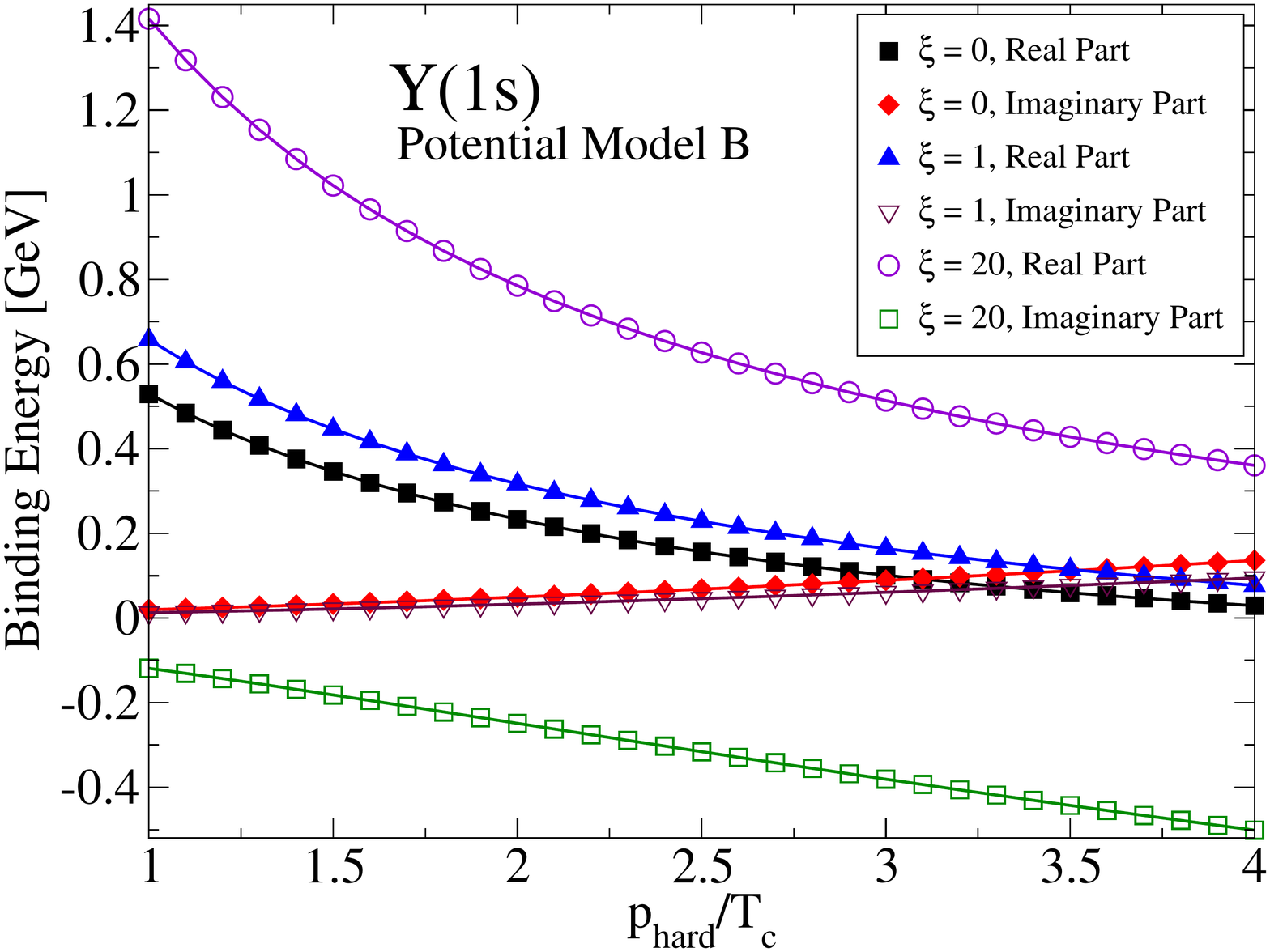}\\
\includegraphics[width=8.1cm]{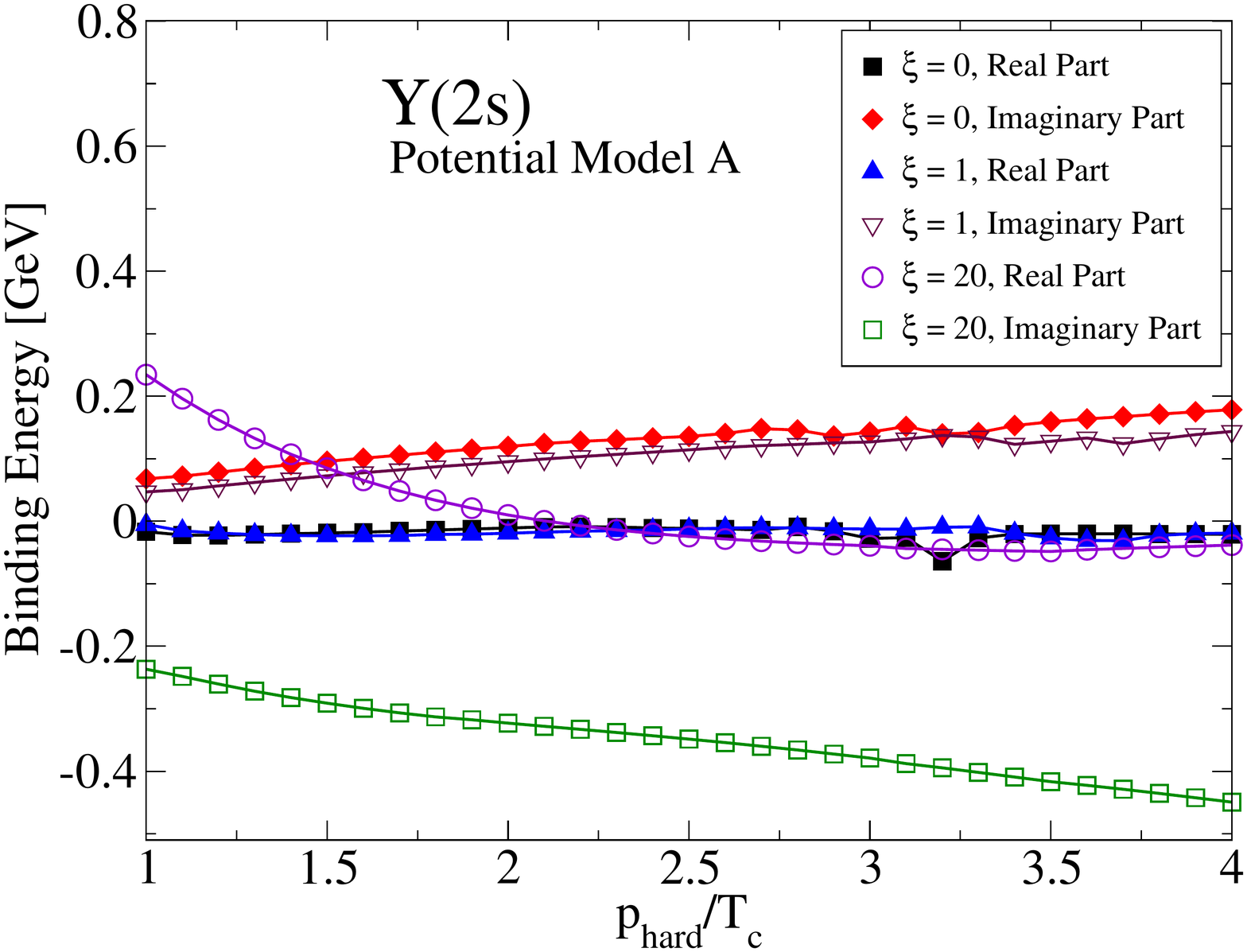}
\includegraphics[width=8.1cm]{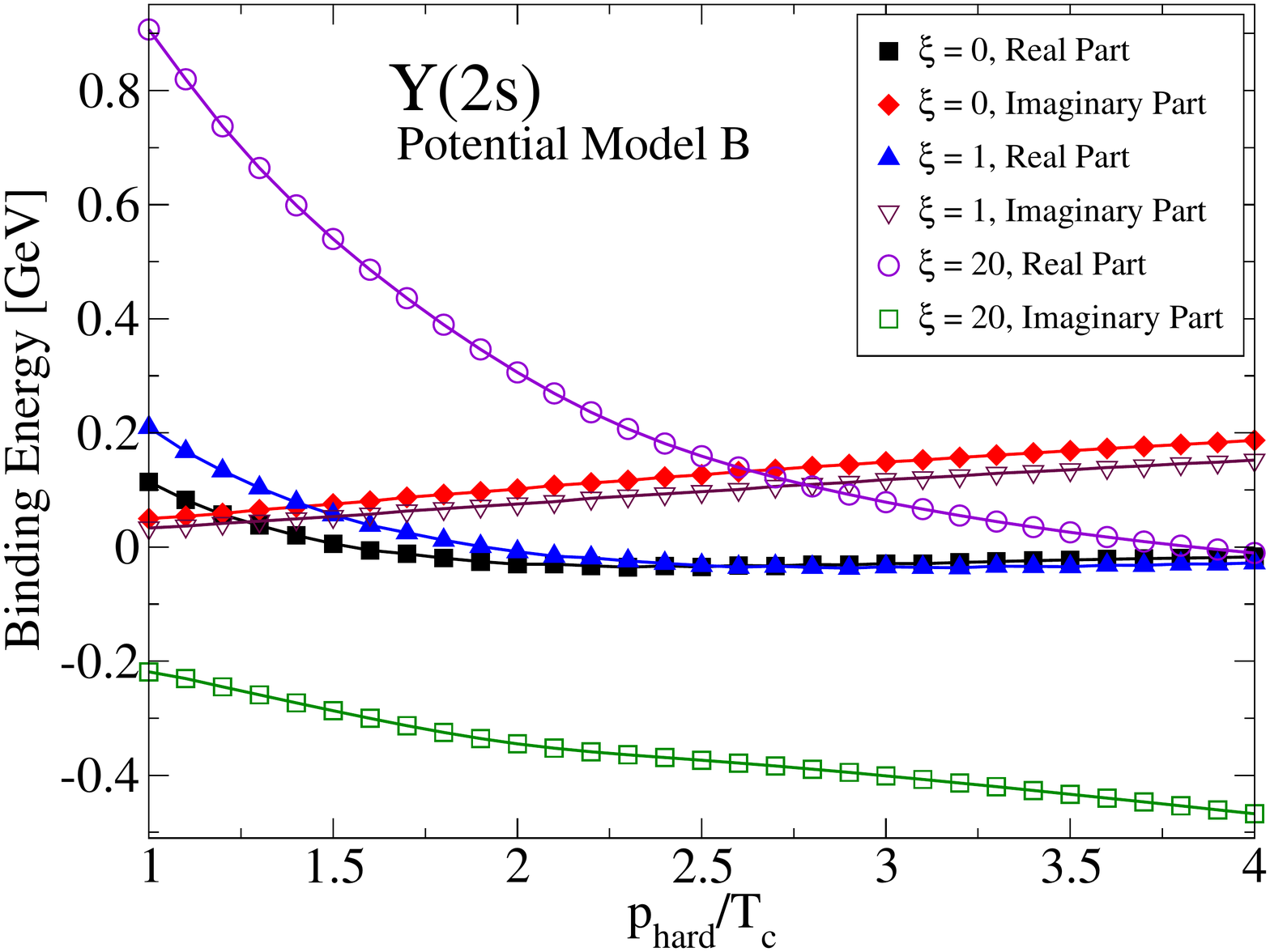}\\
\includegraphics[width=8.1cm]{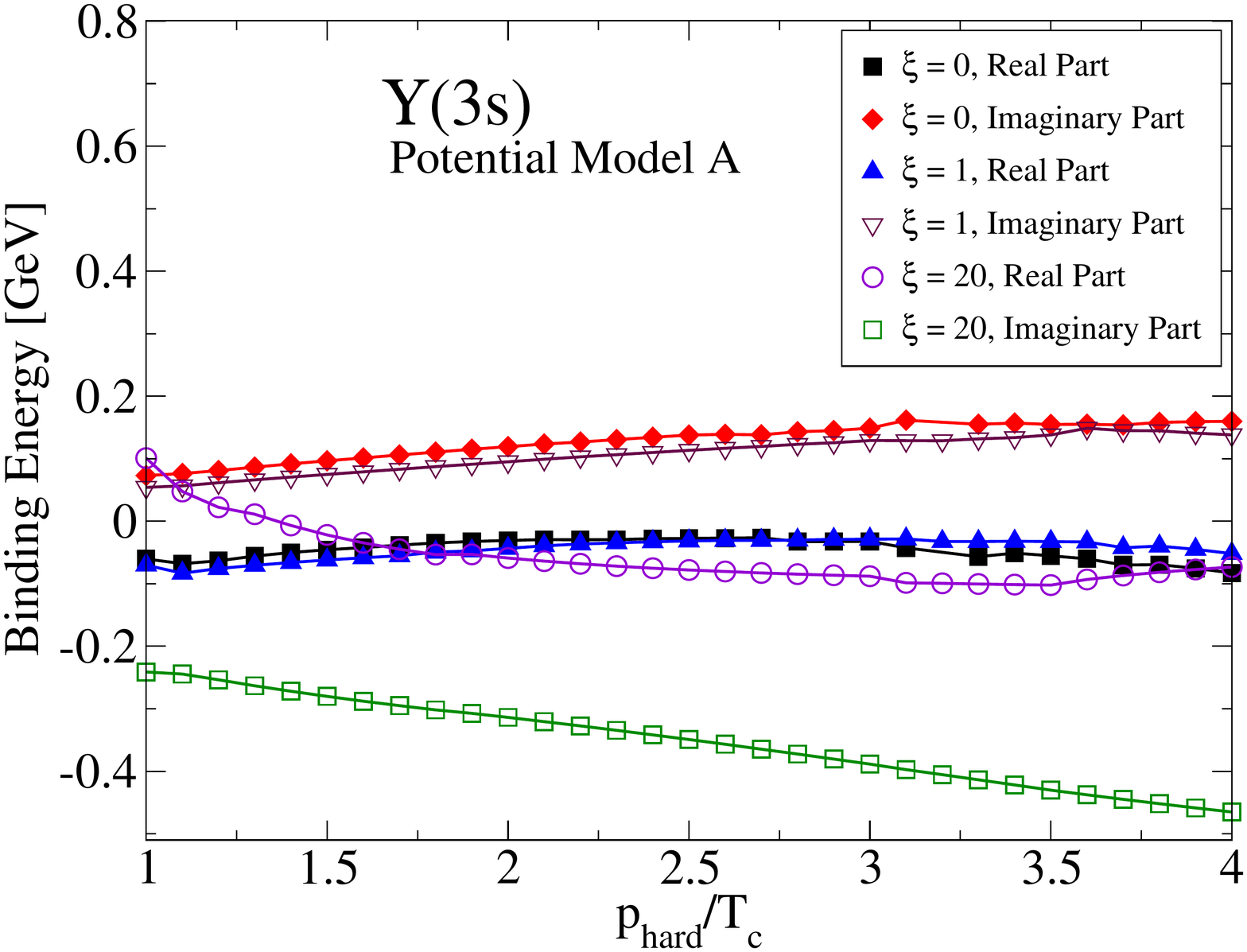}
\includegraphics[width=8.1cm]{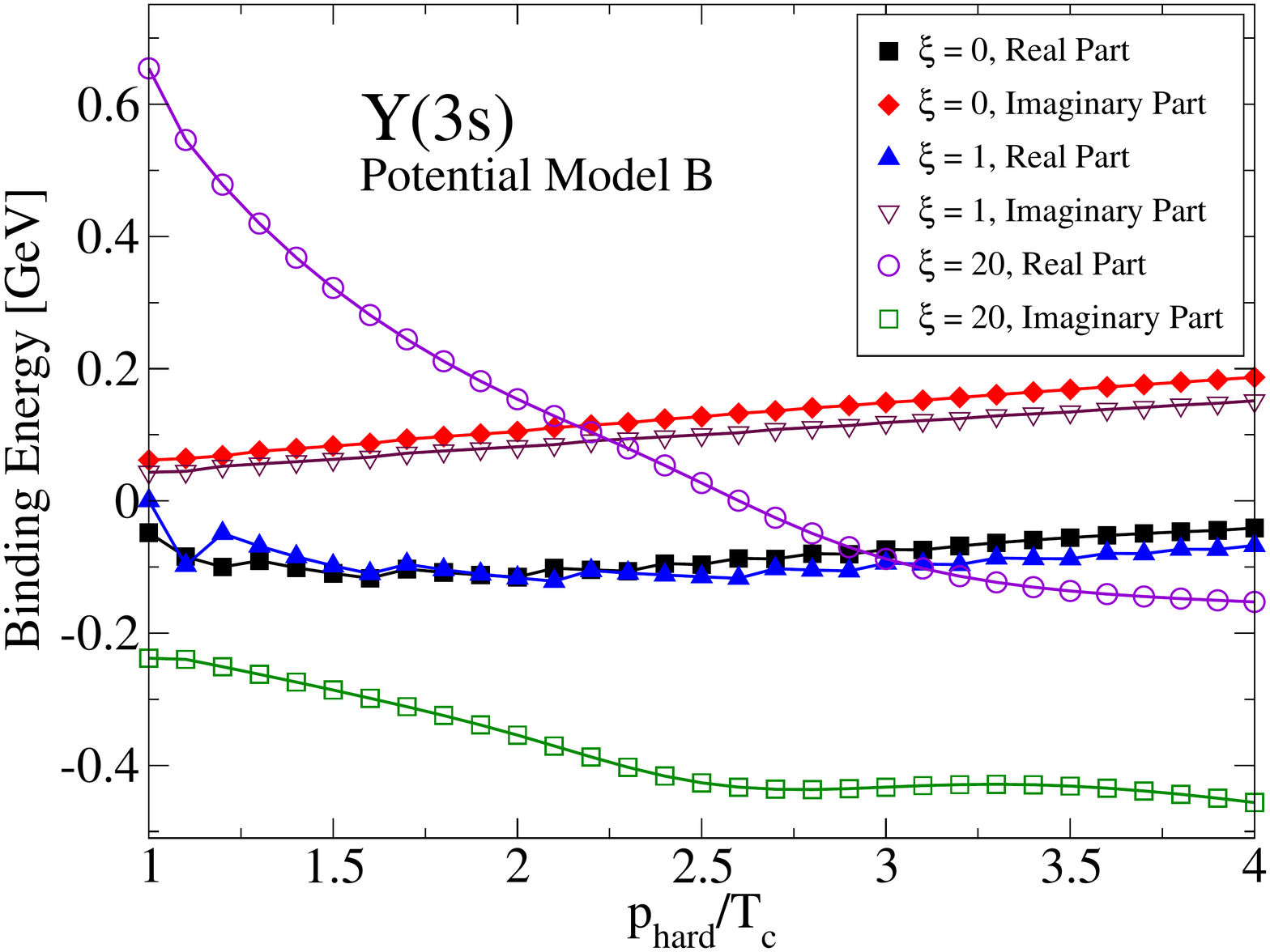}
\end{center}
\vspace{-8mm}
\caption{Real and imaginary parts of the $\Upsilon(1s)$, $\Upsilon(2s)$, and $\Upsilon(3s)$ binding energies 
as a function of the hard momentum
scale, $p_{\rm hard}$.  The left panels show results 
obtained with potential model A (\ref{eq:potmodela}) and the right panels show results from 
potential model B (\ref{eq:potmodelb}).}
\label{fig:binding-y}
\end{figure}

\begin{figure}[th]
\begin{center}
\includegraphics[width=8.1cm]{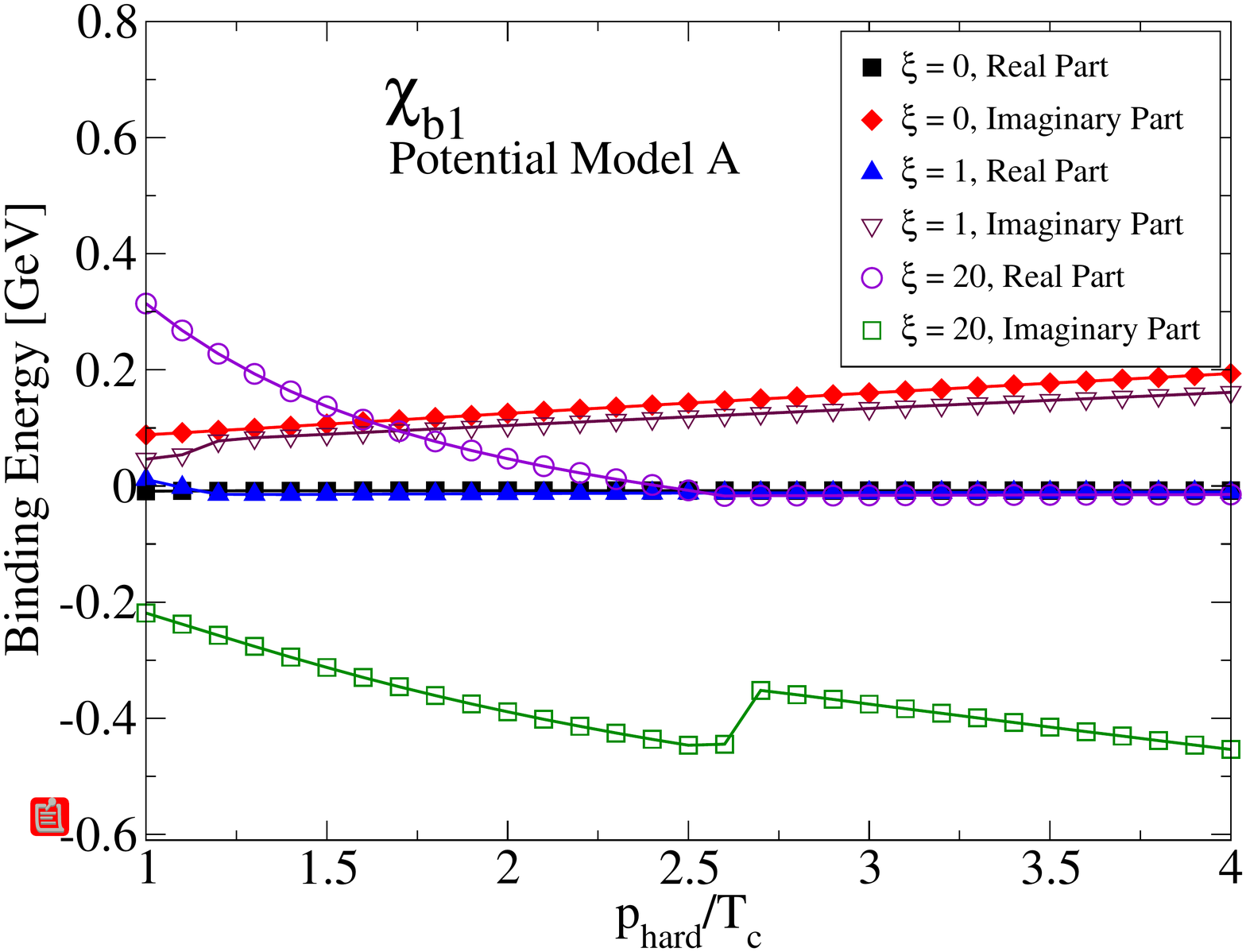}
\includegraphics[width=8.1cm]{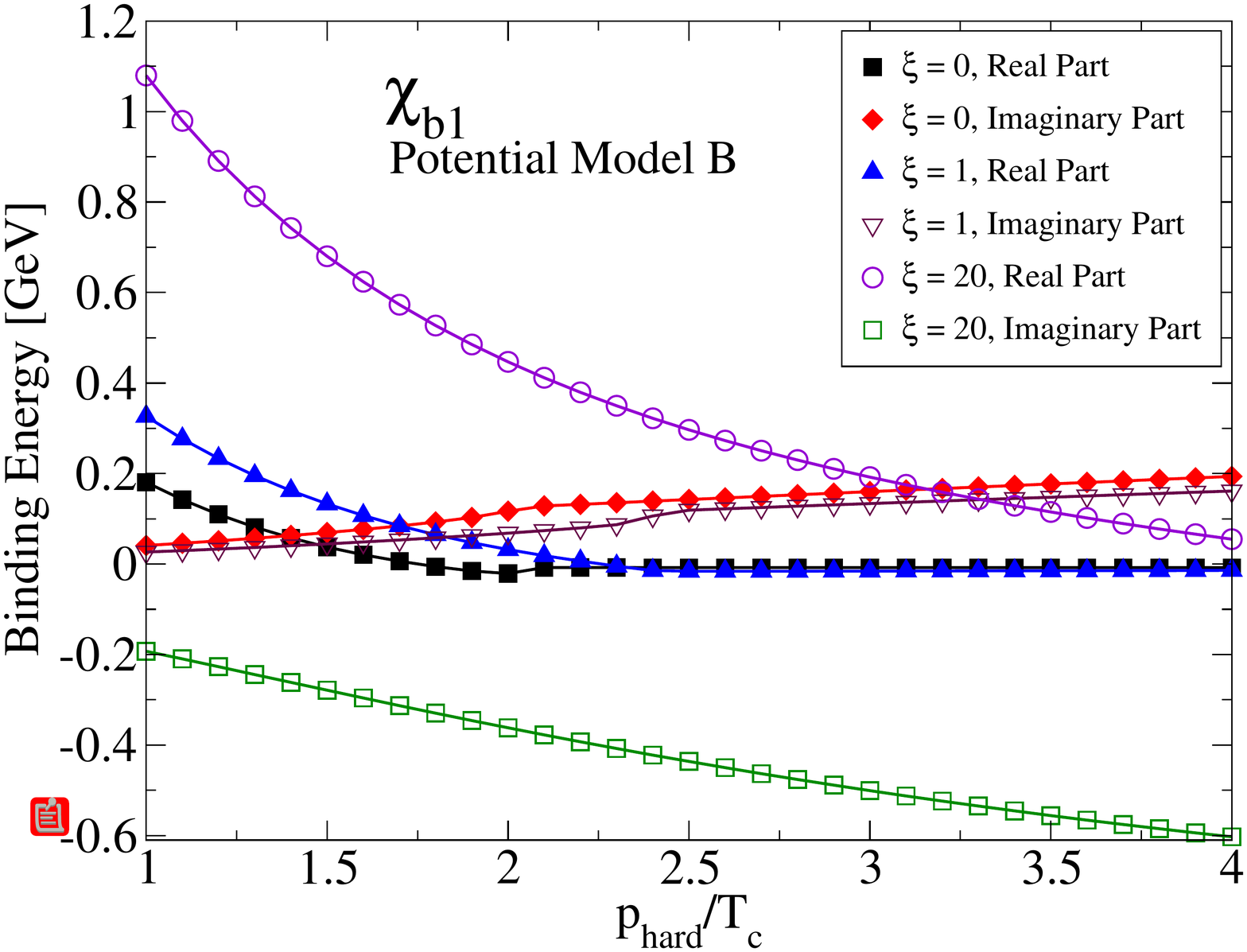}\\
\includegraphics[width=8.1cm]{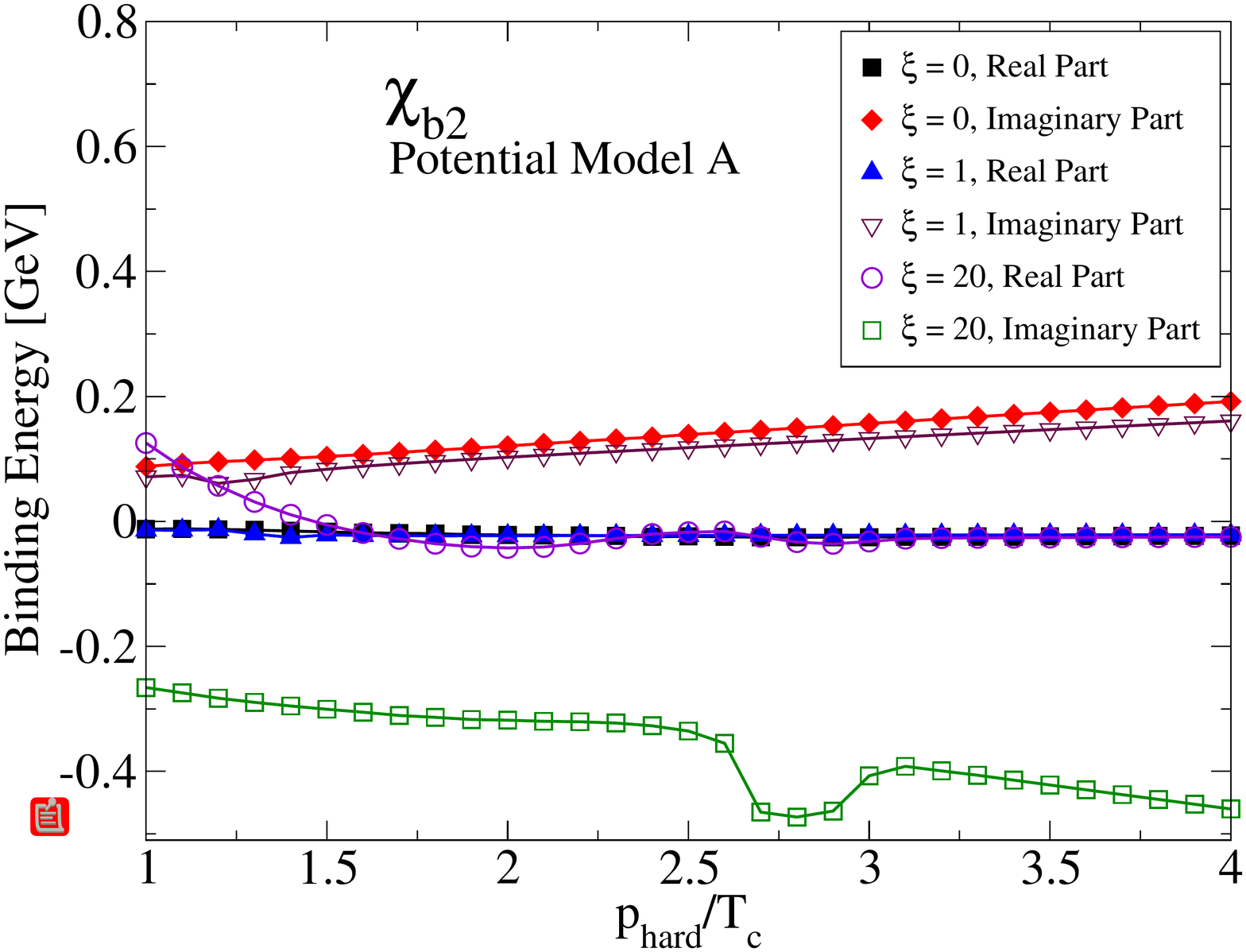}
\includegraphics[width=8.1cm]{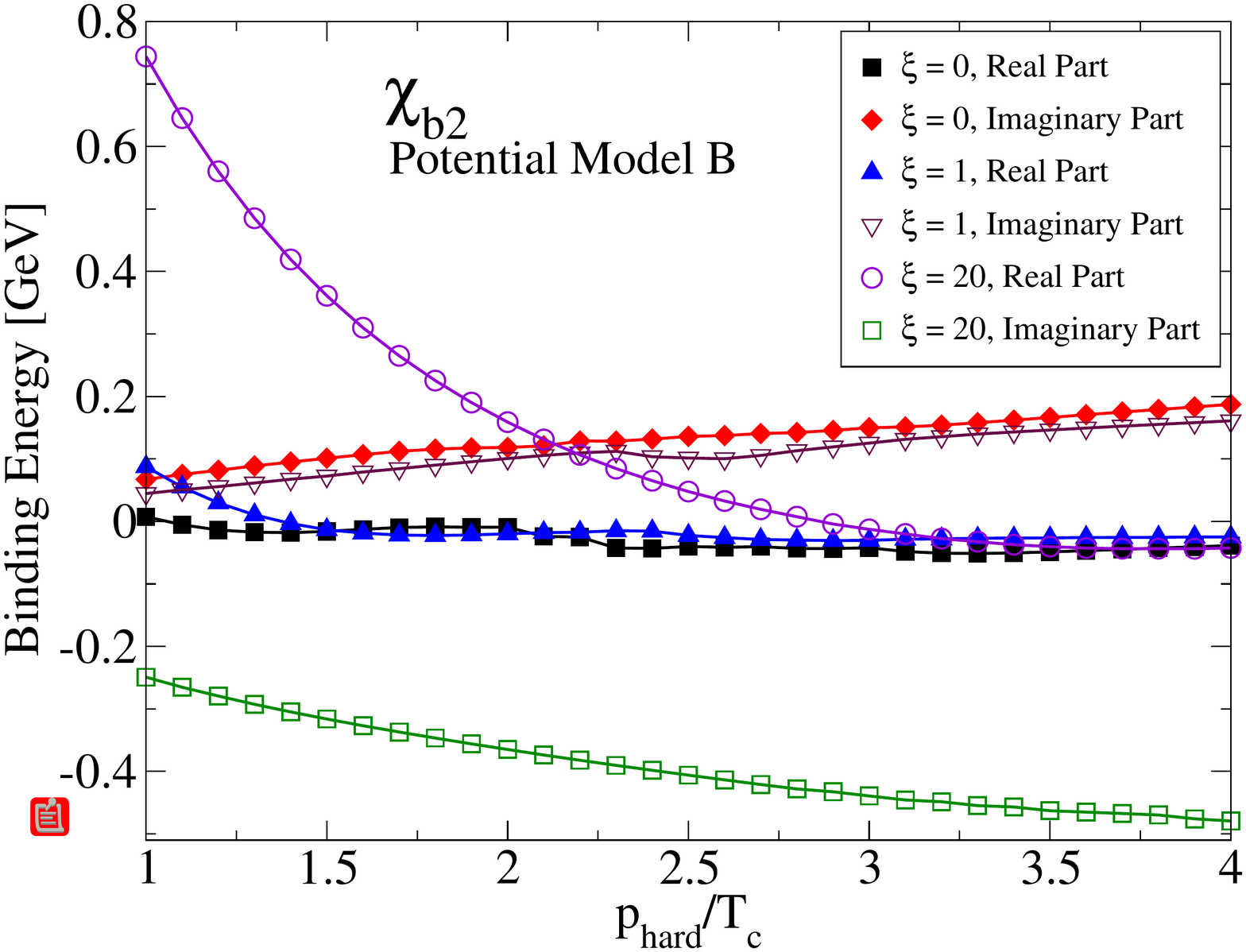}
\end{center}
\vspace{-8mm}
\caption{Real and imaginary parts of the $\chi_{b1}$ and $\chi_{b2}$ binding energies as a 
function of the hard momentum
scale, $p_{\rm hard}$.  The left panels shows results 
obtained with potential model A (\ref{eq:potmodela}) and the right panels show results from 
potential model B (\ref{eq:potmodelb}).}
\label{fig:binding-chib}
\end{figure}

To obtain the binding energy of a state,
$E_{\upsilon,\text{bind}}$, we subtract the quark masses and
the real part of the potential at infinity
\beq
E_{\upsilon,\text{bind}} \equiv -\left( E_\upsilon - m_1 - m_2 -
\frac{\langle \phi_\upsilon | V_\infty(\theta) | \phi_\upsilon
\rangle}{\langle \phi_\upsilon | \phi_\upsilon \rangle} \right) \; ,
\label{bsbindingenergy}
\eeq
where
\beq
V_\infty(\theta) \equiv \lim_{|{\bf r}|\to\infty} \Re[V(\theta,{\bf r})] \, ,
\eeq
which is a purely real quantity.
For an isotropic potential $V_\infty$ is independent of the
quantum numbers $\upsilon$ and equal to either $\sigma/m_D$ or $2\sigma/m_D$
for potential models A and B, respectively. In the
anisotropic case, however, this is no longer true since the operator
$V_\infty(\theta)$ carries angular dependence. 
Its expectation value is, of course, independent of $\theta$ but
does depend on the anisotropy parameter $\xi$.

\subsection{Finding the excited states}

The basic method for finding excited states is to first evolve the
initially random wavefunction to large imaginary times, find the
ground state wavefunction, $\phi_0$, and then project this state out
from the initial wavefunction and re-evolve the partial-differential
equation in imaginary time. However, there are (at least) two more
efficient ways to accomplish this. The first is to record snapshots of
the 3d wavefunction at a specified interval $\tau_{\text{snapshot}}$
during a single evolution in $\tau$. After having obtained the ground
state wavefunction, one can go back and extract the excited
states by projecting out the ground state wavefunction from the
recorded snapshots of $\psi({\bf x},\tau)$ \cite{Sudiarta:2007,Strickland:2009ft}.

An alternative way to select different excited states is to impose a
symmetry condition on the initially random wavefunction which cannot
be broken by the Hamiltonian evolution \cite{Strickland:2009ft}. For example, one can select
the first p-wave excited state of the (anisotropic) potential by
anti-symmetrizing the initial wavefunction around either the $x$, $y$,
or $z$ axes.  In the anisotropic case this trick can be used to
separate the different excited state polarizations in the
quarkonium system and to determine their energy eigenvalues with high
precision.  This high precision allows one to more accurately
determine the splitting between polarization states which are
otherwise degenerate in the isotropic Debye-screened Coulomb potential.
Whichever method is used, once the wavefunction of an excited state
has been determined, one can again use the general
formulas~(\ref{bsenergy}) and~(\ref{bsbindingenergy}) to determine 
the state's binding energy. 

\subsection{Results for the Binding Energies of Bottomonium States}
\label{sec:bindingresults}

In Figs.~\ref{fig:binding-y} and \ref{fig:binding-chib} we show the real and imaginary parts of the $\Upsilon(1s)$, 
$\Upsilon(2s)$, $\Upsilon(3s)$, $\chi_{b1}$, and $\chi_{b2}$ binding energies as a function of the hard momentum 
scale, $p_{\rm hard}$, for $\xi \in \{0,1,20\}$.  
The left panels show results obtained with potential model A (\ref{eq:potmodela}) and the right 
panels show results from potential model B (\ref{eq:potmodelb}).  In each case we show three different values of $\xi$.
For the bottom quark mass we used $m_b = 4.7$ GeV.  For the $\Upsilon(1s)$, $\Upsilon(2s)$, and $\Upsilon(3s)$ states
we used a lattice size of $N^3 = 256^3$ with a lattice spacing of $a = 0.125\;{\rm GeV}^{-1} \simeq 0.025\;{\rm fm}$
giving a lattice size of $L = Na \simeq 6.3\;{\rm fm}$.  For the $\chi_{b1}$ and $\chi_{b2}$ states
we used a lattice size of $N^3 = 256^3$ with a lattice spacing of $a = 0.15\;{\rm GeV}^{-1} \simeq 0.03\;{\rm fm}$ 
giving a lattice size of $L = Na \simeq 7.6\;{\rm fm}$.  Note that the fluctuations seen in some of the data points occur
at values of $\phard$ where the state is unbound.  These fluctuations are due to poor convergence of the Schr\"odinger 
equation algorithm for unbound states.  However, such fluctuations do not enter into our final results because, when the 
states are unbound (have a
negative real part of their binding energy), then we use a large fixed decay rate for these states.  Details of the precise
prescription will be provided in Section~\ref{sec:supfac}.

Defining the disassociation scale as the value of $p_{\rm hard}$ at which the real and imaginary parts of the binding
energy become equal, one finds the values listed in Table~\ref{tab:dissociation}.  As can be seen from the figures and table 
one finds that the dissociation scale increases with increasing $\xi$ such that bottomonium states persist longer in a 
momentum-space anisotropic plasma.  Binding energy data such as those presented in Figs.~\ref{fig:binding-y} and 
\ref{fig:binding-chib} will be used as input to our suppression calculation.

\begin{table}[t]
\begin{center}
\begin{tabular}{|l||l|l||l|l|}
\hline
&\multicolumn{2}{c||}{$\xi$=0}
&\multicolumn{2}{c|}{$\xi$=1}\\
\hline
\hline
{\bf State} & {\bf Potential A} & {\bf Potential B} & {\bf Potential A} & {\bf Potential B} \\\hline
$\Upsilon(1s)$ & 298 MeV &  593 MeV & 373 MeV &  735 MeV\\\hline
$\Upsilon(2s)$ &  $<$ 192 MeV & 228 MeV & $<$ 192 MeV  &  290 MeV \\\hline
$\Upsilon(3s)$ &  $<$ 192 MeV & $<$ 192 MeV & $<$ 192 MeV  &  $<$ 192 MeV  \\\hline
$\chi_{b1}$ &  $<$ 192 MeV & 265 MeV & $<$ 192 MeV  &  351 MeV \\\hline
$\chi_{b2}$ &  $<$ 192 MeV & $<$ 192 MeV & $<$ 192 MeV  &  213 MeV\\\hline
\end{tabular}
\end{center}
\caption{Isotropic and anisotropic dissociation scales for the  $\Upsilon(1s)$, 
$\Upsilon(2s)$, $\Upsilon(3s)$, $\chi_{b1}$, and $\chi_{b2}$.  Dissociation values were determined
by finding the value of $p_{\rm hard}$ 
when the real and imaginary parts of the state's binding energy become equal.}
\label{tab:dissociation}
\end{table}

\section{Dynamical Model}
\label{sec:dynmodel}

In  order to describe the space-time evolution of the system we use ``anisotropic
hydrodynamics'' ({\sc aHydro}) which extends traditional viscous hydrodynamical treatments 
to cases in which the local momentum-space anisotropy of the plasma can be large
\cite{Martinez:2010sc,Martinez:2010sd}.
The result is a dynamical framework that reduces to
2nd order viscous hydrodynamics for weakly anisotropic plasmas, but can
better describe highly anisotropic plasmas.  In this paper we ignore the transverse expansion of
the matter and model the system as a collection of decoupled (1+1)-dimensional systems with different
initial temperatures; however, we allow for the breaking of boost invariance.
For such effectively one-dimensional dynamics which is homogeneous 
in the transverse directions, the {\sc aHydro} approach provides the temporal and spatial rapidity 
evolution of the typical hard momentum of the plasma partons, $p_{\rm hard}$, the plasma
anisotropy, $\xi$, and the four-velocity of the rest frame via a hyperbolic angle $\vartheta$.

We briefly state the setup and final results of Ref.~\cite{Martinez:2010sd} for completeness.  The
starting point for the dynamical equations is to assume the same ansatz (\ref{eq:f_aniso}) 
for the momentum-space 
anisotropic distribution distribution as was used to compute the heavy quark potential
in the previous section.  In the
local rest frame of the plasma the ansatz has two parameters $\phard$ and $\xi$.  
In the boost invariant case $\phard$ and $\xi$ would
be functions only of proper time; however, in the case of broken boost invariance both $\phard$
and $\xi$ becomes functions of proper time, $\tau$, and spatial rapidity, $\varsigma$.  The 
necessary dynamical equations can be obtained by taking moments of the Boltzmann equation
in the relaxation time approximation \cite{Martinez:2010sd}.  The breaking of boost invariance
requires that, in addition to $\phard$ and $\xi$, one must also specify the hyperbolic angle of
the local rest frame of the flow.  This can be accomplished by introducing two four-vectors,
one of which specifies the four velocity of the local rest frame in lab frame, $u^\mu$, 
and an additional four-vector, $v_\mu$, which is orthogonal to $u^\mu$, i.e. $u^\mu v_\mu =0$.
This can be accomplished by introducing a hyperbolic angle $\vartheta$ such that
\begin{subequations}
\begin{align}
u^\mu &= (\cosh \vartheta(\tau,\varsigma),0,0,\sinh \vartheta(\tau,\varsigma)) \, , \\
v^\mu &= (\sinh \vartheta(\tau,\varsigma),0,0,\cosh \vartheta(\tau,\varsigma)) \, ,
\end{align}
\end{subequations}
where $\vartheta(\tau,\varsigma)$ is the hyperbolic angle associated with the velocity of the 
local rest frame as measured in the lab frame \cite{Ryblewski:2011aq}.  If the system were exactly boost-invariant
then we would have $\vartheta(\tau,\varsigma) = \varsigma$ at all times.

\begin{figure}[t]
\begin{center}
\includegraphics[width=16.4cm]{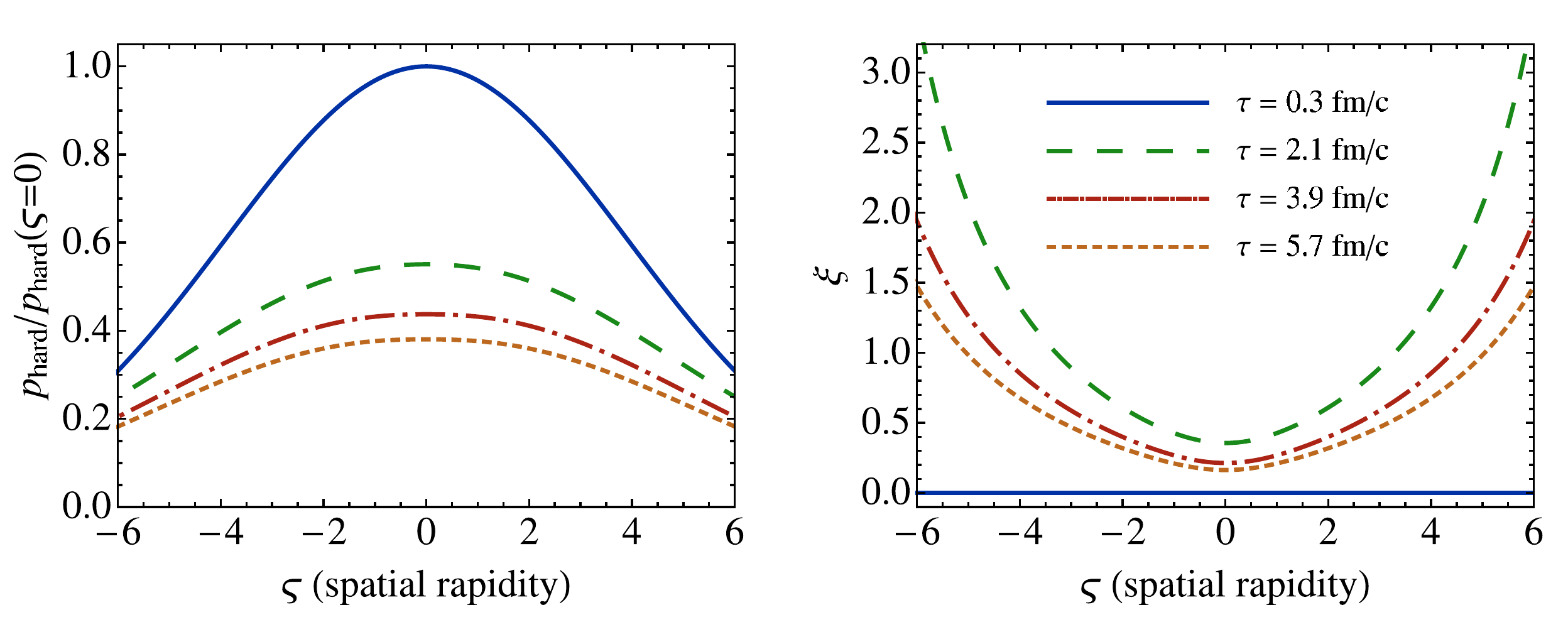}
\end{center}
\vspace{-6mm}
\caption{
Dynamical parameters as a function of spatial rapidity using a strong coupling value of $4\pi\eta/{\cal S}=1$.  
Shown are $\phard$ (left) and $\xi$ (right) with initial conditions $\xi(\tau_0,\varsigma) = 0$ and 
$\phard(\tau_0,\varsigma=0) =$ 540 MeV with $\tau_0 = 0.3$ fm/c .  The initial $\phard$ rapidity 
dependence is given by a Gaussian profile specified in Eq.~(\ref{eq:yprofile}).  Profiles at proper times 
$\tau \in \{0.3,2.1,3.9,5.7\}$ fm/c are shown. 
}
\label{fig:evolution-sc-540}
\end{figure}

\begin{figure}[t]
\begin{center}
\includegraphics[width=16.4cm]{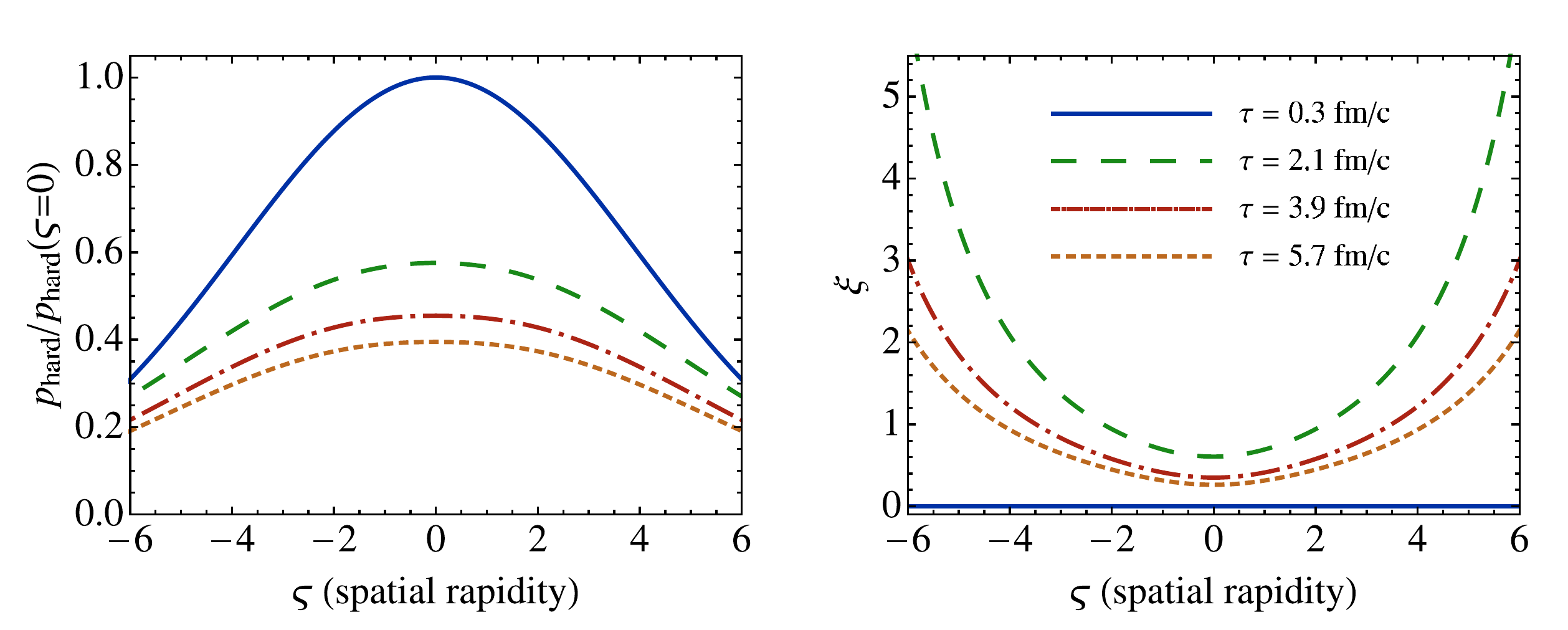}
\end{center}
\vspace{-6mm}
\caption{
Dynamical parameters as a function of spatial rapidity using a strong coupling value of $4\pi\eta/{\cal S}=1$.  
Shown are $\phard$ (left) and $\xi$ (right) with initial conditions $\xi(\tau_0,\varsigma) = 0$ and 
$\phard(\tau_0,\varsigma=0) =$ 350 MeV with $\tau_0 = 0.3$ fm/c .  The initial $\phard$ rapidity 
dependence is given by a Gaussian profile specified in Eq.~(\ref{eq:yprofile}).  Profiles at proper times 
$\tau \in \{0.3,2.1,3.9,5.7\}$ fm/c are shown. 
}
\label{fig:evolution-sc-350}
\end{figure}

\begin{figure}[t]
\begin{center}
\includegraphics[width=16.4cm]{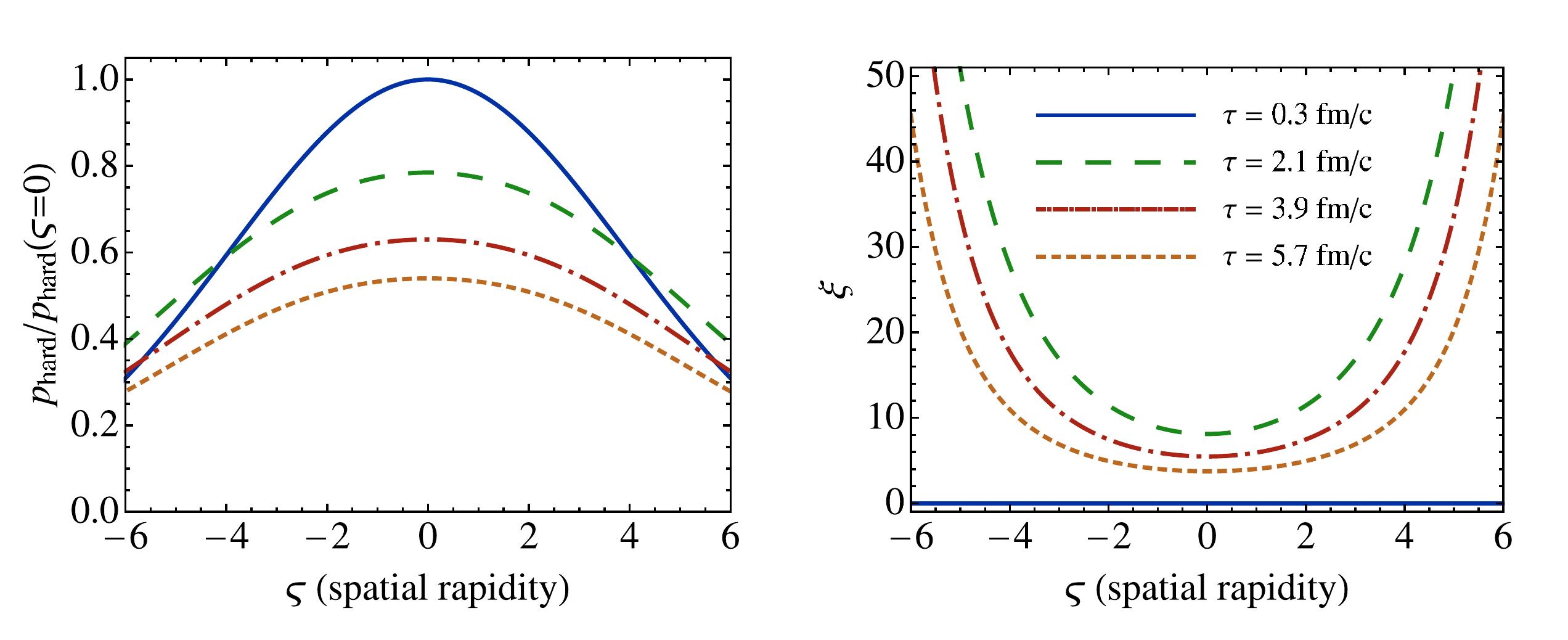}
\end{center}
\vspace{-6mm}
\caption{
Dynamical parameters as a function of spatial rapidity using a strong coupling value of $4\pi\eta/{\cal S}=10$.  
Shown are $\phard$ (left) and $\xi$ (right) with initial conditions $\xi(\tau_0,\varsigma) = 0$ and 
$\phard(\tau_0,\varsigma=0) =$ 540 MeV with $\tau_0 = 0.3$ fm/c .  The initial $\phard$ rapidity 
dependence is given by a Gaussian profile specified in Eq.~(\ref{eq:yprofile}).  Profiles at proper times 
$\tau \in \{0.3,2.1,3.9,5.7\}$ fm/c are shown. 
}
\label{fig:evolution-wc-540}
\end{figure}

\subsection{Moments of the Boltzmann Equation}

In order to obtain the necessary dynamical equations for $\phard$ and $\xi$
we follow \cite{Martinez:2010sd} and
take moments of the Boltzmann equation.  For non-boost-invariant (1+1)-dimensional dynamics 
it suffices to take the zeroth and first moments and project the first moment with either $u^\mu$
or $v^\mu$.  The result is three coupled partial differential equations which give the proper-time
and spatial-rapidity evolution  of $\phard$, $\xi$, and $\vartheta$:

\begin{subequations}
\begin{align}
\label{eq:zerothmoment}
&\frac{1}{1+\xi}\Bigl( \partial_\tau \xi- \frac{2(1+\xi)}{\tau}\,\partial_\varsigma\vartheta\Bigr) 
- \frac{6}{\phard} \partial_\tau \phard = 2 \lambda \left[ 1 - {\cal R}^{3/4}(\xi) \sqrt{1+\xi} \right] \, , \\
&\frac{{\cal R}'(\xi)}{{\cal R}(\xi)}\,\partial_\tau \xi\,+\,4\,\frac{\partial_\tau \phard}{\phard}\,+\,
\frac{\tanh (\vartheta-\varsigma)}{\tau}\,\biggl(\frac{{\cal R}'(\xi)}{{\cal R}(\xi)}\,\partial_\varsigma \xi+\,4\,
\frac{\partial_\varsigma \phard}{\phard}\biggr)\nn\\
&\hspace{4.5cm}=-\Bigl(1+\frac{1}{3}\frac{{\cal R}_L(\xi)}{{\cal R}(\xi)}\Bigr)\Bigl(\tanh (\vartheta-\varsigma)\,\partial_\tau+
\frac{\partial_\varsigma}{\tau}\Bigr)\vartheta\,,\label{eq:1stmoment-1}\\
&\tanh (\vartheta-\varsigma)\Biggl(\frac{{\cal R}'_L(\xi)}{{\cal R}_L(\xi)}\,\partial_\tau \xi\,+\,4\,
\frac{\partial_\tau \phard}{\phard}\Biggr)+\frac{1}{\tau}\biggl(\frac{{\cal R}'_L(\xi)}{{\cal R}_L(\xi)}\,\partial_\varsigma \xi\,+\,
4\,\frac{\partial_\varsigma \phard}{\phard}\biggr)
\nn\\
&\hspace{4.5cm}= -\Bigl(3\,\frac{{\cal R}(\xi)}{{\cal R}_L(\xi)}+1\Bigr)\Bigl(\partial_\tau+
\frac{\tanh (\vartheta-\varsigma)}{\tau}\,\partial_\varsigma \Bigr)\vartheta\,,\label{eq:1stmoment-2}
\end{align}
\label{eq:dynamicalmodel}
\end{subequations}

\noindent
where ${\cal R}(\xi)$ and ${\cal R}_L(\xi)$ are defined in Eqs.~(\ref{energyaniso}) and~(\ref{longpressaniso}),
respectively.  Note that in the derivation of the above equations it was assumed that the system consists
of a plasma of massless particles which results in a conformal equation of state, i.e. ${\cal E}_{\rm iso} = 
3 {\cal P}_{\rm iso}$.

The relaxation rate $\lambda$ appearing in the first equation (\ref{eq:zerothmoment}) is fixed by requiring 
that the equations reduce to the evolution equations of second order viscous hydrodynamics in the limit of 
small $\xi$.  Doing so gives \cite{Martinez:2010sc}
\beq
\lambda = \frac{2T(\tau)}{5\bar\eta} = \frac{2{\cal R}^{1/4}(\xi)\phard}{5\bar\eta} \, ,
\label{eq:gammamatch}
\eeq
where $\bar\eta = \eta/{\cal S}$ is the ratio of the plasma shear viscosity to entropy density 
and we have mapped the equilibrium temperature to $\phard$ and $\xi$ by 
requiring that the anisotropic and isotropic energy densities are the same, i.e. ${\cal E}_{\rm aniso}(\phard,\xi) 
= {\cal E}_{\rm iso}(T)$, which upon using Eq.~(\ref{energyaniso}) gives $T = {\cal R}^{1/4}(\xi)\phard$.

We note, importantly, that since the relaxation rate $\lambda$ is proportional to $\phard$, one expects
that the relaxation to isotropic equilibrium is slower in regions where $\phard$ is lower.  In addition,
we see that the relaxation rate is inversely proportional to $\bar\eta$ which tells us that when the shear
viscosity is small we expect to see larger plasma momentum-space anisotropies developing.  In order to 
illustrate the dependence on initial temperature, in Figs.~\ref{fig:evolution-sc-540} and \ref{fig:evolution-sc-350} 
we show the evolution of $\phard$ and $\xi$ in the case of a strong coupling shear viscosity of $\bar\eta=1/4\pi$ 
for two different assumed initial central temperatures of 540 MeV and 350 MeV, respectively.  As can be
seen from these two figures, as the initial temperature decreases, one sees larger momentum-space anisotropy
as expected from Eq.~(\ref{eq:gammamatch}).  In order to illustrate the dependence on the assumed value of 
$\bar\eta$ in Fig.~\ref{fig:evolution-wc-540} we show the case of $\bar\eta=10/4\pi$ with an initial central 
temperature of 540 MeV.  Comparing Figs.~\ref{fig:evolution-sc-540} and  Fig.~\ref{fig:evolution-wc-540}
we see that there is a dramatic increase in the developed momentum-space anisotropy when changing
$\bar\eta$ from $1/4\pi$ to $10/4\pi$.  The result of these two dependences will be that we will see
less suppression of the bottomonium states when $\phard$ is low or $\bar\eta$ is large.

\section{Initial Conditions}
\label{sec:initialconditions}

In this section we specify the type of initial conditions we use.  We study both RHIC and LHC energies,
therefore in this section we will present the general formulae which can be used in both cases.  In the results 
section we will specify the specific initial temperatures, collision energies, starting proper times, etc. that
we use in each specific case.

\subsection{Transverse Coordinate Dependence}
\label{subsec:transinit}

In this paper we will consider collisions of symmetric nuclei, each containing $A$ nucleons.  We will
study both participant and binary collision type initial conditions \cite{Bialas:1976ed} using a Woods-Saxon 
distribution for each nuclei's transverse profile \cite{Glauber:1970jm}.  For an individual nucleon we 
take the nucleon density to be
\beq
n_A(r) = \frac{n_0}{1 + e^{(r-R)/d}} \, ,
\eeq
where $n_0 = 0.17\;{\rm fm}^{-3}$ is the central nucleon 
density, $R = (1.12 A^{1/3} - 0.86 A^{-1/3})\;{\rm fm}$ 
is the nuclear radius, and $d = 0.54\;{\rm fm}$ is the ``skin depth''.  The density is normalized such that 
$\lim_{A\to\infty} \int d^3r \, n_A(r) = A$, where $A$ is the total number of nucleons in the nucleus.  
The normalization condition fixes $n_0$ to the value specified above.
From the nucleon
density we first construct the thickness function in the standard way by integrating over the longitudinal 
direction, i.e.
\beq
T_A(x,y) = \int_{-\infty}^{\infty} dz \, n_A(\sqrt{x^2+y^2+z^2}) \, .  
\eeq
With this in hand we can construct the  overlap density between two nuclei whose centers are 
separated by an impact parameter vector $\vec{b}$ which we choose to point along the $\hat{x}$ 
direction, i.e. $\vec{b} = b \hat{x}$.  We choose to locate the origin of our coordinate
system to lie halfway between the center of the two nuclei such that the overlap density can be
written as
\beq
n_{AB}(x,y,b) =  T_A(x+b/2,y) T_B(x-b/2,y) \, .
\label{eq:nab}
\eeq
The overlap density will be used later as the probability weight for bottomonium production and 
our ``two-component'' initial condition.
Another quantity of interest is the participant density which is given by 
\bqa
n_{\rm part}(x,y,b) &=& T_A(x+b/2,y) \left[ 1- \left(1-\frac{\sigma_{NN}\,T_B(x-b/2,y)}{B}\right)^{\!\!B} \right]
\nonumber \\
&& \hspace{2cm} + \; T_B(x-b/2,y) \left[ 1- \left(1-\frac{\sigma_{NN}\,T_A(x+b/2,y)}{A}\right)^{\!\!A} \right] \, .
\eqa
For LHC collisions at $\sqrt{s_{NN}}=2.76$ TeV we use $\sigma_{NN}$ = 62 mb and for RHIC
collisions at $\sqrt{s_{NN}}=200$ GeV we use $\sigma_{NN}$ = 42 mb.
From the participant density we construct our first possible initial condition for the 
transverse $\phard$ profile at central rapidity by taking the third root of the rescaled
$n_{\rm part}$
\beq
p_{\rm hard,0}^{\rm part} = T_0 \left[\frac{n_{\rm part}(x,y,b)}{n_{\rm part}(0,0,0)}\right]^{1/3} \, ,
\eeq
where $T_0$ is the central temperature obtained in a central collision between the two nuclei.

\begin{wrapfigure}{r}{0.45\textwidth}
\begin{center}
\includegraphics[width=7.2cm]{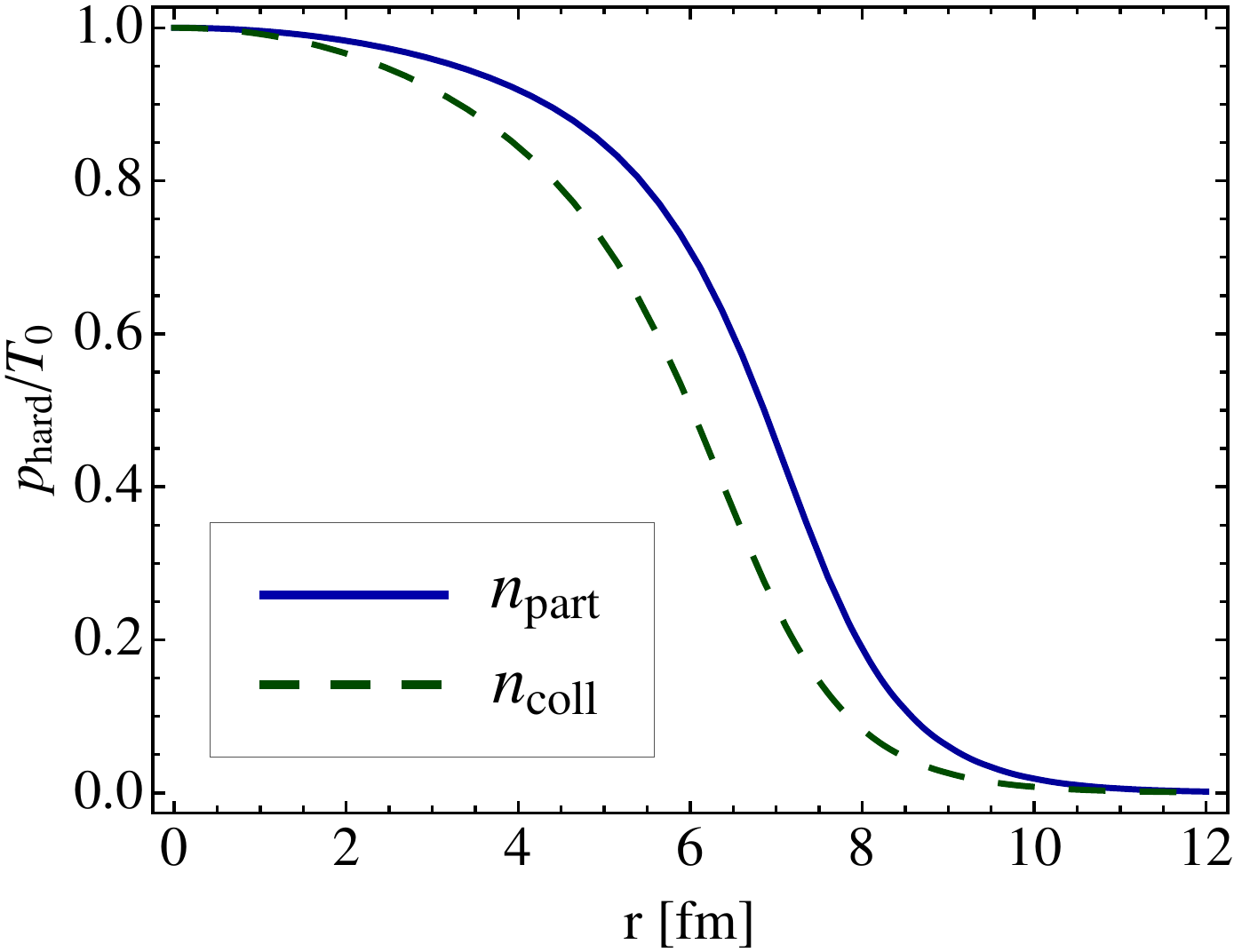}
\end{center}
\vspace{-6mm}
\caption{
Comparison of initial transverse $\phard$ profile given by $n_{\rm part}$ scaling and $n_{\rm coll}$ scaling.  A value of
$\sigma_{NN} = $ 62 mb was used and we show the case of a central collision, i.e. $b=0$.
}
\label{fig:initComp}
\end{wrapfigure}

As an alternative initial condition for $\phard$ one could use the number of binary collisions 
which is defined as
\beq
n_{\rm coll}(x,y,b) = \sigma_{NN} \, n_{AB}(x,y,b) \, .
\eeq
Comparisons with RHIC data show that it is necessary to add an admixture of $n_{\rm coll}$ to
the participant, or wounded-nucleon, scaling.  We will consider such an admixture as our second
possibility by defining
\bqa
n_{\rm mix}(x,y,b) &=& \frac{1}{2}(1-\alpha) \, n_{\rm part}(x,y,b)  \nonumber \\
&& \hspace{5mm} + \alpha \, n_{\rm coll}(x,y,b) \, ,
\eqa
with $\alpha = 0.145$ as fit by the PHOBOS Collaboration \cite{Back:2004dy}.
This gives a second possibility for the initial condition for $\phard$ at central rapidity
\beq
p_{\rm hard,0}^{\rm mix} = T_0 \left[\frac{n_{\rm mix}(x,y,b)}{n_{\rm mix}(0,0,0)}\right]^{1/3} \, .
\eeq
Note that $T_0$ should be adjusted so that both initial conditions give the same particle density at
central rapidity when integrated over the transverse plane.  For $\alpha  = 0.145$ we find that $T_0^{\rm mix}
= 1.079 \, T_0^{\rm part}$ at LHC energies and $T_0^{\rm mix}
= 1.065 \, T_0^{\rm part}$ at RHIC energies.  These values will be used in the results section when
we discuss the initial condition dependence of our results.

\subsection{Spatial Rapidity Dependence}
\label{subsec:rapinit}

In the previous subsection we fixed two possible prescriptions for the transverse temperature profile.  
Since we allow for the breaking of boost-invariance, 
we also need to give the spatial-rapidity dependence in order to complete our specification of the full
three-dimensional initial temperature profile.
For the number density profile in spatial rapidity ($\varsigma$) we use a Gaussian which successfully describes experimentally observed pion 
rapidity spectra from AGS to RHIC energies \cite{Bearden:2004yx,Park:2001gm,Back:2005hs,Veres:2008nq,Bleicher:2005tb}
and extrapolate this result to LHC energies.  The parametrization we use is
\begin{equation}
\label{eq:yprofile}
 n(\varsigma)= n_0 \exp \Biggl(-\frac{\varsigma^2}{2\sigma_\varsigma^2}\Biggr) \; ,
\end{equation}
with
\begin{equation}
\label{eq:width}
 \sigma_\varsigma^2=0.64 \cdot \frac{8}{3}\frac{c_s^2}{(1-c_s^4)}\ln \left(\sqrt{s_{NN}}/2 m_p\right) \; , 
\end{equation}
where $c_s$ is the sound velocity, $m_p = 0.938$ GeV is the proton mass, $\sqrt{s_{NN}}$ is the
nucleon-nucleon center-of-mass energy, and $n_0$ is the number density at central rapidity.
We have added a multiplicative factor of $0.64$ to adjust for broadening of the distribution
in rapidity as a function of proper time since the fits, e.g. from \cite{Bleicher:2005tb}, were to 
the final state spectra rather than initial state spectra.
In this paper we will use an ideal (conformal) equation of state for which $c_s = 1/\sqrt{3}$ in natural units.

\subsection{Full Three-Dimensional Initial Conditions}

We can use Eq.~(\ref{eq:yprofile}) to determine the initial $\phard$ rapidity dependence 
by taking the third root of the number density.  Putting this together with the two possibilities for the 
transverse temperature dependence determined in the Section \ref{subsec:transinit} we can now specify 
the full three-dimensional initial temperature profile.  Depending on whether we use the number
of participants ($n_{\rm part}$) or two component model ($n_{\rm mix}$) scaling we have two possible initial 
$p_{\rm hard}$ profiles:
\bqa 
p_{\rm hard,0}^{\rm I} = T_0 \left[\frac{n_{\rm part}(x,y,b) \, 
e^{-\varsigma^2/(2\sigma_\varsigma^2)}}{n_{\rm part}(0,0,0)}  \right]^{1/3} \quad;\quad {\rm Initial\;Condition\;I}
\; ,
\label{eq:phardinit-I}
\eqa
\bqa 
p_{\rm hard,0}^{\rm II} = T_0 \left[\frac{n_{\rm mix}(x,y,b) \, 
e^{-\varsigma^2/(2\sigma_\varsigma^2)}}{n_{\rm mix}(0,0,0)}  \right]^{1/3} \quad;\quad {\rm Initial\;Condition\;II}
\; .
\label{eq:phardinit-II}
\eqa

\subsection{Allowing for initial momentum-space anisotropy}

If the initial momentum-space anisotropy is assumed to be zero, i.e. $\xi_0=0$, then Eqs.~(\ref{eq:phardinit-I}) and
(\ref{eq:phardinit-II}) can be used without modification.  However, if $\xi_0 \neq 0$ one should require
that the same initial density profile is obtained.  Using the fact that $n(\phard,\xi) = n_{\rm iso}(\phard)/\sqrt{1+\xi}
\propto \phard^3/\sqrt{1+\xi}$
one finds that this requires $T_0(\xi_0) = (1+\xi)^{1/6} T_{0,\rm iso}$.

We must note, however, for completeness sake, that one could also have a non-trivial dependence of the 
initial anisotropy on the transverse direction and spatial rapidity.  In fact, one expects that towards
the transverse and longitudinal edges of the plasma that the initial momentum-space anisotropies should
be larger; however, at this point in time there is no first principles calculation of the ${\bf x}_\perp$ and
$\varsigma$ dependence of $\xi$ at the earliest times after the collision, 
so here we will choose the simplest possibility, which is that it is a constant and equal to zero.  We will
explore the possibility of finite initial momentum-space anisotropy in future works.

\section{Computing the suppression factor}
\label{sec:supfac}

The {\sc aHydro} time evolution gives us $\phard$ and $\xi$ as a function of proper time, transverse coordinate 
${\bf x}_\perp$, and spatial rapidity $\varsigma$.  Solution of the Schr\"odinger equation gives us the real and 
imaginary parts of the binding energy of a given state as a function of $\phard$ and $\xi$.  Putting this together 
gives us the real and imaginary parts of the binding energy 
as a function of proper time, transverse coordinate ${\bf x}_\perp$, and 
spatial rapidity $\varsigma$: $\Re[E_{\rm bind}(\tau,{\bf x}_\perp,\varsigma)]$ and 
$\Im[E_{\rm bind}(\tau,{\bf x}_\perp,\varsigma)]$, respectively.  

If the real part of the binding energy is positive, then the state 
is bound.  If the real part of the binding energy is negative, then the
state is unbound.  The imaginary part of the binding energy will give
us information about the decay rate of the state in question.  To see
the exact relationship we can compute the quantum mechanical occupation
number as a function of proper time
\bqa
n_\upsilon(\tau) &=& \langle \phi_\upsilon^*(\tau,{\bf x}) \phi_\upsilon(\tau,{\bf x}) \rangle \nonumber \, , \\
&=&  \langle \left(\phi_\upsilon({\bf x}) e^{-i E \tau}\right)^* \left(\phi_\upsilon({\bf x}) e^{-i E \tau}\right)  \rangle
\nonumber \, , \\
&=&  \langle\phi_\upsilon^*({\bf x})\phi_\upsilon({\bf x}) \rangle e^{2 \Im[E] \tau} \, , \nonumber \\
&=&  n^0_\upsilon \, e^{2 \Im[E] \tau} \, ,
\eqa
where in the last line we have identified $n^0_\upsilon  = 
\langle\phi_\upsilon^*({\bf x})\phi_\upsilon({\bf x}) \rangle$.  In order to
connect this to the decay rate, $\Gamma$, we note that $\Gamma$
is defined empirically through $n_\upsilon(t) = 
n^0_\upsilon \, \exp(-\Gamma \tau)$ so that we can identify $\Gamma = -2 \Im[E]$.
Finally, from Eq.~(\ref{bsbindingenergy})  we have $\Im[E_{\rm bind}] = - \Im[E]$ 
so that
\beq
\Gamma(\tau,{\bf x}_\perp,\varsigma) = 
\left\{
\begin{array}{ll}
2 \Im[E_{\rm bind}(\tau,{\bf x}_\perp,\varsigma)]  & \;\;\;\;\; \Re[E_{\rm bind}(\tau,{\bf x}_\perp,\varsigma)] >0 \\
10\;{\rm GeV}  & \;\;\;\;\; \Re[E_{\rm bind}(\tau,{\bf x}_\perp,\varsigma)] \le 0 \\
\end{array}
\right.
\eeq
The value of 10 GeV in the second case is chosen to be large in order to quickly suppress states which are
fully unbound.  We have checked the sensitivity of our results to this value and find that there is very little
dependence on this number as long as it is greater than 1 GeV such that the states are suppressed quickly within
the plasma lifetime.  In addition, we set the width to zero if the imaginary part of the binding energy is
less than zero.  Negative values of the imaginary part of the binding energy occur only at large values 
of $\xi$ and are a result of the small-$\xi$ expansion being applied outside of its range of applicability.  
Since large $\xi$ corresponds to a (nearly) free streaming plasma, one expects that the widths should return 
to their vacuum values ($\sim$ keV) justifying this choice.

We can integrate the instantaneous decay rate, $\Gamma$, 
over proper-time to extract the dimensionless logarithmic suppression factor
\beq
\zeta(p_T,{\bf x}_\perp,\varsigma) \equiv \Theta(\tau_f-\tau_{\rm form}(p_T)) \int_{{\rm max}(\tau_{\rm form}(p_T),\tau_0)}^{\tau_f} 
d\tau\,\Gamma(\tau,{\bf x}_\perp,\varsigma) \, ,
\label{eq:zeta}
\eeq
where $\tau_{\rm form}(p_T)$ is the lab-frame formation time of the state in question.
The formation time of a state in its local rest frame 
can be estimated by the inverse of its vacuum binding energy~\cite{Karsch:1987uk}.
In the lab frame the formation time depends
on the transverse momentum of the state via the gamma factor $\tau_{\rm form}(p_T) = \gamma 
\tau_{\rm form}^0 = E_T \tau_{\rm form}^0/M$ where $M$ is the mass of the relevant state and
$\tau_{\rm form}^0$ is the formation time of the state in its local rest frame.
For the formation times for the $\Upsilon(1s)$, $\Upsilon(2s)$, $\Upsilon(3s)$, $\chi_{b1}$ and
$\chi_{b2}$ states we take $\tau_{\rm form}^0$ = 0.2 fm/c, 0.4 fm/c, 0.6 fm/c, 0.4 fm/c, and
0.6 fm/c, respectively.

We take the initial proper time $\tau_0$ for plasma evolution to be $\tau_0 =$ 0.3 fm/c at both RHIC and LHC energies.
The final time, $\tau_f$, is defined to be the proper time when the local energy density becomes less than that 
of an $N_c=3$ and $N_f=2$ ideal gas of quark and gluons with a temperature
of $T = 192$ MeV. At this energy density, plasma screening effects are assumed to decrease
rapidly due to the transition to the hadronic phase and the widths of the states will become
approximately equal to their vacuum widths.

From $\zeta$ obtained via Eq.~(\ref{eq:zeta}) we can directly compute the suppression factor $R_{AA}$
\beq
R_{AA}(p_T,{\bf x}_\perp,\varsigma) = e^{-\zeta(p_T,{\bf x}_\perp,\varsigma)} \, .
\eeq
For averaging over transverse momenta and implementing any cuts necessary we assume that 
all states have a $1/E_T^4$ spectrum which is consistent with the high-$p_T$ spectra measured by CDF
\cite{Acosta:2001gv}.  Integrating over transverse momentum given $p_T$-cuts $p_{T,\rm min}$ and
$p_{T,\rm max}$ we obtain the $p_T$-cut suppression factor
\beq
R_{AA}({\bf x}_\perp,\varsigma) \equiv \frac{\int_{p_{T,\rm min}}^{p_{T,\rm max}} 
dp_T^2 \, R_{AA}(p_T,{\bf x}_\perp,\varsigma) /(p_T^2 + M^2)^2}{\int_{p_{T,\rm min}}^{p_{T,\rm max}}dp_T^2/(p_T^2 + M^2)^2} \, .
\eeq
For implementing cuts in centrality we compute $R_{AA}$ for finite impact parameter $b$ and map centrality
to impact parameter in the standard manner.  For the cuts over centrality and rapidity, we use a 
flat distribution.

In order to compare with experimental observations we should finally average $R_{AA}({\bf x}_\perp,\varsigma)$
over ${\bf x}_\perp$.  For this operation we use a production probability distribution which is set by the overlap density
specified in Eq.~(\ref{eq:nab})
\beq
\langle R_{AA}(\varsigma) \rangle \equiv 
\frac{\int_{{\bf x}_\perp} \! d{\bf x}_\perp \, n_{AA}({\bf x}_\perp)\,%
R_{AA}({\bf x}_\perp,\varsigma)}{\int_{{\bf x}_\perp} \! d{\bf x}_\perp \,%
n_{AA}({\bf x}_\perp)} \, .
\label{eq:geoaverage}
\eeq

\begin{table}[t]
\begin{center}
\begin{tabular}{|l|l|l|}
\hline
\multicolumn{3}{|c|}{$\Upsilon(1s)$ Production}\\\hline
\hline
{\bf Mechanism} & {\bf \% $\pm$ Stat $\pm$ Sys} \cite{Affolder:1999wm} & {\bf $f_i$ used herein} \\\hline
Direct Production & 50.9 $\pm$ 8.2 $\pm$ 9.0 &  0.51 \\\hline
$\Upsilon(2s)$ decay&  10.7 $\pm$ 7.7 $\pm$ 4.8 & 10.7 \\\hline
$\Upsilon(3s)$ decay&  0.8 $\pm$ 0.6 $\pm$ 0.4 & 0.8 \\\hline
$\chi_{b1}$ decay&  27.1 $\pm$ 6.9 $\pm$ 4.4 & 27\\\hline
$\chi_{b2}$ decay&  10.5 $\pm$ 4.4 $\pm$ 1.4 & 10.5\\\hline
\end{tabular}
\end{center}
\caption{Feed down fractions extracted from experiment \cite{Affolder:1999wm} including errors (middle column) 
and the value chosen for use herein (right column).  Values of $f_i$ are constrained such that $\sum_i f_i = 1$.}
\label{tab:feeddown}
\end{table}

\section{Excited State Feed Down}
\label{sec:feeddown}

Since a certain fraction of $\Upsilon(1s)$ states produced in high energy collisions come
from the decay of excited states, when computing the full (inclusive) $R_{AA}$ for the $\Upsilon(1s)$
one must also consider the suppression of the excited states which decay or ``feed down'' to it.
In order to fix the feed down fractions we use data from $\sqrt{s} = $ 1.8 TeV pp collisions at 
CDF \cite{Affolder:1999wm} with a cut $p_T^\Upsilon > $ 8.0 GeV/c.  The resulting 
feed down fractions are listed in Table~\ref{tab:feeddown}.

Based on these numbers, we can construct the full (or inclusive) $\Upsilon(1s)$ $R_{AA}$ including the effect
of the suppression of excited states via
\beq
R_{AA}^{\rm full}[\Upsilon(1s)] = \sum_{i\,\in\,{\rm states}} f_i \,R_{i,AA} \, ,
\eeq
where $R_{i,AA}$ is the direct suppression of the $i^{\rm th}$ state and the production fractions, $f_i$, 
are given in Table~\ref{tab:feeddown}.

\section{Results for $R_{AA}$}
\label{sec:results}

In this section we present our main results which consist of the suppression factors $R_{AA}$ for the 
$\Upsilon(1s)$, $\Upsilon(2s)$, and $\Upsilon(3s)$, $\chi_{b1}$ and $\chi_{b2}$.  We will present
each state's suppression factor as a function of centrality (number of participants) and rapidity.
We will then compute the inclusive $R_{AA}$ for the $\Upsilon(1s)$ including the feed effect as
described in Section~\ref{sec:feeddown}.  To close the section we will present the inclusive  $R_{AA}$ for the $\Upsilon(1s)$
as a function of transverse momentum and investigate the sensitivity to the choice of the type of initial conditions
used.

\begin{figure}[t]
\begin{center}
\includegraphics[width=8cm]{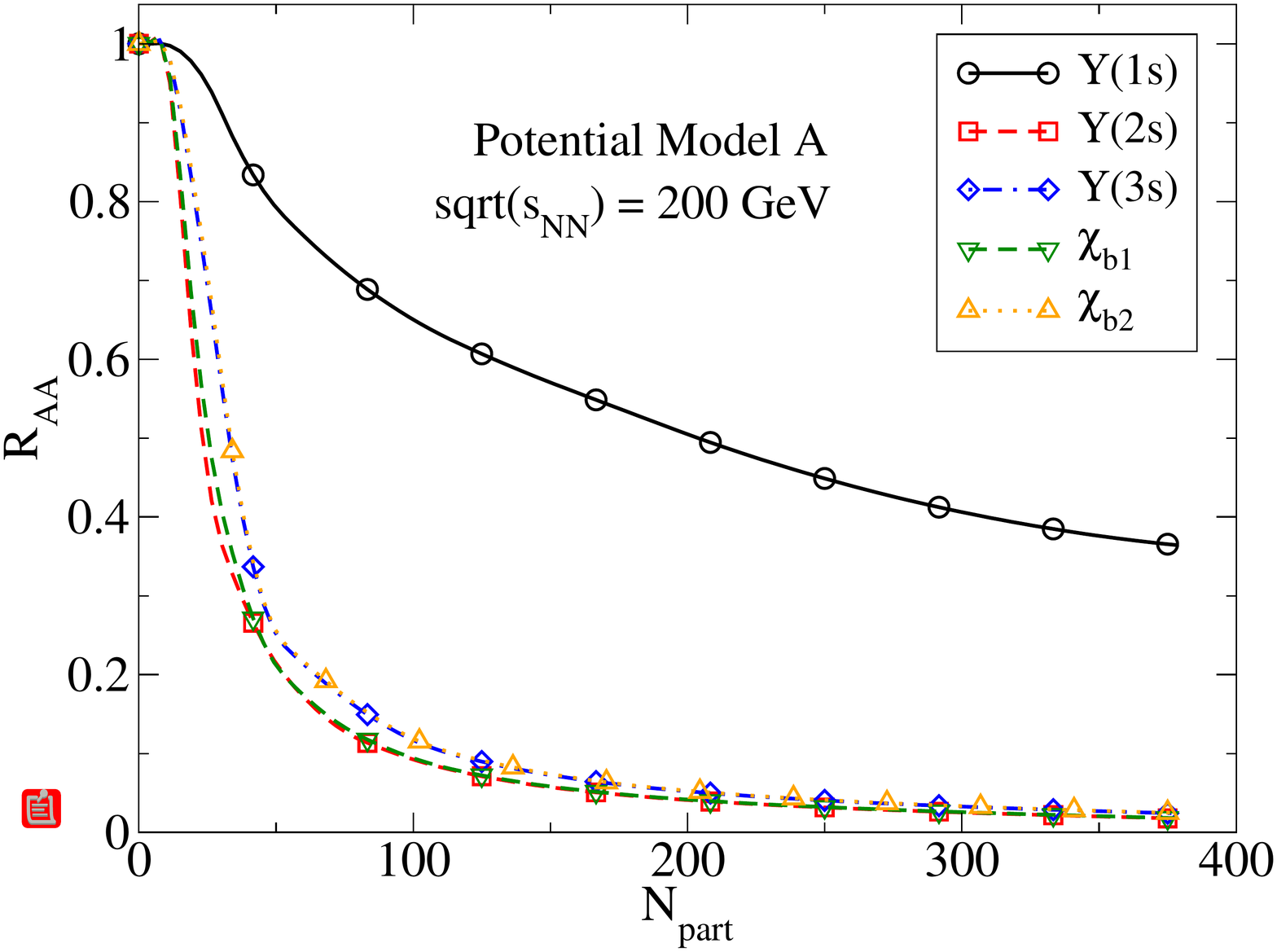}
\includegraphics[width=8cm]{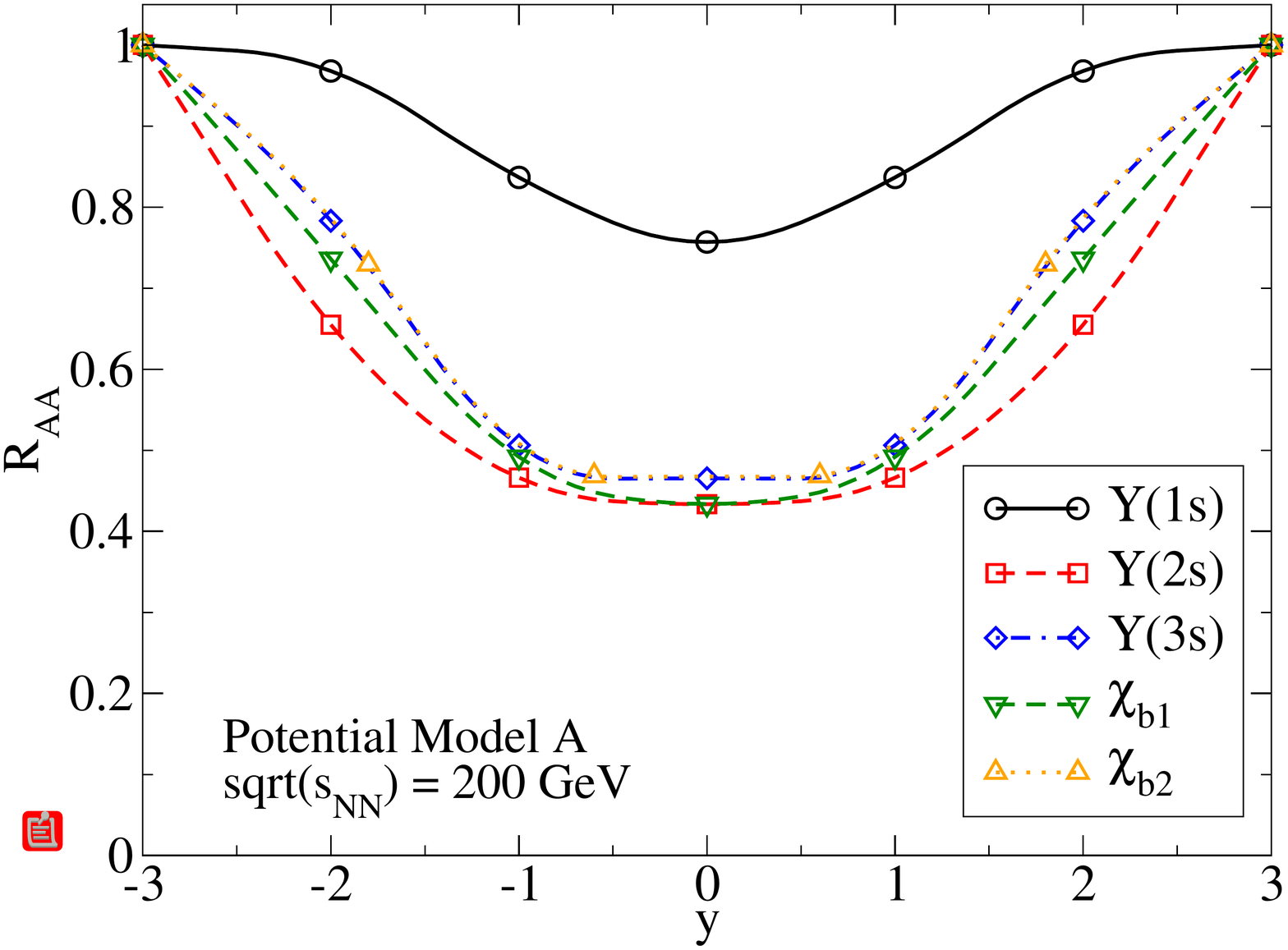}
\\
\includegraphics[width=8cm]{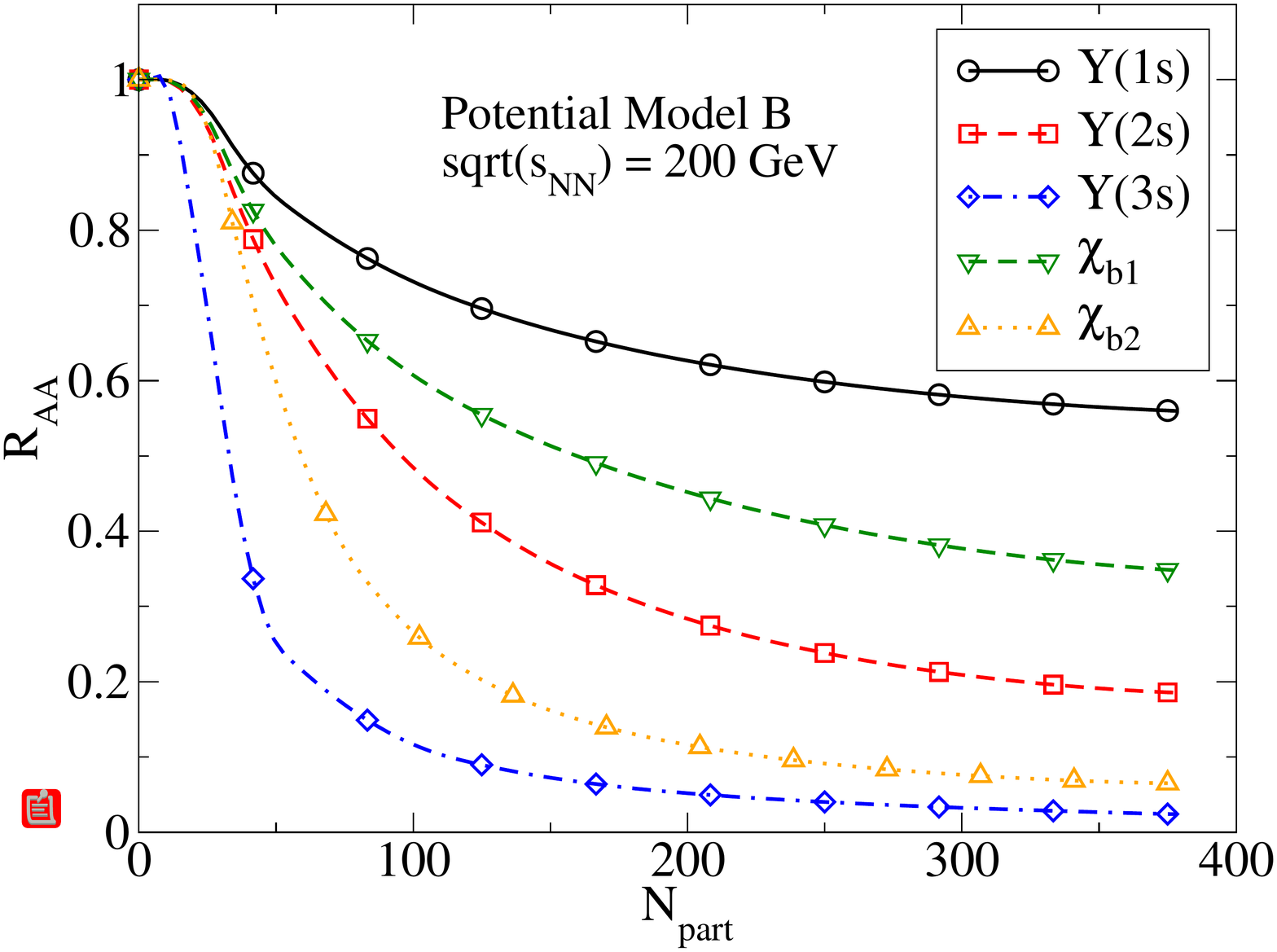}
\includegraphics[width=8cm]{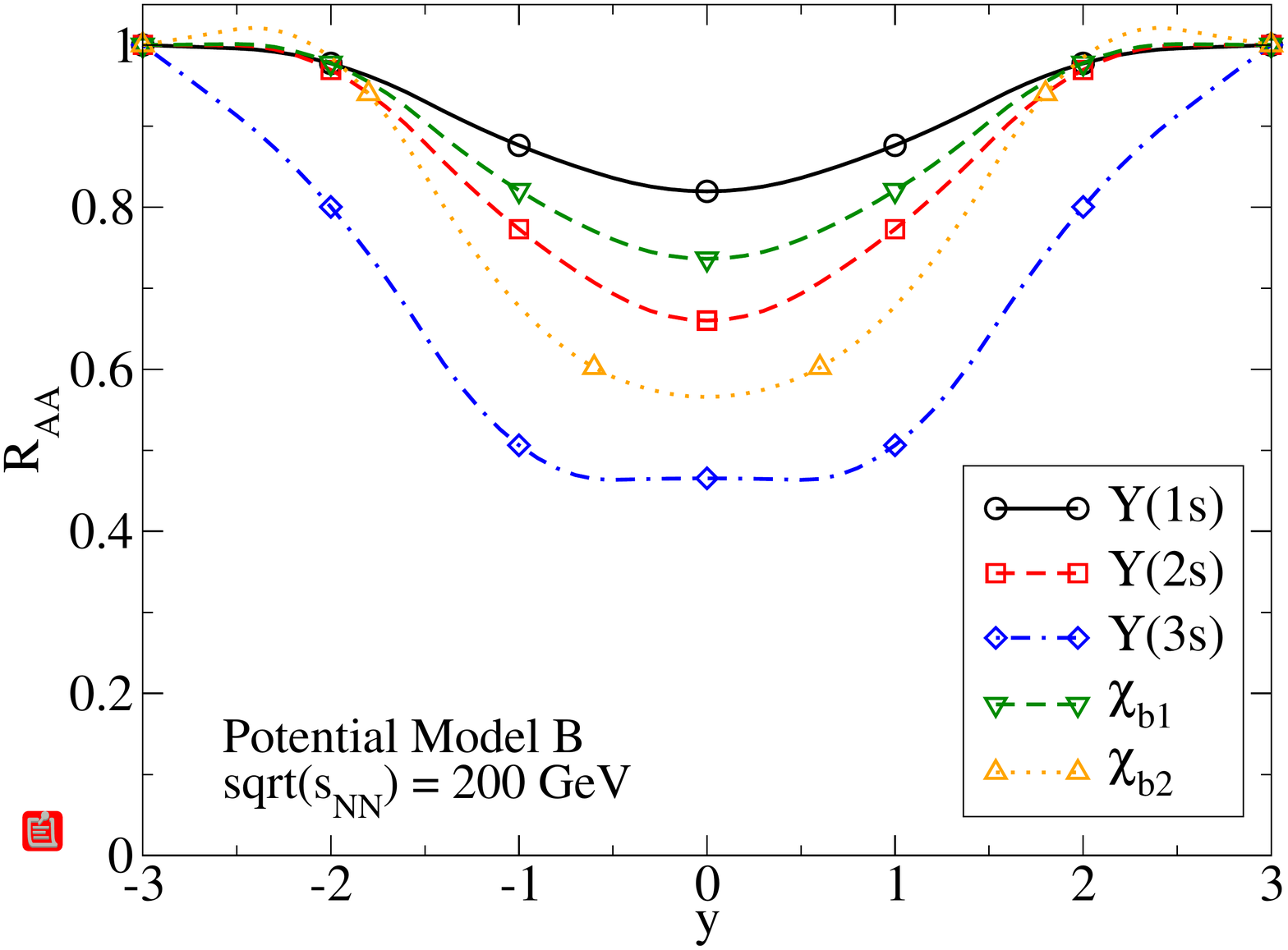}
\end{center}
\vspace{-6mm}
\caption{
RHIC suppression factor $R_{AA}$ for the $\Upsilon(1s)$, $\Upsilon(2s)$, $\Upsilon(3s)$, $\chi_{b1}$, and $\chi_{b2}$ 
states as a function of the number of participants (left) and rapidity (right).  The top row uses potential model A 
(\ref{eq:potmodela}) and the bottom row uses potential model B (\ref{eq:potmodelb}).  In all plots we used 
$\sqrt{s_{NN}} = 200$ GeV, assumed a shear viscosity to entropy density ratio of $4 \pi \eta/{\cal S} = 1$, and 
implemented cuts of $0 < p_T < 20$ GeV and (left) rapidity $|y| < 0.5$ (right) centrality 0-100\%.
}
\label{fig:raa-rhic-states}
\end{figure}

\begin{figure}[t]
\begin{center}
\includegraphics[width=8cm]{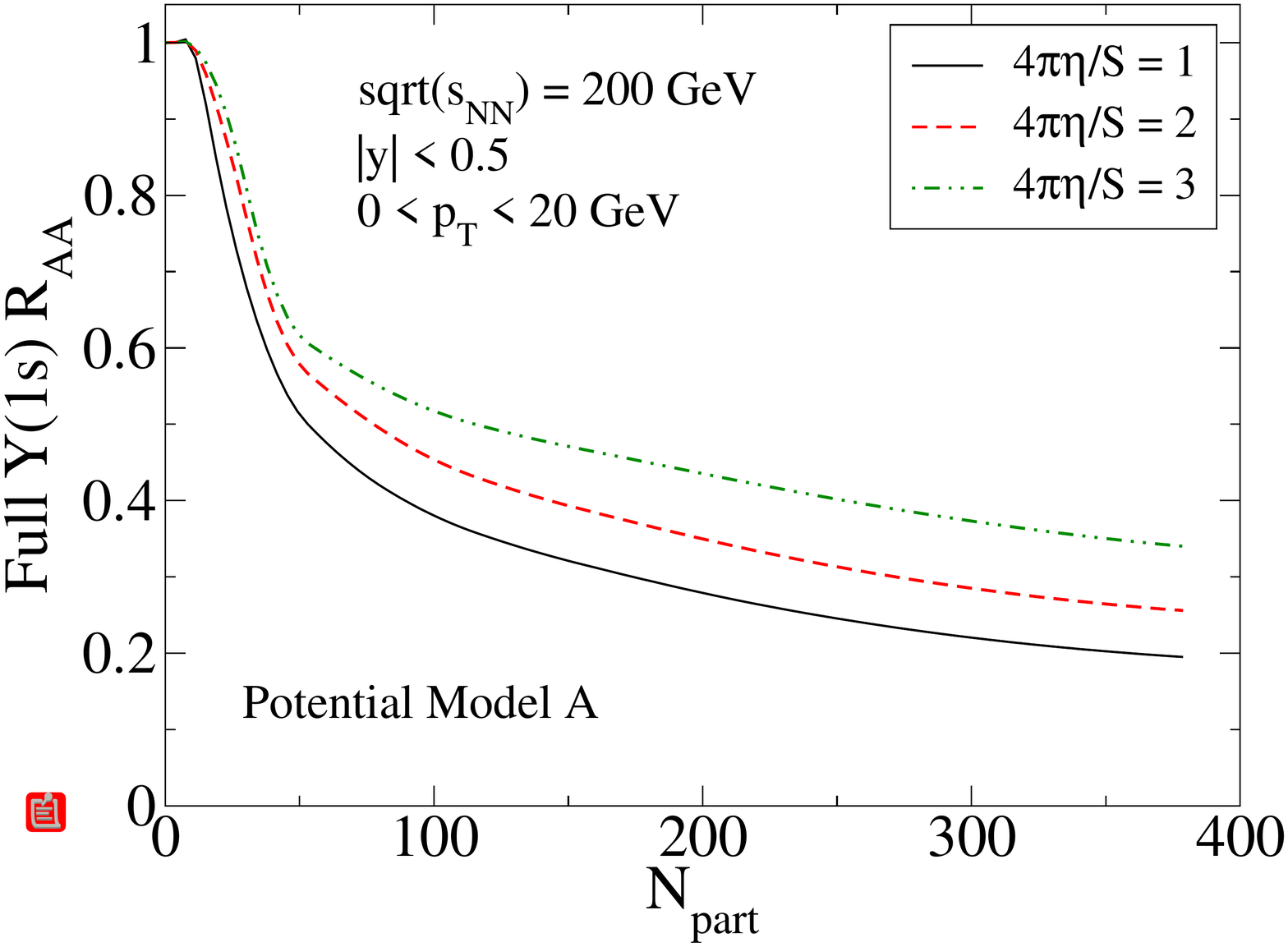}
\includegraphics[width=8cm]{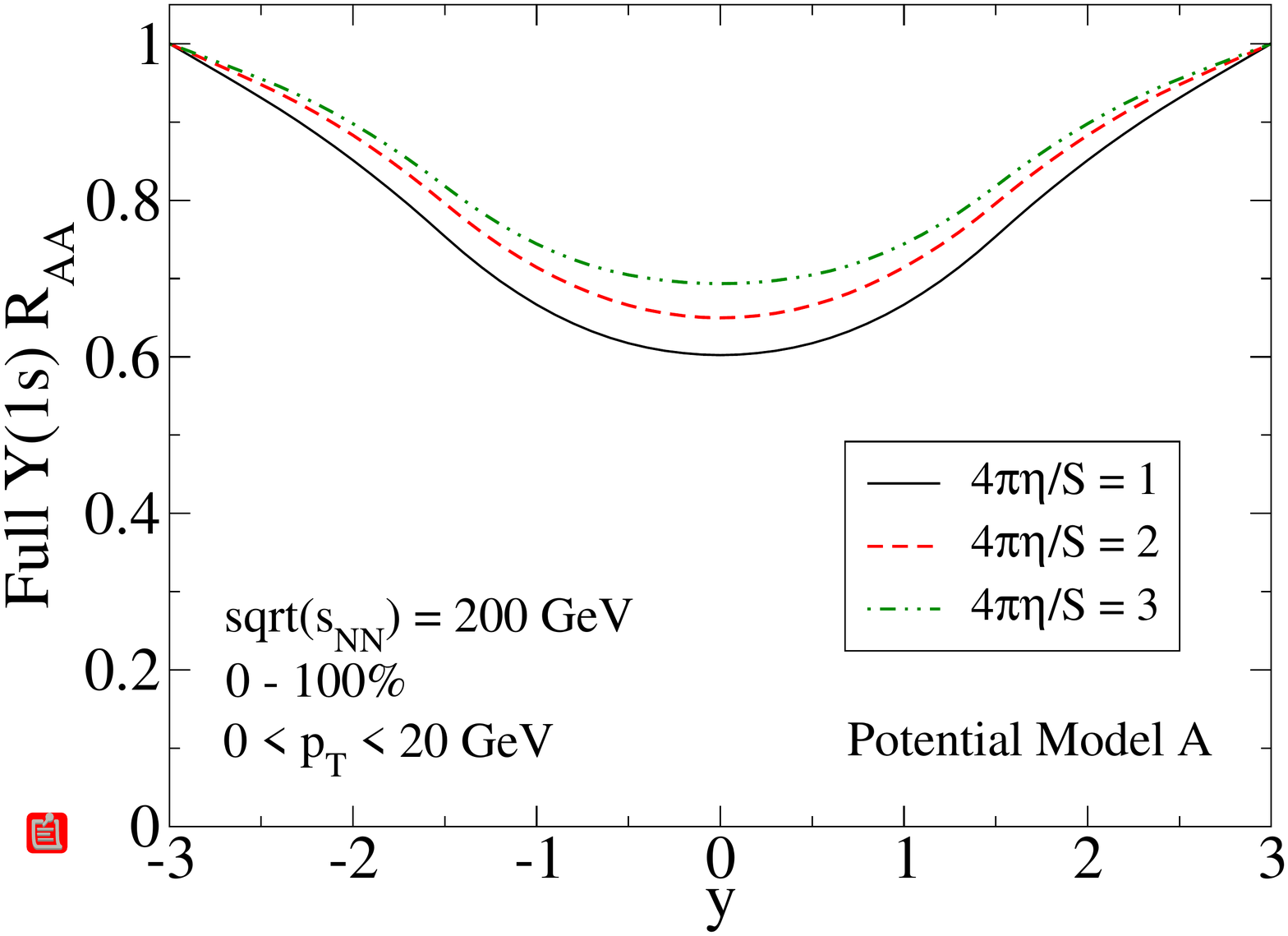}\\
\includegraphics[width=8cm]{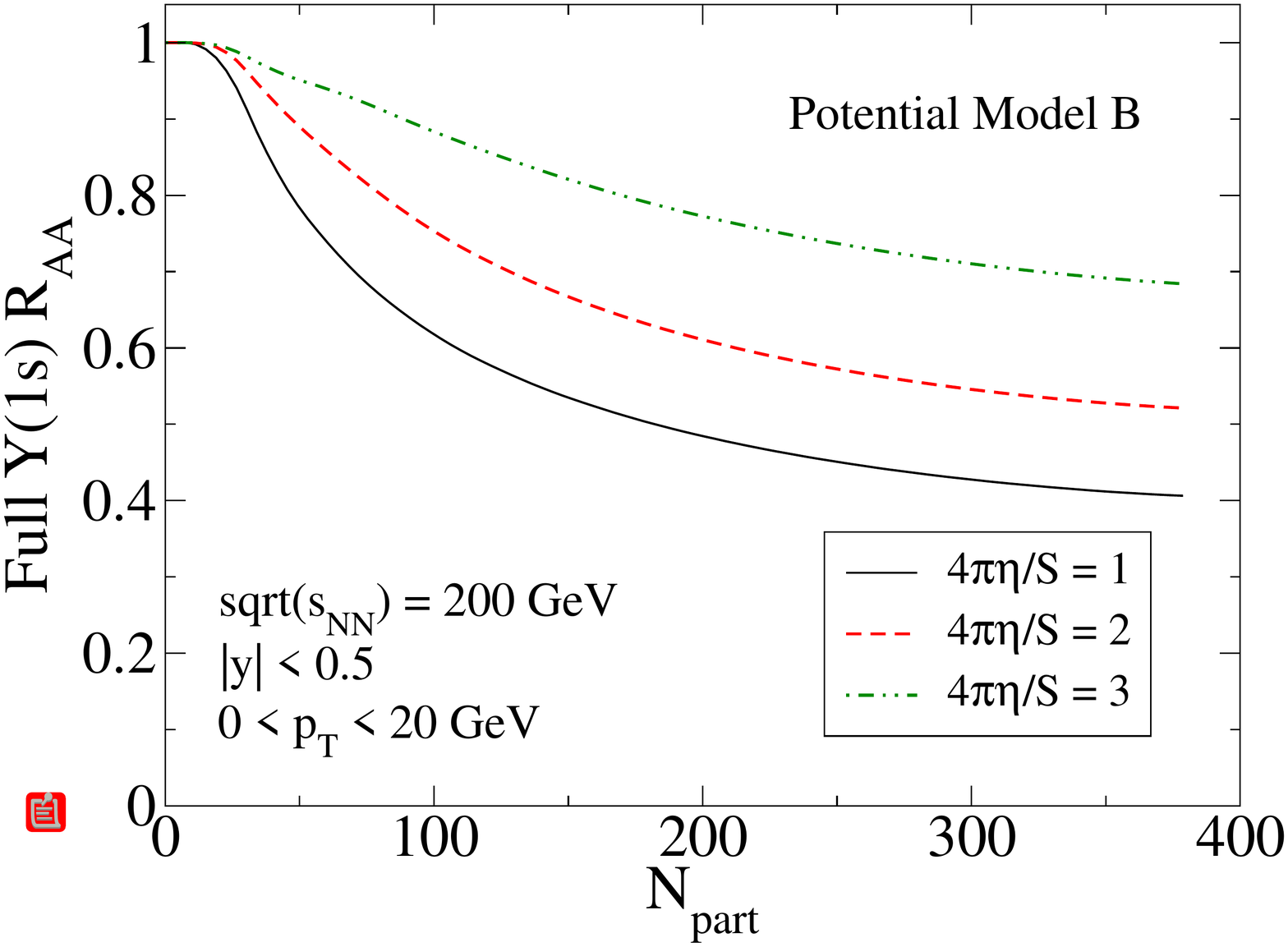}
\includegraphics[width=8cm]{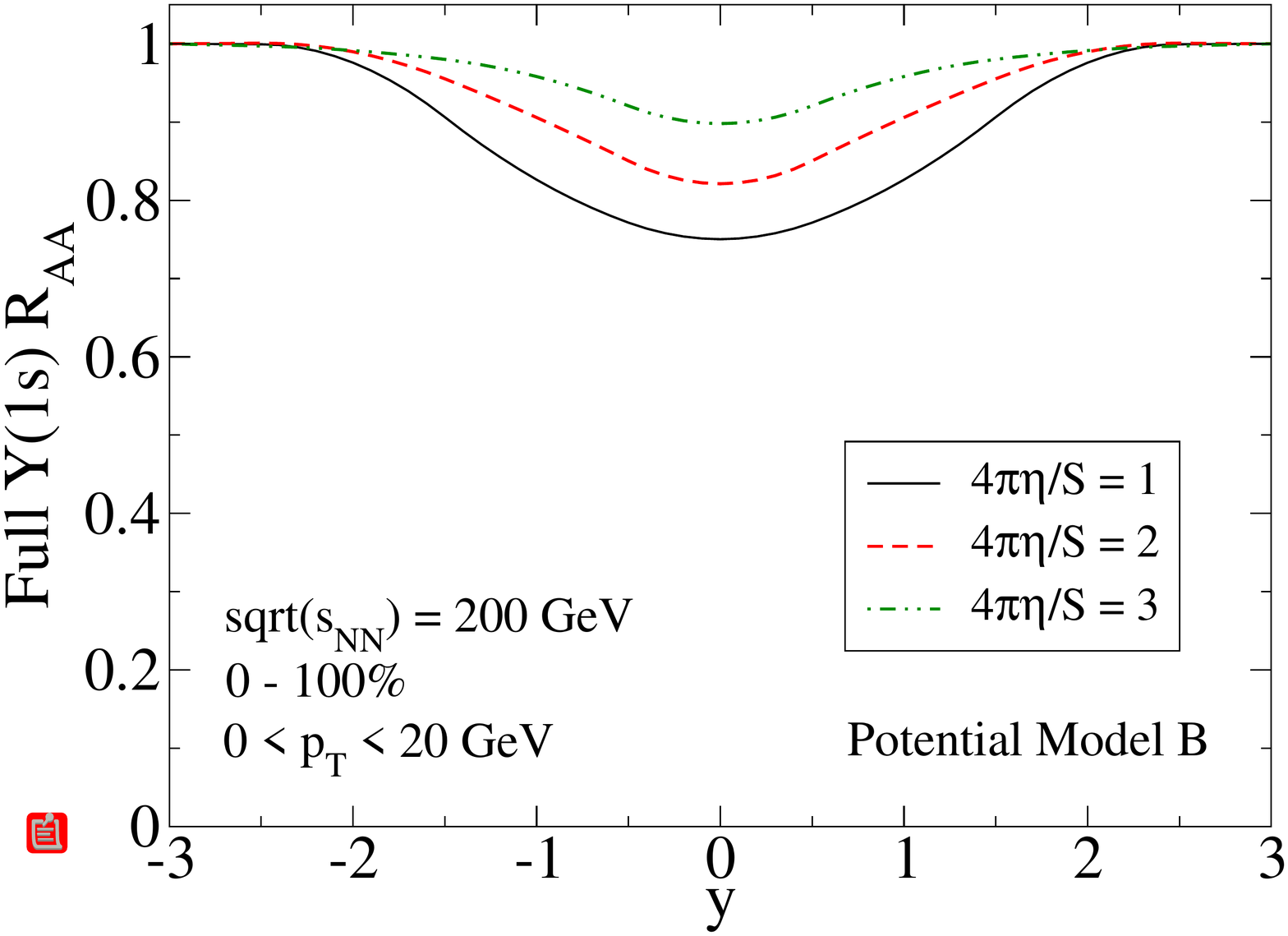}
\end{center}
\vspace{-6mm}
\caption{
RHIC inclusive or ``full'' suppression factor $R_{AA}$ for the $\Upsilon(1s)$ including feed down effects.  The
three different lines correspond to different assumptions for the shear viscosity to entropy ratio 
$4 \pi \eta/{\cal S} \in \{1,2,3\}$.  In all plots we used $\sqrt{s_{NN}} = 200$ GeV and 
implemented cuts of $0 < p_T < 20$ GeV and and (left) rapidity $|y| < 0.5$ (right) centrality 0-100\%.
}
\label{fig:y1s-rhic}
\end{figure}

\subsection{Suppression at RHIC Energies}
\label{sec:rhicresults}

The highest energy RHIC runs collide gold nuclei at a collision energy of $\sqrt{s_{NN}} = 200$ GeV.  In this
subsection we will focus on the resulting using wounded-nucleon (or participant) scaling for the initial condition
with $\sigma_{NN} = $ 42 mb.
Fixing the initial time for the {\sc aHydro} evolution to $\tau_0 = $ 0.3 fm/c and requiring that the final
charged particle multiplicity is fixed to $dN_{\rm ch}/dy = 620$, we find that for $4\pi\eta/{\cal S} = \{1,2,3\}$ 
we must fix the initial central temperature for a central collision to be $T_0 = \{ 442, 433, 428 \}$ MeV.  The decrease 
of the initial central temperature with increasing $\eta/{\cal S}$ is a result of the fact that one has
more entropy generation as $\eta/{\cal S}$ increases.  As a result, it is necessary to lower the initial temperature in
order to allow for particle production.

In Fig.~\ref{fig:raa-rhic-states} we show the predicted suppression factor $R_{AA}$ for the $\Upsilon(1s)$, $\Upsilon(2s)$, 
$\Upsilon(3s)$, $\chi_{b1}$, and $\chi_{b2}$ states as a function of the number of participants (left) and rapidity (right).
The top row uses potential model A  (\ref{eq:potmodela}) and the bottom row uses potential model B (\ref{eq:potmodelb}).  
In all plots we used  $\sqrt{s_{NN}} = 200$ GeV, assumed a shear viscosity to entropy density ratio of 
$4 \pi \eta/{\cal S} = 1$, and implemented cuts of $0 < p_T < 20$ GeV and and (left) rapidity $|y| < 0.5$ (right) centrality 
0-100\%.  As can be seen from this
figure, potential model A (\ref{eq:potmodela}) provides much more suppression than potential model B (\ref{eq:potmodelb}),
both as a function of number of participants and rapidity.  In both cases we see clear signs of sequential
suppression, with the higher excited states having stronger suppression than the ground state.  However, we note
that even for states that are melted at relatively low central temperatures, we still obtain a non-vanishing suppression
factor for these states.  This is due to the fact that near the edges, where the temperature is lower, one does not
see suppression of the states.  Upon performing the geometrical average prescribed in Eq.~(\ref{eq:geoaverage}) we see
that a large fraction of the states produced can survive even when the central temperature of the plasma is above 
their naive dissociation temperature.

In Fig.~\ref{fig:y1s-rhic} we show the inclusive suppression factor $R_{AA}^{\rm full}[\Upsilon(1s)]$ obtained using
the feed down prescription presented in Section~\ref{sec:feeddown}.  As can be seen from these figures,
potential model A (free energy) predicts much stronger suppression than potential model B (internal energy).  
As we can see the result has a significant dependence on the assumed shear viscosity to entropy density ratio.  This
could, in principle, be used to constrain $\eta/{\cal S}$ from RHIC data on bottomonium suppression.  

\begin{figure}[t]
\begin{center}
\includegraphics[width=8.1cm]{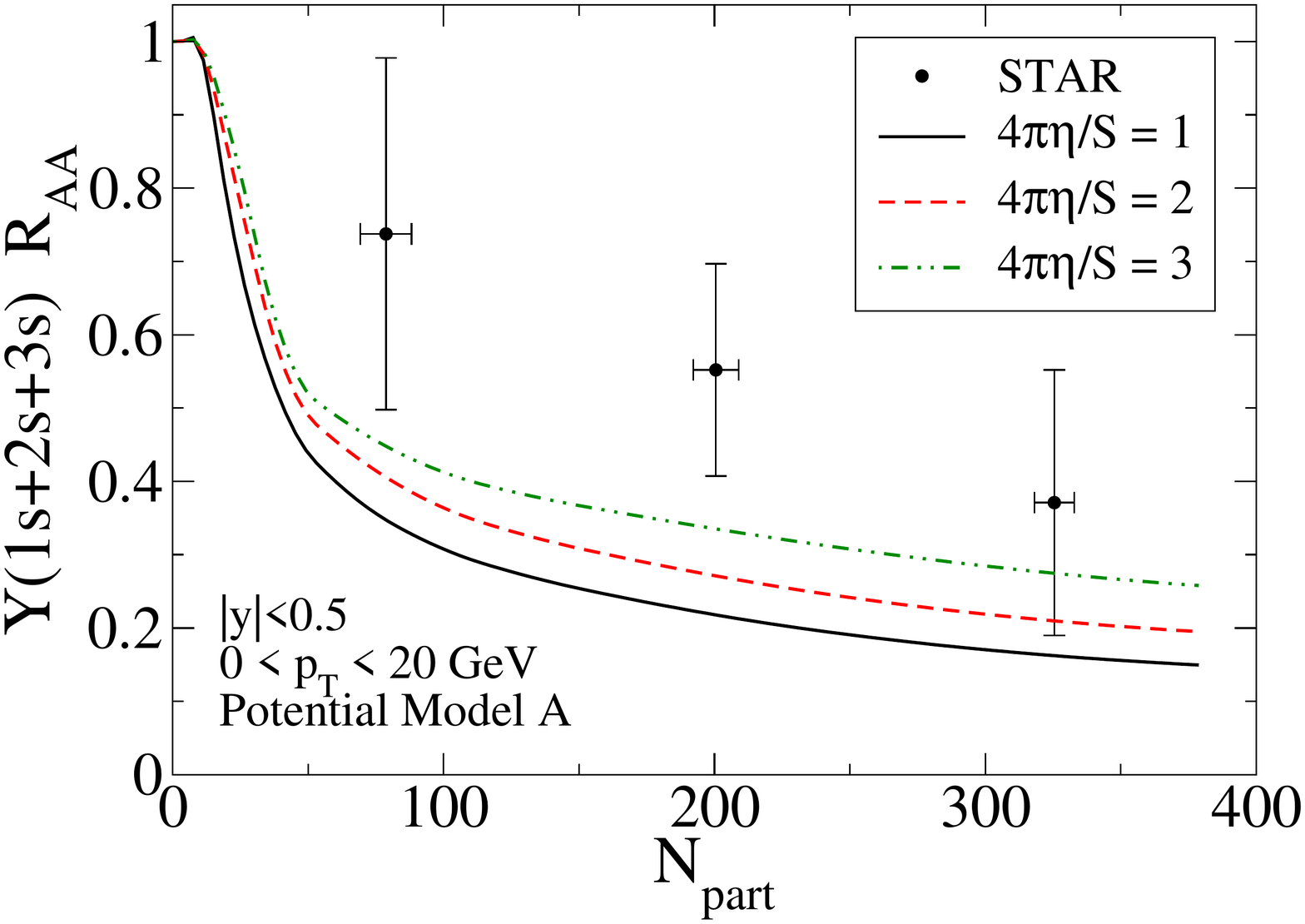}
\includegraphics[width=8.1cm]{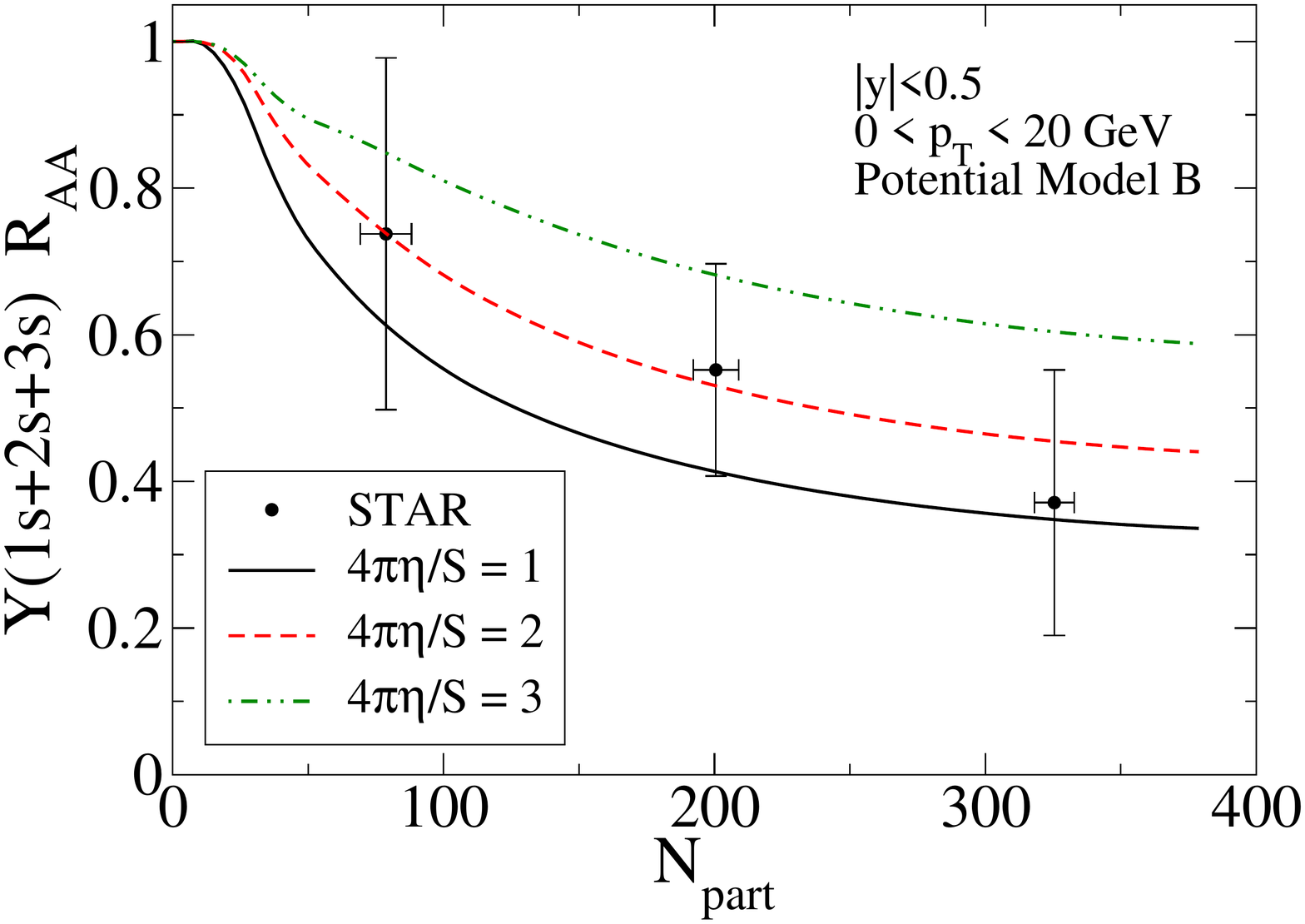}
\end{center}
\vspace{-6mm}
\caption{
RHIC $\Upsilon(1s+2s+3s)$ suppression factor determined via Eq.~(\ref{eq:raa1s2s3s}) compared with 
experimental data from the STAR Collaboration \cite{Reed:2011fr}.  The
three different lines correspond to different assumptions for the shear viscosity to entropy ratio 
$4 \pi \eta/{\cal S} \in \{1,2,3\}$.  In all plots we used $\sqrt{s_{NN}} = 200$ GeV and 
implemented cuts of $0 < p_T < 20$ GeV and $|y| < 0.5$.
}
\label{fig:y1s2s3s-rhic-npart}
\end{figure}

\subsubsection{$R_{AA}$ for $\Upsilon(1s+2s+3s)$ and comparison to STAR data}

Due to limited statistics and resolution the STAR Collaboration does not report separate suppression
factors for the $\Upsilon(1s)$, $\Upsilon(2s)$, and $\Upsilon(3s)$ states.  Instead, they compute an effective 
total suppression of all three
states by integrating the counts in a dielectron-invariant mass window which encompasses all
three states
\beq
R_{AA}[\Upsilon(1s+2s+3s)] \equiv 
\frac{\int_{m_-}^{m_+} d m_{\mu\mu} \, n_{\mu\mu}^{AA}}{n_{part}\int_{m_-}^{m_+} d m_{\mu\mu} \, n_{\mu\mu}^{pp}} \, ,
\eeq
where $m_-$ and $m_+$ are the dielectron pair invariant masses which cover the $\Upsilon(1s)$, $\Upsilon(2s)$, and $\Upsilon(3s)$
spectral peaks, e.g. $m_-$ = 8.5 GeV and $m_+ = 11$ GeV.  If the spectral peaks have approximately the same width and
are well separated, as is the case with these three states, then one finds that to good approximation
\beq
R_{AA}[\Upsilon(1s+2s+3s)] \simeq 
\frac{R_{AA}[\Upsilon(1s)] + c_{2s} R_{AA}[\Upsilon(2s)]  + c_{3s} R_{AA}[\Upsilon(3s)]}{1+c_{2s}+c_{3s}} \, ,
\label{eq:raa1s2s3s}
\eeq
where $c_{2s}$ and $c_{3s}$ are the ratios of the $\Upsilon(2s)$ and $\Upsilon(3s)$ states' background subtracted p-p 
peak heights to the $\Upsilon(1s)$ state's background subtracted p-p peak height, respectively.  From preliminary LHCb
results \cite{Callot:1390408} in the dimuon channel one finds $c_{2s} \simeq 0.24$ and $c_{3s} \simeq 0.11$.  These
values are consistent with CMS measurements of the $\Upsilon(2s)/\Upsilon(1s)$ and 
$\Upsilon(3s)/\Upsilon(1s)$ cross section ratios \cite{Khachatryan:2010zg}.  
We will use these values assuming that they are a good approximation to the relative p-p peak heights in the dielectron channel.

In Fig.~\ref{fig:y1s2s3s-rhic-npart} we plot $R_{AA}[\Upsilon(1s+2s+3s)] $ as determined using Eq.~(\ref{eq:raa1s2s3s})
and compare with 
experimental data from the STAR Collaboration \cite{Reed:2011fr}.
As can be seen from this figure, potential model A (free energy) gives too much suppression when compared to RHIC data.
One could argue that there could be some enhancement from regeneration; however, at RHIC, in particular, the number
of bottom and anti-bottom quarks generated on an event-by-event basis is incredibly small and therefore regeneration due
to recombination of the bottom and anti-bottom quarks is highly improbable.  Potential model B, on the other hand, does a 
very good job of reproducing the existing STAR data for $R_{AA}[\Upsilon(1s+2s+3s)]$.  From the right 
panel we can obtain an estimate of $\eta/{\cal S}$: $0.08 < \eta/{\cal S} < 0.24$.  Unfortunately, a more accurate 
determination will require more data with reduced statistical errors.

\begin{figure}[t]
\begin{center}
\includegraphics[width=8cm]{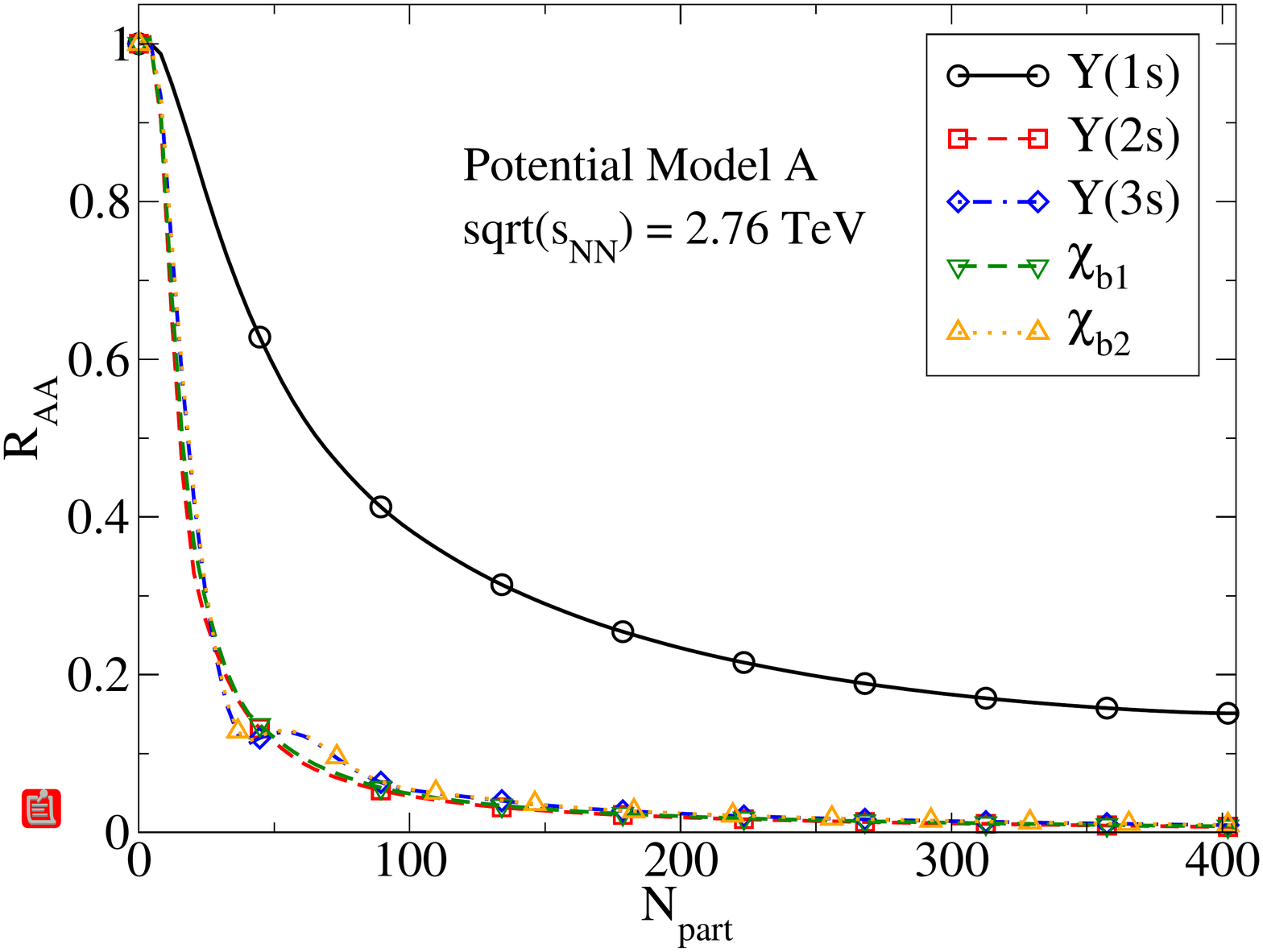}
\includegraphics[width=8cm]{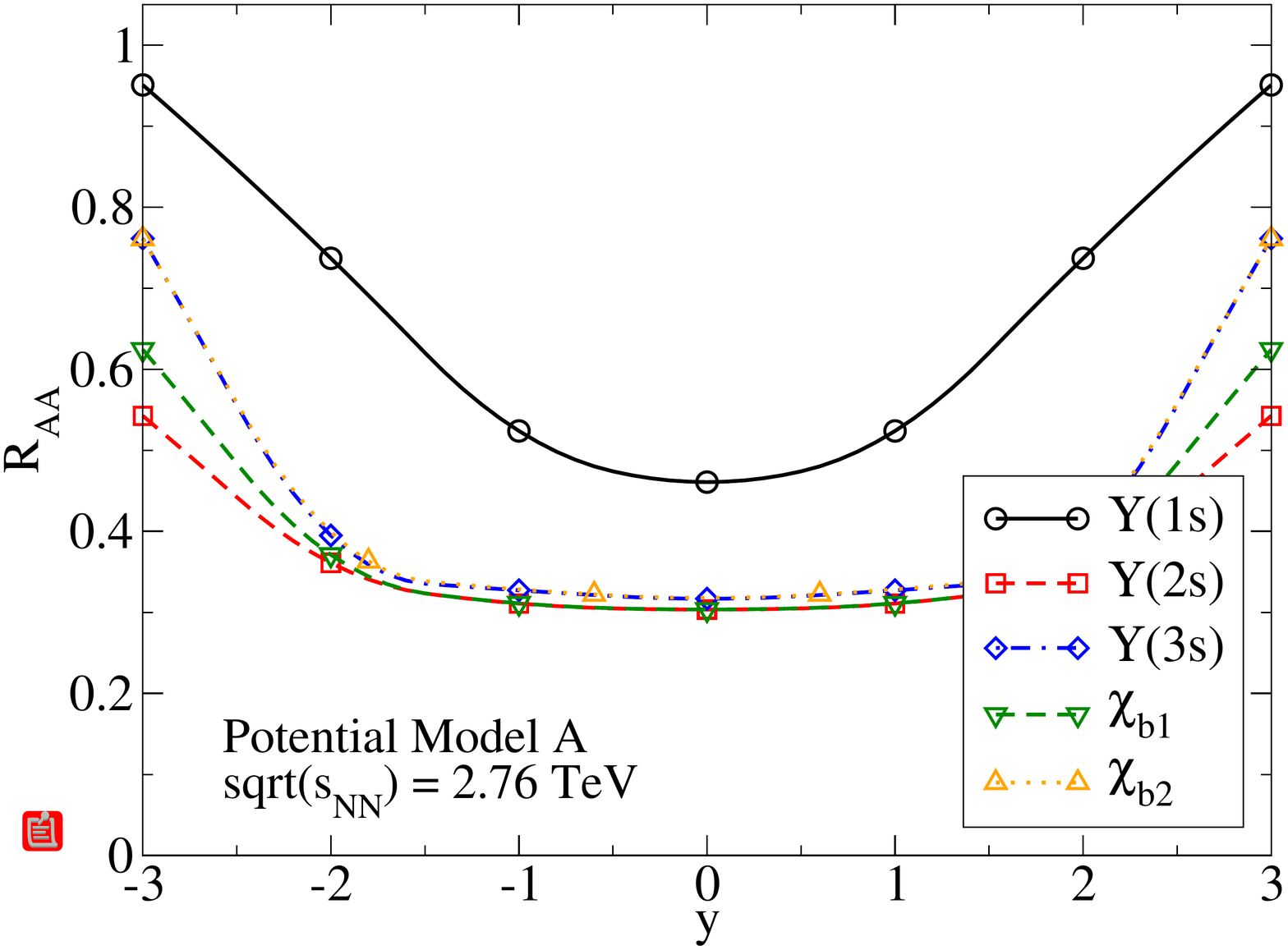}
\\
\includegraphics[width=8cm]{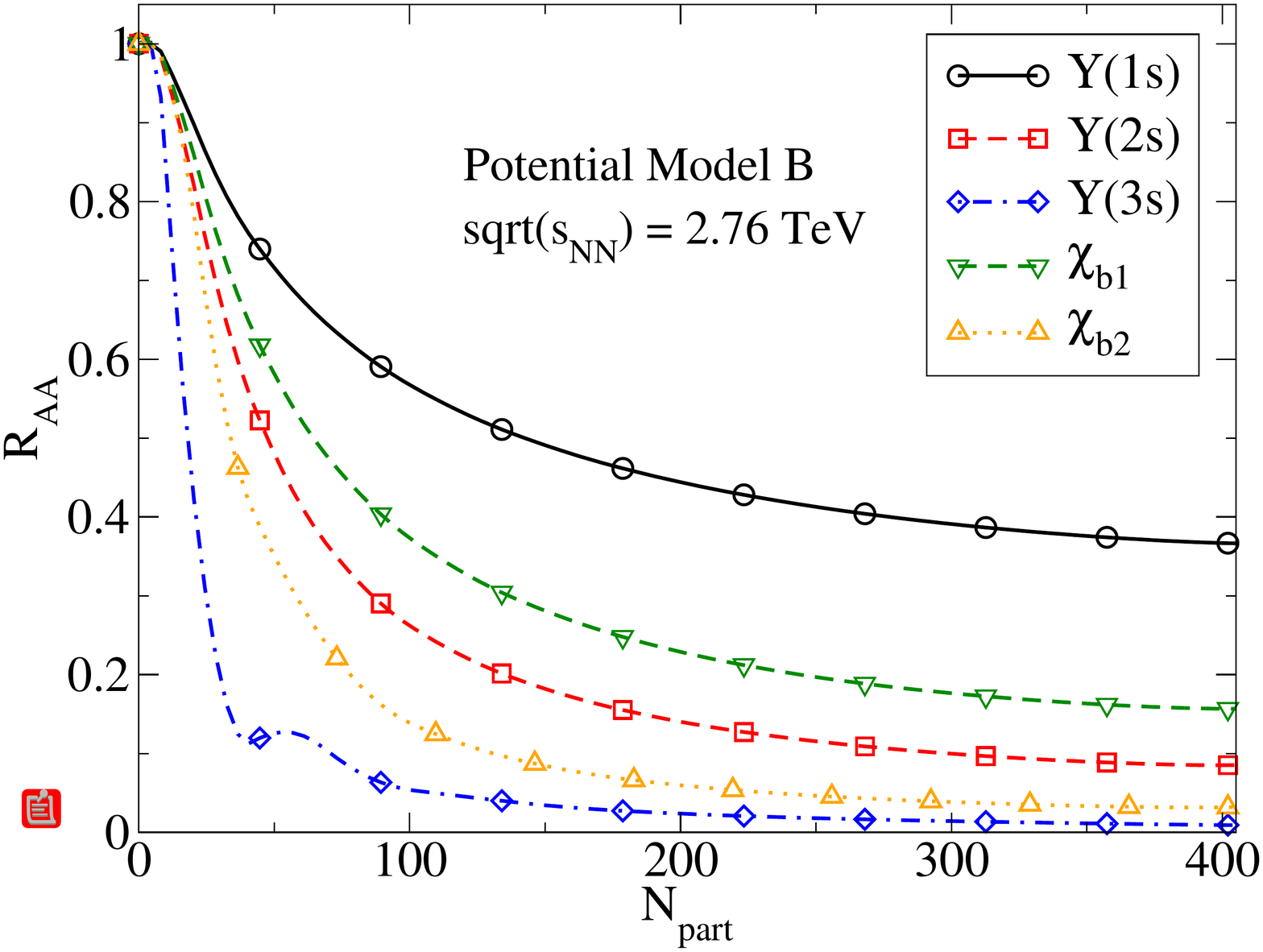}
\includegraphics[width=8cm]{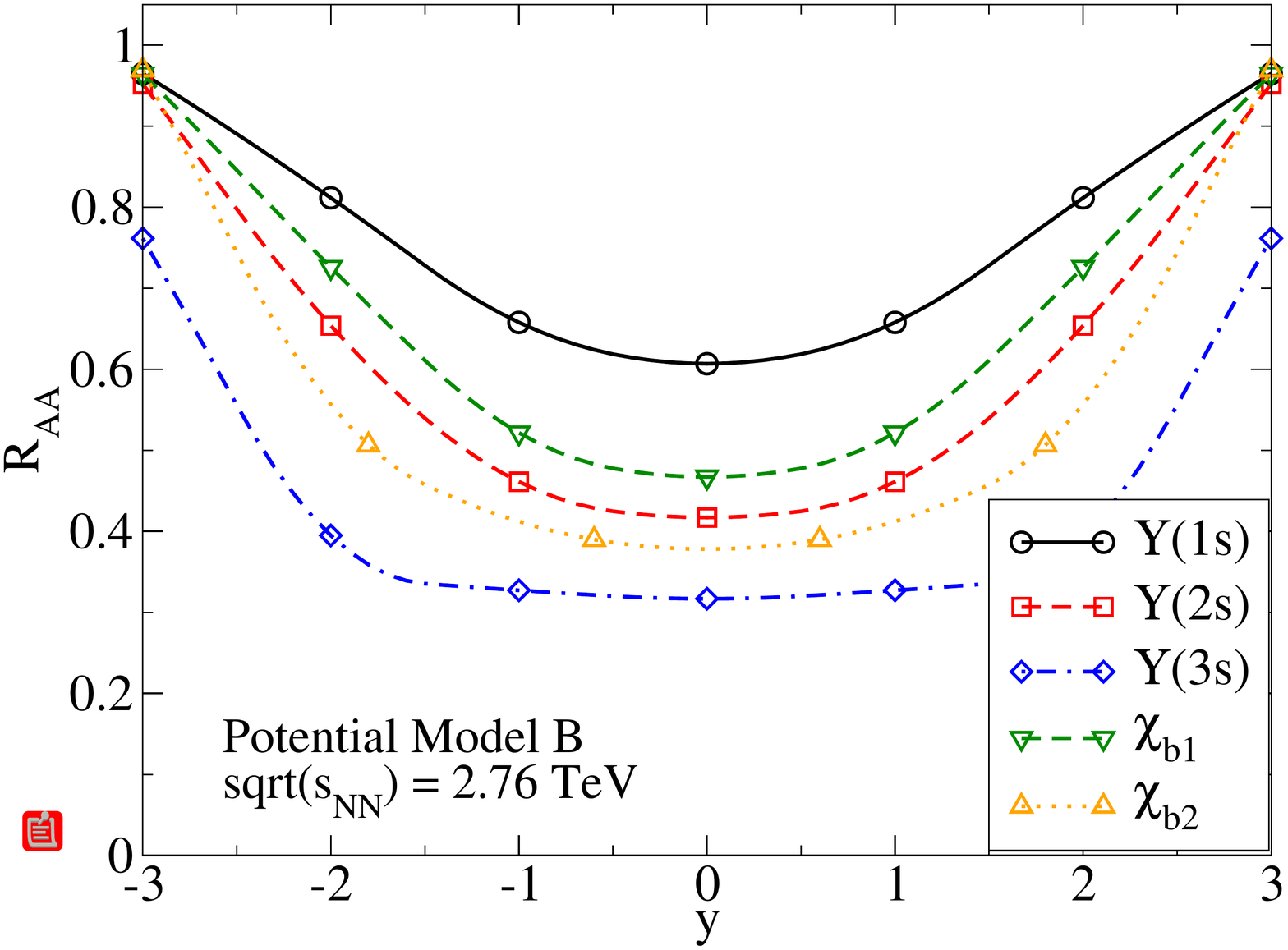}
\end{center}
\vspace{-6mm}
\caption{
LHC suppression factor $R_{AA}$ for the $\Upsilon(1s)$, $\Upsilon(2s)$, $\Upsilon(3s)$, $\chi_{b1}$, and $\chi_{b2}$ 
states as a function of the number of participants (left) and rapidity (right).  The top row uses potential model A 
(\ref{eq:potmodela}) and the bottom row uses potential model B (\ref{eq:potmodelb}).  In all plots we used 
$\sqrt{s_{NN}} = 2.76$ TeV, assumed a shear viscosity to entropy density ratio of $4 \pi \eta/{\cal S} = 1$, and 
implemented cuts of $0 < p_T < 20$ GeV and (left) rapidity $|y| < 2.4$ (right) centrality 0-100\%.
}
\label{fig:raa-lhc-states}
\end{figure}

\subsection{Suppression at LHC Energies}
\label{sec:lhcresults}

The current LHC runs collide lead nuclei at a collision energy of $\sqrt{s_{NN}} = 2.76$ TeV.  In this
subsection we will focus on the resulting using wounded-nucleon (or participant) scaling for the initial condition
with $\sigma_{NN} = $ 62 mb.
Fixing the initial time for the {\sc aHydro} evolution to $\tau_0 = $ 0.3 fm/c and requiring that the final
charged particle multiplicity is fixed to $dN_{\rm ch}/dy = 1400$, we find that for $4\pi\eta/{\cal S} = \{1,2,3\}$ 
we must fix the initial central temperature for a central collision to be $T_0 = \{ 567, 550, 539 \}$ MeV.  As before, 
the decrease of the initial central temperature with increasing $\eta/{\cal S}$ is a result of the fact that one has
more entropy generation as $\eta/{\cal S}$ increases.

\begin{figure}[t]
\begin{center}
\includegraphics[width=8cm]{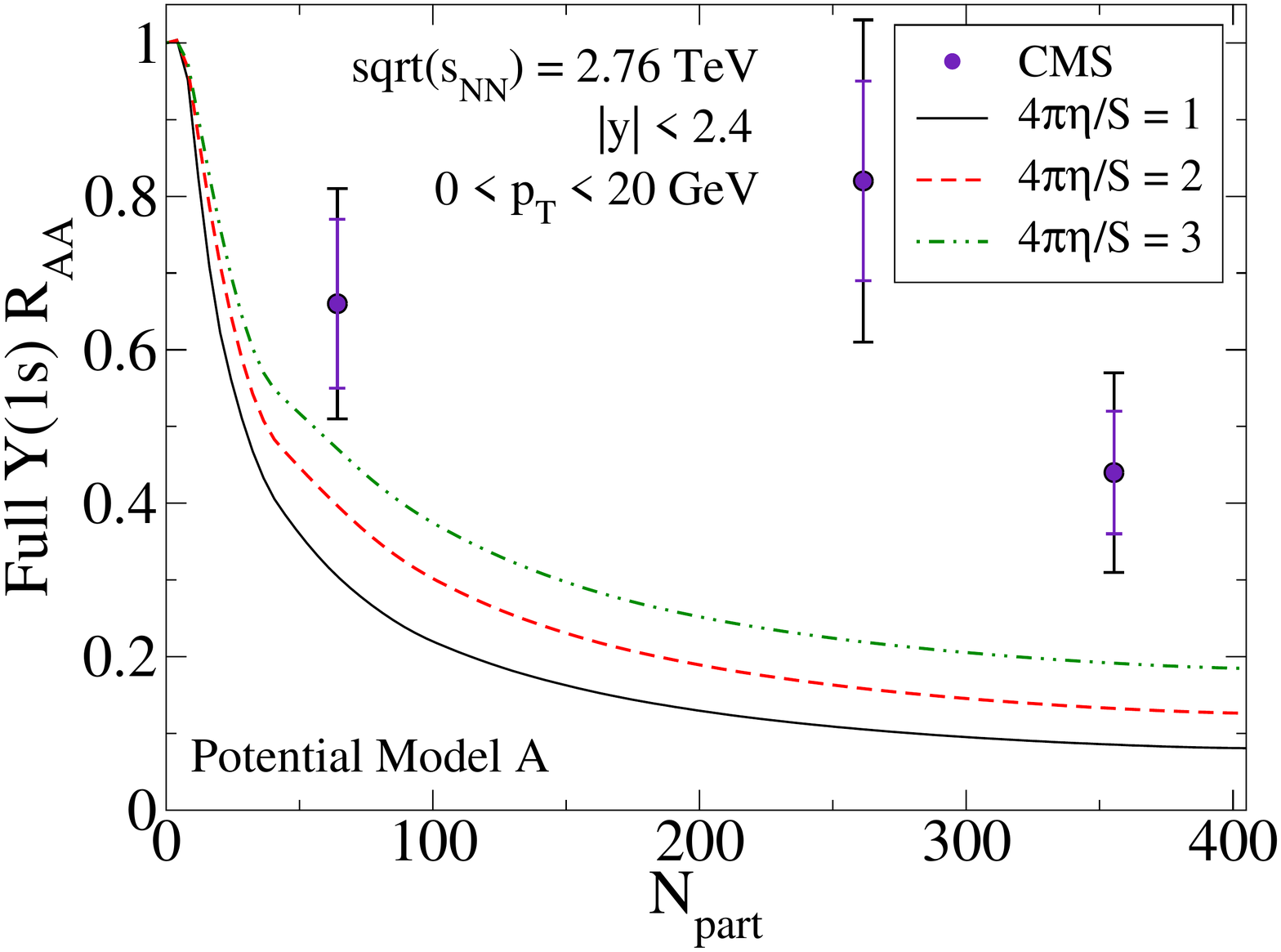}
\includegraphics[width=8cm]{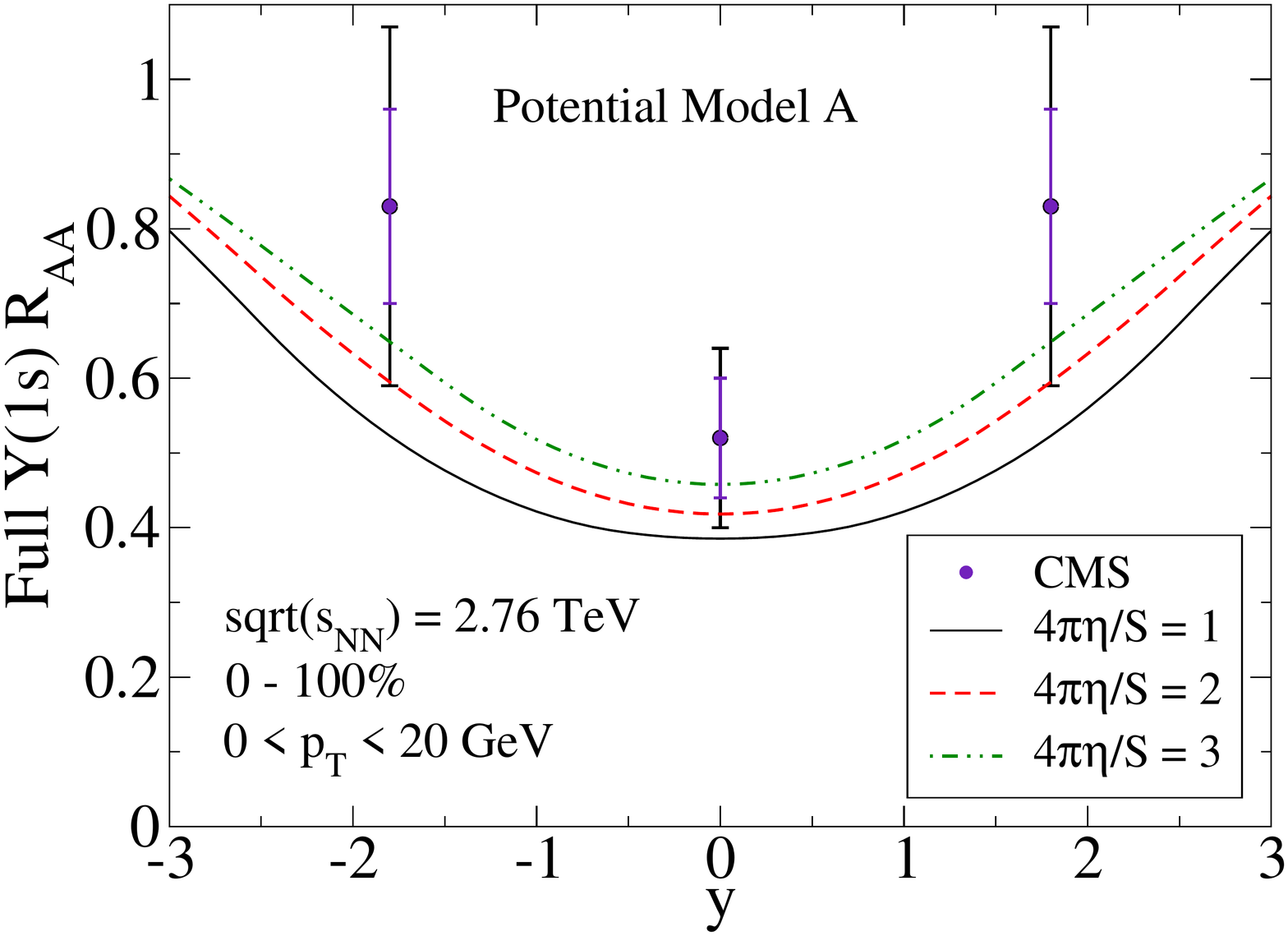}\\
\includegraphics[width=8cm]{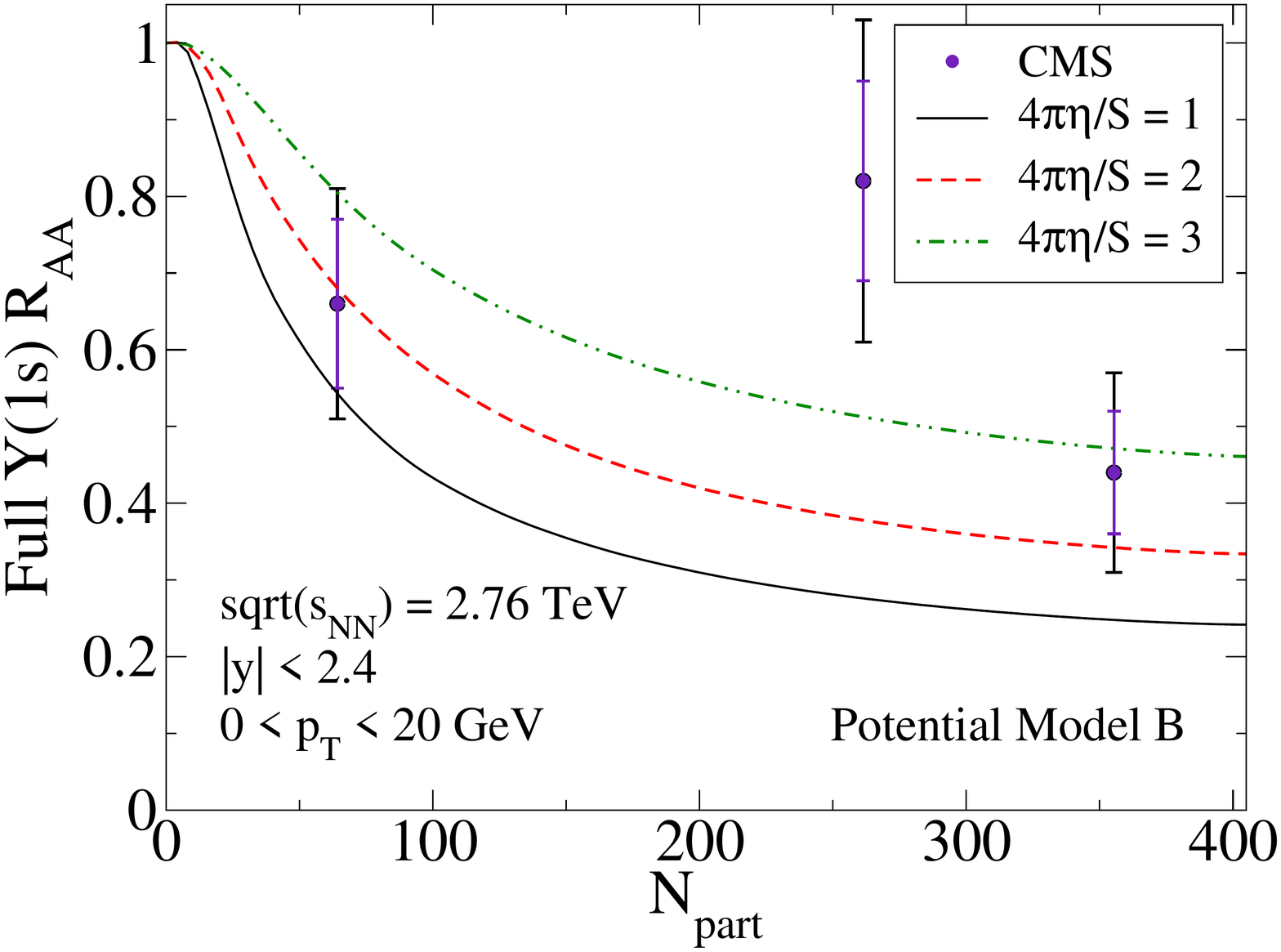}
\includegraphics[width=8cm]{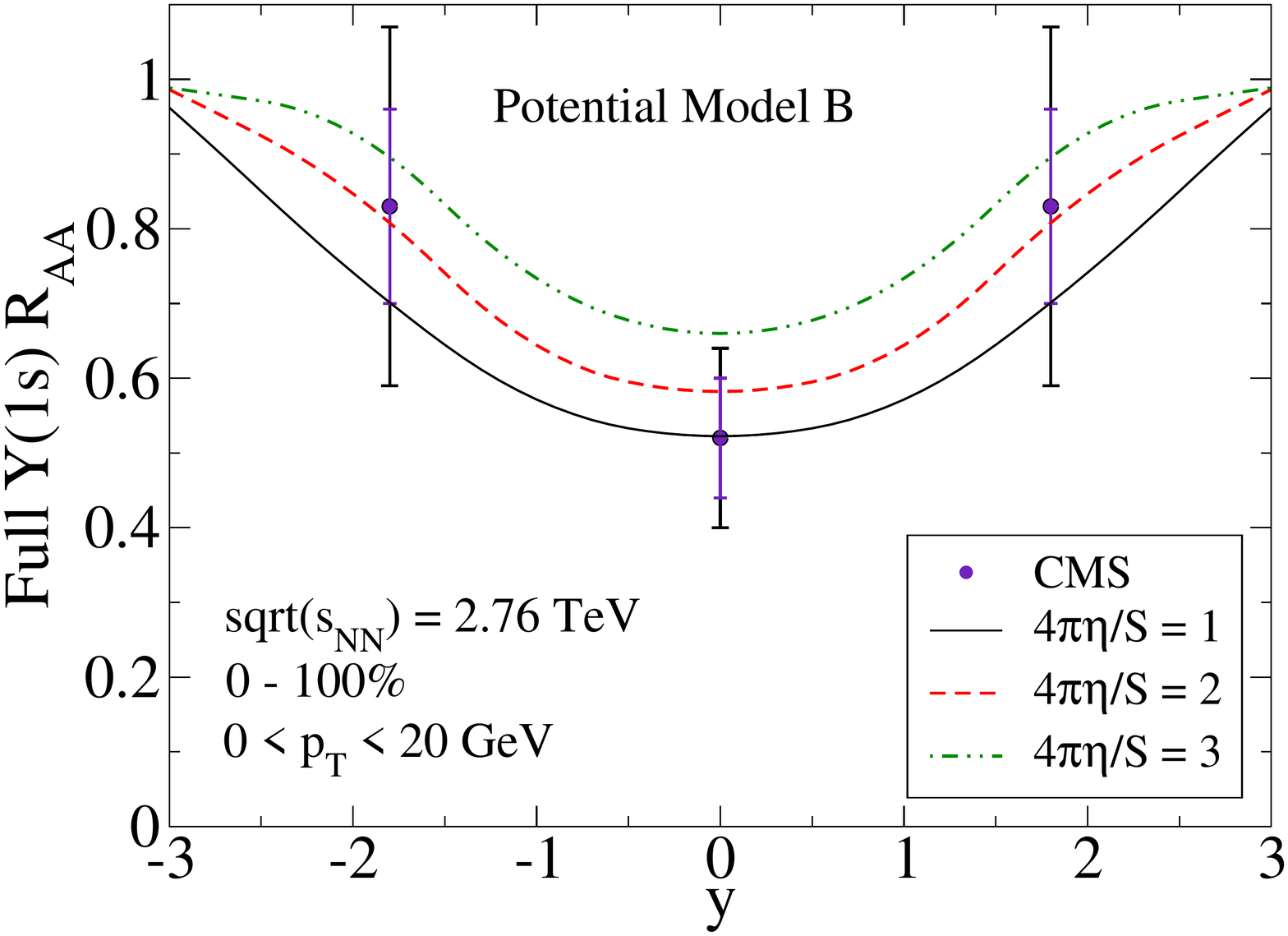}
\end{center}
\vspace{-6mm}
\caption{
LHC inclusive or ``full'' suppression factor $R_{AA}$ for the $\Upsilon(1s)$ including feed down effects compared to
experimental data are from the CMS Collaboration \cite{HIN-10-006}.  The
three different lines correspond to different assumptions for the shear viscosity to entropy ratio 
$4 \pi \eta/{\cal S} \in \{1,2,3\}$.  In all plots we used 
$\sqrt{s_{NN}} = 2.76$ TeV and  implemented cuts of $0 < p_T < 20$ GeV and (left) rapidity 
$|y| < 2.4$ (right) centrality 0-100\%.}
\label{fig:y1s-lhc}
\end{figure}

In Fig.~\ref{fig:raa-lhc-states} we show the predicted suppression factor $R_{AA}$ for the $\Upsilon(1s)$, $\Upsilon(2s)$, 
$\Upsilon(3s)$, $\chi_{b1}$, and $\chi_{b2}$ states as a function of the number of participants (left) and rapidity (right).
The top row uses potential model A  (\ref{eq:potmodela}) and the bottom row uses potential model B (\ref{eq:potmodelb}).  
In all plots we used  $\sqrt{s_{NN}} =  2.76$ TeV, assumed a shear viscosity to entropy density ratio of 
$4 \pi \eta/{\cal S} = 1$, and implemented cuts of $0 < p_T < 20$ GeV and (left) rapidity $|y| < 2.4$ (right) centrality 
0-100\%. As can be seen from this
figure, as was the case at RHIC energies, 
potential model A (\ref{eq:potmodela}) provides much more suppression than potential model B (\ref{eq:potmodelb}),
both as a function of number of participants and rapidity.  In both cases we see clear signs of sequential
suppression, with the higher excited states having stronger suppression than the ground state.  However, we once
again note
that even for states that are melted at relatively low central temperatures, we still obtain a non-vanishing suppression
factor for these states.  This is due to the fact that near the edges, where the temperature is lower, one does not
see suppression of the states.  Upon performing the geometrical average prescribed in Eq.~(\ref{eq:geoaverage}) we see
that a large fraction of the states produced can survive even when the central temperature of the plasma is above 
their naive dissociation temperature.

In Fig.~\ref{fig:y1s-lhc} we show the inclusive suppression factor $R_{AA}^{\rm full}[\Upsilon(1s)]$ obtained using
the feed down prescription presented in Section~\ref{sec:feeddown}.  As can be seen from these figures,
potential model A (free energy) predicts much stronger suppression than potential model B (internal energy).  
Comparing to the available CMS data \cite{HIN-10-006} we see that, as was the case at RHIC energies, potential
model B (internal energy) does a much better job of reproducing the data than potential model A (free energy)
both as a function of centrality and rapidity.  Using the potential model B results we can obtain an estimate for
$\eta/{\cal S}$ at LHC energies:  $0.08 < \eta/{\cal S} < 0.24$ which is the same range obtained from the STAR
data obtained with gold-gold collisions at lower energies.  As before, more precisely determining $\eta/{\cal S}$ 
will require more data from the LHC which should be forthcoming in the near future.

\begin{figure}[t]
\begin{center}
\includegraphics[width=10cm]{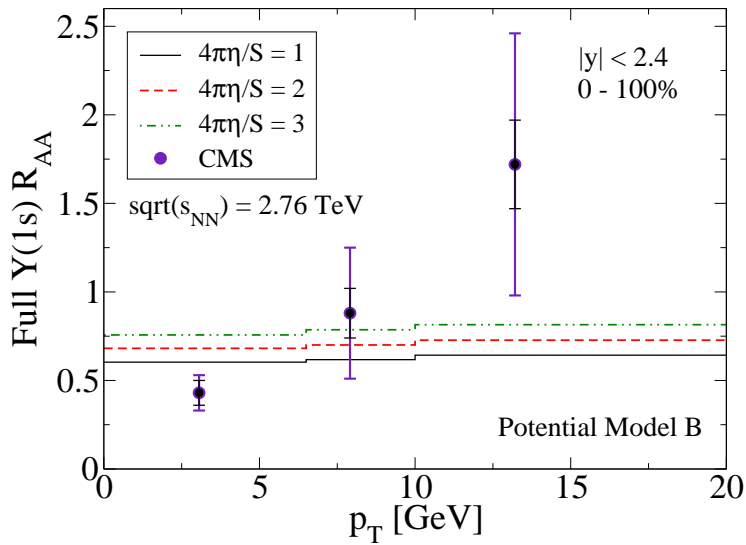}
\end{center}
\vspace{-6mm}
\caption{
LHC inclusive or ``full'' suppression factor $R_{AA}$ for the $\Upsilon(1s)$ including feed down effects as a function of 
transverse momentum compared to experimental data are from the CMS Collaboration \cite{HIN-10-006}.  The
three different lines correspond to different assumptions for the shear viscosity to entropy ratio 
$4 \pi \eta/{\cal S} \in \{1,2,3\}$.  For the plot we used 
$\sqrt{s_{NN}} = 2.76$ TeV and  implemented cuts of $0 < p_T < 20$ GeV, $|y| < 2.4$, and centrality 0-100\%.}
\label{fig:ptdep}
\end{figure}

\subsection{Transverse momentum dependence}
\label{sec:ptdep}

In Fig.~\ref{fig:ptdep} we plot the minimum bias (centrality 0-100\%) full suppression factor for the $\Upsilon(1s)$ including 
feed down effects as a function of transverse momentum.  Since we ignore the transverse expansion of the matter
created in the heavy ion collision, the only $p_T$ dependence which is included comes from the formation time 
effect.  One expects based on this that higher $p_T$ states will have weaker suppression since, in the lab frame,
they are formed at a later proper-time when the plasma is cooler.  This expectation is borne out by Fig.~\ref{fig:ptdep};
however, as can be seen from this figure there is only a weak $p_T$-dependence of the result.  This is to be 
contrasted with the relatively much larger $p_T$ dependence of the CMS results.  Looking forward we note that
there are two additional places where a momentum dependence could enter the final results:  (1) an intrinsic velocity
dependence of the damping rate itself and (2) the effect of heavy quark states being nearly free streaming in the
soft background.  It has been shown \cite{Escobedo:2011ie} that adding a finite velocity relative to the medium 
affects the heavy quark potential, so this would indeed be something that one will need to investigate in future work
and may help to improve agreement with the experimental data.  The second effect will require the simultaneous
solution of transport equations for nearly free streaming heavy quark states and the soft sector.

\begin{figure}[t]
\begin{center}
\includegraphics[width=8cm]{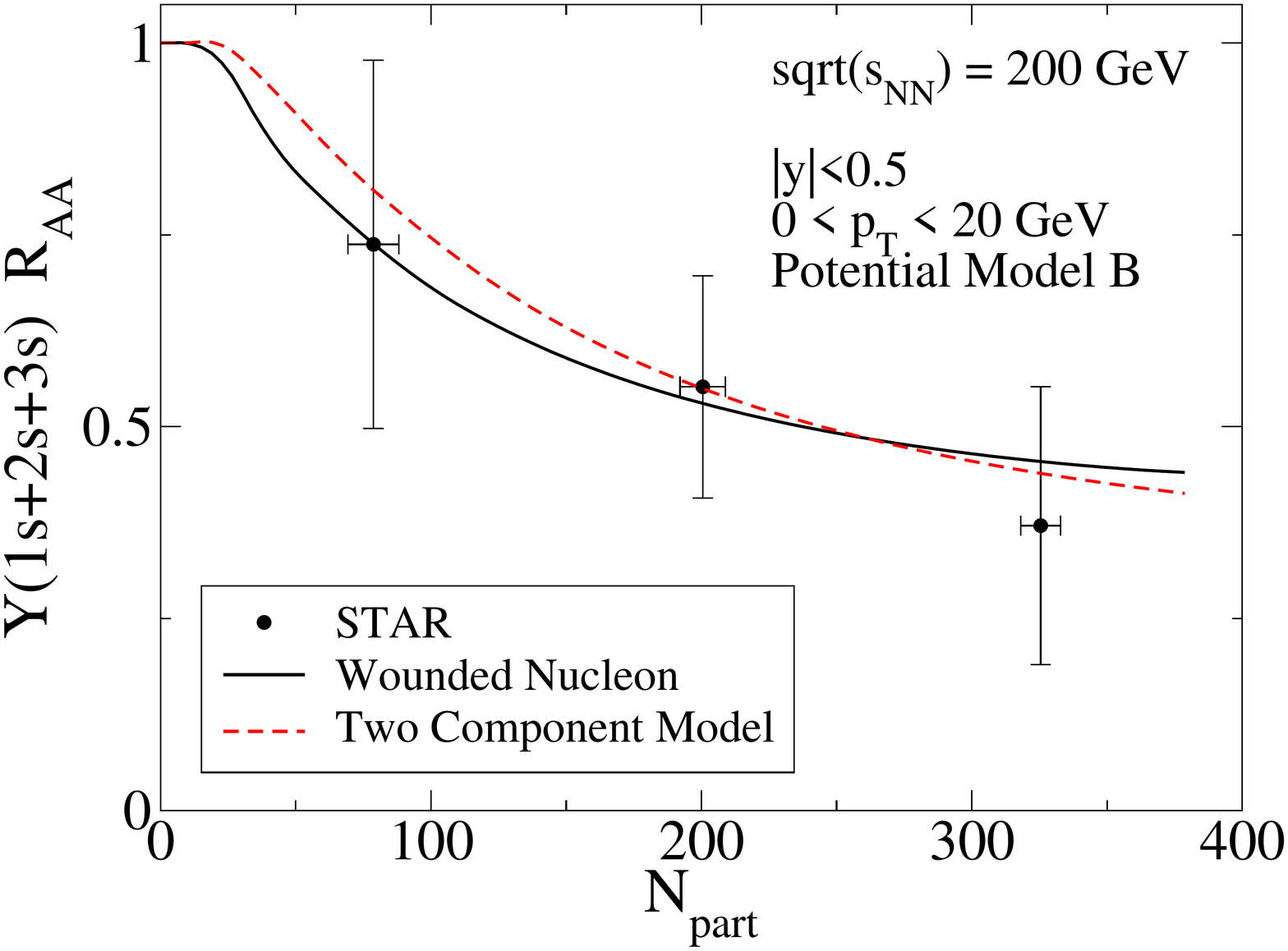}
\includegraphics[width=8cm]{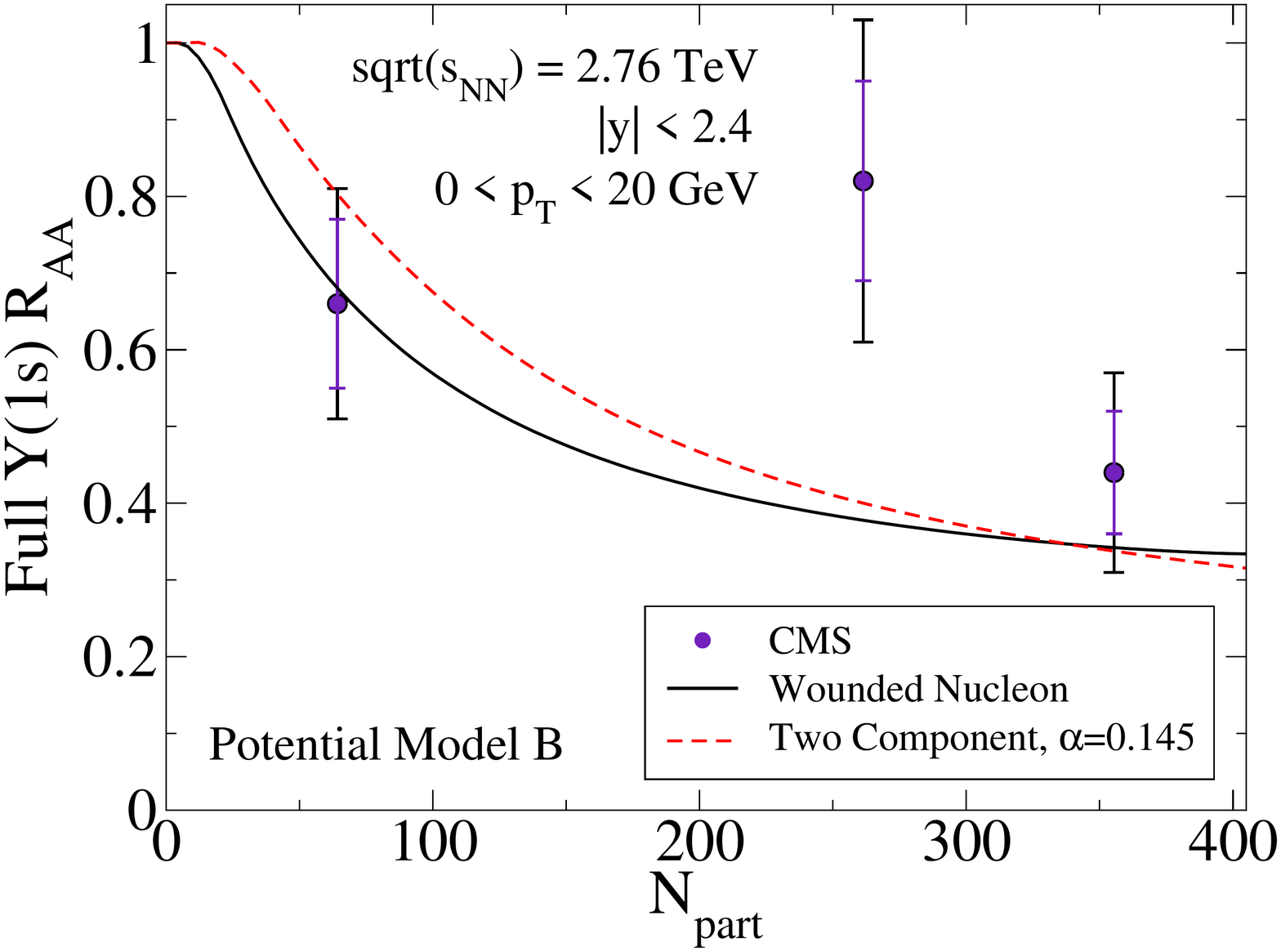}
\end{center}
\vspace{-6mm}
\caption{
RHIC (left) and LHC (right) inclusive suppression factor $R_{AA}$ for the $\Upsilon(1s)$ including feed down effects
compare to STAR \cite{Reed:2011fr} and CMS \cite{HIN-10-006} data.  
In both plots we have fixed $4 \pi \eta/S = 2$.  Collision energies and cuts applied
are indicated in each figure.  The solid black line is the result obtained assuming wounded nucleon initial conditions
and the dashed red line is the result obtained used a two component model with $\alpha=0.145$.}
\label{fig:icComp}
\end{figure}

\subsection{Dependence on the choice of initial condition type}
\label{sec:icvar}

As detailed in Section~\ref{sec:initialconditions} we consider two types of initial conditions:  (I) the wounded nucleon
model specified in Eq.~(\ref{eq:phardinit-I}) and (II) a two-component model which consists of an admixture of participant 
and binary scaling specified in Eq.~(\ref{eq:phardinit-II}).  In Fig.~\ref{fig:icComp} we show the results obtained
for $R_{AA}[\Upsilon(1s+2s+3s)]$ at RHIC energies and the full (or inclusive) $R_{AA}$ for the $\Upsilon(1s)$.
In both plots we have assumed $4 \pi \eta/S = 2$.
Because changing the initial condition type affects particle multiplicities we have adjusted the initial temperature
at RHIC energies from 433 MeV to 461 MeV and at LHC from 567 MeV to 612 MeV in order to keep the charged particle 
multiplicity fixed at $dN_{ch}/dy = 620$ and $dN_{ch}/dy = 1400$, respectively.  As can be seen from Fig.~\ref{fig:icComp},
for peripheral collisions there is a larger dependence on the choice of initial condition type, while for central collisions
the result obtained is not much affected by the choice of initial condition.  This is to be contrasted with the dependence
of the result on the assumed value of $\eta/{\cal S}$ which affects the suppression at all centralities.  This leaves hope
that one can disentangle the initial condition effect and the effect of the assumed value of $\eta/{\cal S}$.

\section{Conclusions and Outlook}
\label{sec:conclusions}

In this paper we considered the suppression of bottomonium states in ultrarelativistic heavy ion collisions.
We computed the suppression as a function of centrality, rapidity, and transverse momentum for 
the states $\Upsilon(1s)$, $\Upsilon(2s)$, $\Upsilon(3s)$, $\chi_{b1}$, and $\chi_{b2}$.  Using this information,
we then computed the inclusive $\Upsilon(1s)$ suppression as a function of centrality, rapidity, and transverse 
momentum including feed down effects.  Calculations were performed for both RHIC $\sqrt{s_{NN}} =$ 200 GeV 
Au-Au collisions and LHC $\sqrt{s_{NN}} =$ 2.76 TeV Pb-Pb collisions.  Our calculations build upon a
concerted theoretical effort to understand recently obtained RHIC and LHC data on bottomonium suppression 
\cite{Strickland:2011mw,Brezinski:2011ju,Song:2011nu,Emerick:2011xu,Song:2011ev}.

We studied two different potential models which were based on the heavy quark free energy (A)
and internal energy (B).  We found that the potential based on the free energy gives too much suppression
when compared to the available experimental data at both RHIC and LHC energies.  On the other hand, 
results obtained from
the potential model that was based on the internal energy seem to be in reasonably
good agreement with data obtained at both collision energies.  We are therefore led to conclude that one
should not use potential models based on the free energy.
From the comparison of our theoretical results obtained using the potential based on the internal energy and 
data available from the STAR and CMS Collaborations we were able to constrain the shear viscosity to entropy ratio 
to be in the range $0.08 < \eta/{\cal S} < 0.24$.  We find that our results are consistent with the creation of a high 
temperature quark-gluon plasma at both RHIC and LHC collision energies.  

That being said, it is worrisome that one sees such a strong dependence of the results on the potential model
used.
However, herein we find that at both RHIC and LHC energies a potential based on the internal energy seems
to better describe the available data with values for the shear viscosity to entropy ratio which are consistent
with those determined from bulk collective flow.
The dependence on the potential used emphasizes the need for a concerted theoretical effort to better determine 
the heavy quark potential analytically via finite temperature effective field theory methods and/or numerically via lattice 
QCD studies.  This will require determination of the both the real and imaginary parts of the potential at 
short and long distances and also the dependence on the momentum-space anisotropy of the plasma partons.
The calculation of the short range part of the potential for arbitrary momentum-space anisotropy is currently
underway. 

In future work we also plan to include the effect of allowing heavy quark states to have a flow which 
is decoupled from the soft medium and to include the effect of finite velocities on the heavy quark decay rate.
This will include the addition of full 3+1d {\sc aHydro} evolution so that we can simultaneously describe elliptic 
flow and bottomonium suppression.  It would also be interesting to investigate the behavior of heavy quarkonium
widths near $T_c$ using an AdS/QCD model.  
Finally, it will also be necessary to investigate the possibility of pair recombination due to residual spatial correlations among
suppressed pairs \cite{Young:2008he,Young:2009tj}.
How these future investigations will affect the quoted range for $\eta/{\cal S}$ is a critical open question which
will need to be addressed.
We leave these interesting questions for future work.

\section*{Acknowledgments}
D. Bazow and M. Strickland were supported by NSF grant No. PHY-1068765.
M. Strickland received additional support from the Helmholtz International Center for FAIR 
Landesoffensive zur Entwicklung Wissenschaftlich-\"Okonomischer Exzellenz program.

\bibliographystyle{utphys}
\bibliography{bottomonium}

\end{document}